\documentclass[10pt,aps,prl,amsmath,amssymb,amsfonts,nofootinbib,long]{revtex4}
\usepackage{epsfig,latexsym,bm}

\begin{document}

\title{Evolution of primordial magnetic fields in mean-field approximation}

\author{Leonardo Campanelli$^{1}$}
\email{leonardo.campanelli@ba.infn.it}
\affiliation{$^1$Dipartimento di Fisica, Universit\`{a} di Bari, I-70126 Bari, Italy}

\date{\today}


\begin{abstract}
We study the evolution of phase-transition-generated cosmic magnetic fields coupled to the
primeval cosmic plasma in turbulent and viscous free-streaming regimes. The evolution laws for the
magnetic energy density and correlation length, both in helical and non-helical cases, are
found by solving the autoinduction and Navier-Stokes
equations in mean-field approximation. Analytical results are
derived in Minkowski spacetime and then extended to the case of a Friedmann universe with
zero spatial curvature, both in radiation and matter dominated eras. The three possible
viscous free-streaming phases are characterized by a drag term in the Navier-Stokes equation which depends
on the free-streaming properties of neutrinos, photons, or hydrogen atoms, respectively.
In the case of non-helical magnetic fields,
the magnetic intensity $B$ and the magnetic correlation
length $\xi_B$ evolve asymptotically with the temperature $T$ as
$B(T) \simeq \kappa_B (N_i v_i)^{\varrho_1} (T/T_i)^{\varrho_2}$ and
$\xi_B(T) \simeq \kappa_\xi (N_i v_i)^{\varrho_3} (T/T_i)^{\varrho_4}$.
Here, $T_i$, $N_i$, and $v_i$ are, respectively, the temperature, the number of magnetic domains
per horizon length, and the bulk velocity at the onset of the particular regime. The coefficients $\kappa_B$,
$\kappa_\xi$, $\varrho_1$, $\varrho_2$, $\varrho_3$, and $\varrho_4$,
depend on the index of the assumed initial power-law magnetic spectrum, $p$, and on the particular regime,
with the order-one constants $\kappa_B$ and $\kappa_\xi$ depending also on the cut-off adopted
for the initial magnetic spectrum.
In the helical case, the quasi-conservation of the
magnetic helicity implies, apart from logarithmic corrections and a factor proportional to
the initial fractional helicity,
power-like evolution laws equal to those in the non-helical case, but with $p$ equal to zero.
\end{abstract}


\maketitle


\section{I. Introduction}

All observed galaxies and cluster of galaxies present microgauss, large-scale magnetic fields.
The origin of these cosmic magnetic fields has not yet been fully understood, although
several proposals, especially in the last years, have been put forward to explain
why our universe is magnetized (for reviews on cosmic magnetic fields,
see~\cite{Review1,Review2,Review3,Review5,Review6}).
These proposals belong to two distinct classes of generating mechanisms characterized, roughly speaking, by
the time when they operate.
Astrophysical mechanisms work on at the epoch of
large-scale structure formation or later, and can be based on the Biermann battery
effect~\cite{Biermann} in the first supernova remnants~\cite{Generation1a},
or on the Weibel instability~\cite{Weibel} in intergalactic plasmas~\cite{Generation1b}.
Mechanisms performing in the early universe can, on they part,
classified in mechanisms operating in the very early universe, namely
during an inflationary epoch of the universe~\cite{Generation2}, and mechanisms
operating after inflation~\cite{Harrison,GenerationA1,GenerationA2,GenerationA3},
for example during electroweak (QED) or quark-hadron (QCD) cosmological
phase transitions~\cite{Generation3}.

Magnetic fields generated during large-scale structure
formation through astrophysical processes usually have large
correlation scales but low intensities. Then, the precondition for
explaining the observed fields is an amplification of these seeds. In
principle, dynamo mechanisms operating in gravitationally bound
large-scale structures (such as galaxies and clusters of galaxies)
could substantially increase the intensity of such fields up to
observed values~\cite{Dynamo}. However, astrophysical mechanisms
cannot explain the presence of large-scale magnetic fields recently
detected in cosmic voids~\cite{Blazars}.

Inflation-produced magnetic fields have the potentiality to explain
the origin of cosmic magnetic fields since they possess a
large correlation.
The unpleasant feature of this class of mechanisms is that,
in order to obtain astrophysically interesting magnetic intensities, one has to
introduce some nonstandard (to wit, speculative) interaction term in
the photon field Lagrangian.
A possible exception is represented by a generating
mechanism first noticed in~\cite{Tsagas} and successively analyzed in the series of papers~\cite{Barrow et al}.
Here it is shown that,
starting from the standard Maxwell Lagrangian and assuming a small negative curvature
term (not present in all the other inflationary mechanisms),
a very strong magnetic field can emerge as the result of photons creation from
a varying gravitational field. The validity of the results
in theses works, however, has been put into question
in~\cite{Adamek et al} and~\cite{Shtanov-Sahni}. According to the authors of~\cite{Shtanov-Sahni} indeed,
the intensity of the produced field would be today, in the best case scenario,
as low as $B_0 \sim 10^{-59}$G and thus astrophysically unimportant.
Another exception (the last one to our knowledge) of the use of nonstandard physics for the
case of inflationary magnetic fields is represented by the recent work~\cite{Campanelli1} (see also~\cite{Campanelli1bis}).
Here, criticisms to all previous inflationary generating mechanisms for cosmic magnetic fields
have been drawn. The main criticism concerns the use of unrenormalized, and then unphysical,
vacuum magnetic fluctuations instead of the renormalized ones (the only ones which are physically acceptable).
In~\cite{Campanelli1}, it is shown that in standard Maxwell theory, the actual magnetic field resulting from
inflationary renormalized vacuum magnetic fluctuations is scale independent and has
an intensity which depends only on the scale of inflation. If this scale is around $10^{16}$GeV
(a plausible value for such a scale), that field directly accounts for galactic and galaxy cluster
magnetic fields.

Magnetic fields created during cosmological phase transitions
have small length scales. This is due to the fact that microphysical
processes which participate in the generation of the
magnetic field are necessarily uncorrelated on scales greater than the
Hubble scale which represents an (event) horizon for all
causal physical processes.
Nevertheless, the primordial universe could have been very turbulent at the time
of QED and QCD phase transitions. This means that any magnetic field coupled to the
primeval plasma could have been processed by magnetohydrodynamic (MHD) effects.
Indeed, it is well-known in the literature that turbulence effects increase the
typical magnetic length scale~\cite{Biskamp}. Moreover, if
the magnetic field is helical the growth of magnetic correlation
could be much more significant.
It is interesting to observe that if the large-scale magnetic fields we observe today were
relic helical fields from cosmological phase transitions (or inflation), then
this would be a manifestation of a macroscopic and primeval $P$ and $CP$ violation,
since magnetic helicity is odd under discrete $P$ and $CP$ transformations.
\footnote{Helical magnetic fields are indeed characterized by the fact of
possessing an asymmetry between the number of left-handed and right-handed photon helicity
states (see, e.g.,~\cite{Campanelli2}).}
Moreover, models for generating helical magnetic
fields in the early universe there exist in the literature~\cite{Campanelli2,Generation4}.

It is worth noting that magnetic fields produced in the early universe are
subject to a variety of constraints coming from the fact that
the presence of a primordial magnetic field could spoil the predictions of
the standard cosmological model. The most important limits
came from the study of Big Bang Nucleosynthesis~\cite{B-BBN} and
Cosmic Microwave Background (CMB) anisotropies~\cite{Durrer-Giovannini,Giovannini,B-CMB}.
In particular, large-scale, primordial helical magnetic fields
could leave peculiar signatures in CMB radiation~\cite{CMBH},
since they would introduce a parity-odd cross-correlations of the CMB anisotropies,
not present in the case of a non-helical magnetic field.

Also, primordial helical magnetic fields could affect
the phenomenology of axions since the latter are coupled, through the axion-photon interaction term,
to external magnetic fields.
Indeed, the generation of helical magnetic fields
before the axion coherent oscillations start at a temperature around $few$ GeV, could be in contradiction
with the existence of the axion~\cite{Campanelli3}, since a strong helicity production
would in turn produce too much of an axion relic
abundance, in disagreement with astrophysical observations.

Finally, it is interesting to observe that the presence of large-scale helical magnetic fields
in the actual universe could be directly detected by analyzing the propagation properties
of charged cosmic rays, as shown in~\cite{Kahniashvili1}.

Magnetic fields generated during inflation or cosmological phase transitions can
evolve as the universe expands, since their properties
can be modified by magnetohydrodynamic turbulent effects due to the coupling with
the primeval cosmic plasma.
As recently shown in~\cite{Kahniashvili2}, typical inflation-produced magnetic fields
remain almost unchanged on scales of cosmological interest even in the presence of a turbulent
plasma (as in the case, for example, of a cosmic phase transition)
although on small scales, which are however astrophysically uninteresting, their power gets
gradually suppressed.

The fact that the universe during a cosmological phase transition could be very turbulent
was first realized in~\cite{Brandenburg}, where the importance of MHD turbulent effects
on the evolution of phase-transition-produced magnetic fields was stressed.
Here, in order to study the correlation properties of an evolving cosmic magnetic field, however,
a simple toy model was employed, which replaced the full MHD equations. Since then,
there are been many efforts to study, more accurately, the evolution of turbulent magnetic
fields~\cite{Olesen,Brandenburg1,Shiromizu,Son,Biskamp1,Christensson1,Brandenburg2,Sigl,Kalelkar,Verma,Campanelli4,
Yousef,Christensson2,Campanelli5,Kahniashvili3,Tevzadze,Kahniashvili4}.

However, only in the work~\cite{Jedamzik}, it was realized that, other than
turbulent phases, phase-transition-generated magnetic fields can experience
dissipation processes induced by free-streaming neutrinos, photons, or hydrogen atoms.
Indeed, such cosmic components can have mean free paths which exceed the typical correlation scale
of a primordial magnetic field, so they can be considered as free-streaming species with respect
to the magnetic field itself. In this case, the dissipation of magnetic energy is ruled
by a ``drag'' term in the Navier-Stokes equation, instead of the usual diffusion term
proportional to the kinematic viscosity.

Such a possible phase in MHD, called ``dragged magnetohydrodynamics''~\cite{Banerjee1,Banerjee2},
has been fully analyzed in~\cite{Banerjee1,Banerjee2}, where a systematic study of the evolution
of cosmic magnetic fields has been performed by direct numerical integration
of MHD equations. To our knowledge, numerical simulations of dragged MHD
have so far been performed only in~\cite{Banerjee1,Banerjee2} where, moreover,
a justification of the results has been given by using simple scaling arguments.

The aim of this paper is to analyze the evolution of phase-transition-generated magnetic
fields, in turbulent and viscous free-streaming regimes, by using suitable
mean-field-approximated magnetohydrodynamic equations.

In particular, we will derive the evolution laws
for the magnetic energy and correlation length both in non-helical and
helical cases. We anticipate that our results agree with the numerical results in~\cite{Banerjee1,Banerjee2}.

The paper is organized as follows.
In the next section, we introduce the relevant equations
and the physical quantities of dragged MHD. Also, we find some new exact results in dragged MHD,
and we introduce the transformations
connecting the evolution of a primordial magnetic field in a (static) Minkowski spacetime to that in a
flat Friedmann (expanding) universe.
In section III, we apply some rigorous scaling arguments, first introduced by Olesen
for the study of turbulent MHD, to dragged MHD.
In section IV, we find the equations governing the evolution of the magnetic energy and helicity
by using three different mean-field-theory approaches.
In section V, we solve these equations for some specific initial conditions, and find the
evolution laws of the magnetic intensity and correlation length in radiation dominated universe
for the case of free-streaming neutrinos and photons.
In section VI, the case of free-streaming photons and hydrogen atoms in matter dominated universe
is analyzed.
In section VII, we justify some scaling arguments of~\cite{Banerjee1,Banerjee2} in the light of our exact
analytical results obtained in section II.
In section VIII, we discuss the evolution of freely-decaying turbulent magnetic fields,
comparing the various and often different
results present in the literature. Also, we complete our work started in~\cite{Campanelli5}
by extending our previous results, obtained in Minkowski spacetime, to the case of a flat Friedmann universe.
In section IX, we study the dependence of our results on the choice of the initial cut-off of the
assumed, initial magnetic power spectrum.
In section X, we discuss our results and compare them to the results of~\cite{Banerjee1,Banerjee2}.
In section XI, we draw our conclusions and give possible suggestions for further investigations.
Finally, appendixes A,B, and C contain some technical details of
our analytical computations.


\section{II. Dragged magnetohydrodynamics}

In this section, we introduce the equations and physical quantities of interest for dragged
magnetohydrodynamics. Some new exact results are derived,
some scaling arguments are discussed, and the transformations
relating the evolution of a magnetic field in Minkowski and Friedmann spacetimes are introduced.

\subsection{IIa. Preliminaries}

{\it Full magnetohydrodynamic equations}.-- The full magnetohydrodynamic equations in an expanding
Friedmann universe for a relativistic imperfect fluid
were first derived by Jedamzik, Katalini\'{c}, and Olinto in~\cite{Jedamzik}.
A full analysis of these equations would require the inclusion of effects deriving from inhomogeneities
in the matter density, pressure field, particle number, viscosity, temperature, etc.,
those coming from compressibility of the plasma flow,
as well as the study of relativistic effects in the case of ultra-relativistic fluid motion.

Inhomogeneities in the magnetohydrodynamic variables
generally leads to magnetohydrodynamic modes (e.g., Alfv\'{e}n waves, slow and fast magnetosonic waves~\cite{Jedamzik})
which propagates through the magnetized plasma, and their decay can result in a
dissipation of magnetic energy stored in the small-scale fluctuating part of a cosmic magnetic field.

Compressibility of plasma flow is generally expected to be realized in a cosmological
context. For example, a recent analysis by Giovannini~\cite{Giovannini1}
has argued that, due to adiabatic inhomogeneities of the scalar curvature,
the cosmic plasma flow at large scales could be compressible 
in the period of time between electron-positron annihilation around $T_{e^{+}e^{-}} \! \sim 1$MeV~\cite{Kolb}
and last scattering at $T \sim 0.3$eV~\cite{Kolb}.

Also, ultra-relativistic fluid motion can develop
during radiation era and could play, in principle, in important role in the evolution
of cosmic magnetic fields during radiation era.

Nevertheless, in the following and due to the complexity of the full MHD equations,
we limit ourselves to the study of a simplified version of magnetohydrodynamic
equations. We will work in the hypothesis where the small-fluctuating parts
of MHD variables do not propagate (thus neglecting the development and the damping of the previously
indicated MHD modes), as well as in the limit of incompressibility of the cosmic plasma
and small velocity fields ($v/c \ll 1$, where $v$ is the typical
velocity of bulk fluid motion and $c$ is the speed of light).
The first approximation implies that the decay laws of a cosmic magnetic field we will find
below must be considered as conservative, in the sense that a (slightly) faster magnetic energy
decay could happen as a result of the inclusion in the analysis of propagating MHD modes.
The validity of the second approximation, as discussed in~\cite{Banerjee2},
depends on the initial magnetic field intensity as well as on the epoch considered.
A general expectation is that very strong magnetic fields
render compressible the fluid motion after the decoupling of photons from the cosmic flow.
In the analysis of~\cite{Banerjee2}, however, it is shown that incompressibility is justified
for comoving magnetic fields intensities below the value $6 \times 10^{-11}$G.

Having pointing out the possible limitations of our simplified analysis,
it is worth noticing that we will use the same set of approximate MHD equations used
in~\cite{Banerjee1,Banerjee2} to derive, numerically, the evolution properties
of a cosmic magnetic field.

{\it Reduced magnetohydrodynamic equations}.-- The ``reduced'', incompressible,
\footnote{The kinetic energy associated to a turbulent
fluid is $E_v = (1/2) \rho \, v^2$, where $\rho$ is
the energy density of the universe, and $v$ is the typical
velocity of bulk fluid motion associated to turbulence. Incompressibility
of the primordial plasma
means that the acoustic Mach number, $\text{M}_{\rm s} =
v/v_{\rm s}$, where $v_{\rm s}$ is the speed of sound,
is much less then unity~\cite{Biskamp}. For example, in the
radiation era, when the cosmic fluid is ultra-relativistic and
then $v_{\rm s} = 1/\sqrt{3}$, the incompressibility condition
reads $v \ll 1/\sqrt{3}$.}
Newtonian magnetohydrodynamic equations in the
presence of a drag term and in Minkowski spacetime are~\cite{Banerjee1,Banerjee2}
\begin{eqnarray}
\label{Eq1} && \frac{{\text D}{\textbf v}}{{\text d}t} = (\nabla \times {\textbf v}_A) \times {\textbf v}_A - \alpha {\textbf v}, \\
\label{Eq2} && \frac{\partial {\textbf v}_A}{\partial t}  = \nabla \times
({\textbf v} \times {\textbf v}_A) + \eta \nabla^2 {\textbf v}_A,
\end{eqnarray}
where ${\text D}/{\text dt} = \partial_t + {\textbf v} \cdot \nabla$ is the so-called hydrodynamical derivative,
${\textbf v}$ is the velocity of bulk fluid motion and $\nabla \cdot {\textbf v} = 0$,
%
%
${\textbf v}_A = {\textbf B}/\sqrt{\rho + P}$
is the Alfv\'{e}n velocity, ${\textbf B}$ is the magnetic field, and
$\rho$ and $P$ are the energy density and the thermal pressure of the
fluid, respectively. The frictional coefficient $\alpha$ and the resistivity
$\eta$ are dissipative parameters, and are determined by microscopic physics.

We will consider throughout the paper the case of constant $\rho$ and $P$.
In the case of a Friedmann universe, where $\rho$ and $P$ evolve in time,
the quantities of interest, as we will see in section IIb, are $\tilde{\rho}$ and $\tilde{P}$,
instead of $\rho$ and $P$, with the ``tilded'' quantities being time independent.
For this reason, and for the sake of simplicity, we can take $\rho + P = 1$ in Eq.~(\ref{Eq1}),
yielding ${\textbf v}_A = {\textbf B}$.

The Navier-Stokes equation~(\ref{Eq1}) and the autoinduction equation~(\ref{Eq2})
can then be written, respectively, as
\begin{eqnarray}
\label{Eq1bis} && \frac{{\text D}{\textbf v}}{{\text d}t} = {\textbf F}_L - {\textbf F}_d, \\
\label{Eq2bis} && \frac{\partial {\textbf B}}{\partial t}  = \nabla \times
({\textbf v} \times {\textbf B}) + \eta \nabla^2 {\textbf B},
\end{eqnarray}
where we have introduced the Lorentz force ${\textbf F}_L = {\textbf J} \times {\textbf B}$,
%
%
and the damping force ${\textbf F}_d = \alpha {\textbf v}$,
%
%
with ${\textbf J} = \nabla \times {\textbf B}$ being the magnetic current.
Let us now introduce the relevant physical quantities in (dragged) MHD.
The magnetic and kinetic energy densities are defined as
\begin{eqnarray}
\label{Eq3} && E_B(t) = \frac{1}{2} \int \! d^{\,3} x \,
{\textbf B}^2({\textbf x},t) = \int_{0}^{\infty} \!\!\! dk \,
{\mathcal E}_B(k,t),
\\
\label{Eq4} && E_v(t) = \frac{1}{2} \int \! d^{\,3} x \,
{\textbf v}^2({\textbf x},t) = \int_{0}^{\infty} \!\!\! dk \,
{\mathcal E}_v(k,t),
\end{eqnarray}
where ${\mathcal E}_B = 2\pi k^2 \, |{\textbf B}({\textbf k})|^2$
and ${\mathcal E}_v = 2\pi k^2 \, |{\textbf v}({\textbf k})|^2$
are the magnetic and kinetic energy density spectra. Here,
${\textbf B}({\textbf k})$ and ${\textbf v}({\textbf k})$ are the
magnetic and kinetic fields in Fourier space, with ${\textbf k}$
being the wavenumber and $k = |{\textbf k}|$.
The magnetic and cross helicity densities are
\begin{eqnarray}
\label{Eq5} && H_B(t) = \int\! d^{\,3}x \, {\textbf A}
\cdot {\textbf B} = \int_{0}^{\infty} \!\!\! dk \, {\mathcal
H}_B(k,t),
\\
\label{Eq6} && H_c(t) = \int\! d^{\,3}x \, {\textbf v}
\cdot {\textbf B} = \int_{0}^{\infty} \!\!\! dk \, {\mathcal
H}_c(k,t),
\end{eqnarray}
respectively, where ${\mathcal H}_B = 4\pi k^2 {\textbf A}({\textbf k}) \! \cdot
\! {\textbf B}^*({\textbf k})$ and ${\mathcal H}_c = 4\pi k^2
{\textbf v}({\textbf k}) \! \cdot \! {\textbf B}^*({\textbf k})$
are the magnetic and cross helicity density spectra, ${\textbf A}$ is the
vector potential, and an asterisk denotes complex conjugation.
It is worth noticing that, for all magnetic and kinetic field
configurations, the magnetic and cross helicity spectra
satisfy the following realizability conditions:
$|{\mathcal H}_B| \leq 2k^{-1} {\mathcal E}_B$ and
$|{\mathcal H}_c| \leq 2 ({\mathcal E}_v {\mathcal E}_B)^{1/2}$.
The total energy $E_{\rm tot} = E_B + E_v$ and the cross-helicity are
conserved quantities when $\alpha = \eta = 0$, while the magnetic
helicity is conserved when $\eta = 0$. This follows directly
from their evolution laws, which can be straightforwardly derived
from the MHD equations~(\ref{Eq1bis}) and (\ref{Eq2bis}). They are
\begin{eqnarray}
\label{Eq7} && \frac{\partial{E}_{\rm tot}}{\partial t} = -2 \alpha E_v - \eta \!
\int \! d^{\,3}x \, \textbf{J}^2,
\\
\label{Eq8} && \frac{\partial {H}_c}{\partial t} = -\alpha H_c - \eta \! \int \!
d^{\,3} x \, {\textbf J} \cdot {\boldsymbol \omega},
\\
\label{Eq9} && \frac{\partial {H}_B}{\partial t} = -2\eta \! \int \! d^{\,3} x \,
{\textbf J} \cdot {\textbf B},
\end{eqnarray}
where ${\boldsymbol \omega} = \nabla \times \textbf{v}$ is the so-called vorticity.
We also introduce the relevant length scales in (dragged) MHD,
i.e., the so-called correlation lengths, which are the
characteristic lengths associated with the large magnetic and
kinetic energy eddies of turbulence. These are defined by
\begin{eqnarray}
\label{Eq10} && \xi_B(t) =
\frac{\int_0^{\infty} \! dk k^{-1} {\mathcal E}_B(k)}{\int_0^{\infty} \! dk \, {\mathcal E}_B(k)} \, ,
\\
\label{Eq11} && \xi_v(t) =
\frac{\int_0^{\infty} \! dk k^{-1} {\mathcal E}_v(k)}{\int_0^{\infty} \! dk \, {\mathcal E}_v(k)} \, .
\end{eqnarray}
With the aid of the magnetic and kinetic correlation lengths, it is
straightforward to transform the above ``local'' realizability conditions on the
magnetic and cross helicity spectra, into the following ``global'' realizability conditions:
$|H_B| \leq 2 \xi_B E_B$ and
$|H_c| \leq \min [\, E_{\rm tot},2 \int \! dk \, ({\mathcal E}_v {\mathcal E}_B)^{1/2} \,]$.
Finally, it is useful to define the kinetic Reynolds number, the magnetic Reynolds number,
and the Prandtl number as
${\text{Re}} = v/(l \alpha)$, ${\text{Re}}_B = vl/\eta$, and
${\text{Pr}} =  {\text{Re}}_B/{\text{Re}}$,
%
%
respectively, where $v$ and $l$ are the typical velocity and length scale of the fluid
motion.

\subsection{IIb. The equilibrium state}

A dragged phase develops when the kinetic Reynolds number is small~\cite{Banerjee1,Banerjee2}.
For this reason, we work in the case where ${\text{Re}} \ll 1$. Defining the ``drag time''
$\tau_d = \alpha^{-1}$,
%
%
we have $|{\text D}_t{\textbf v}|/|{\textbf F}_d| \sim
\tau_d / \tau_{\rm eddy} = {\text{Re}}$, where we used $\partial_t \sim 1/\tau_{\rm eddy}$ and
introduced the so-called (kinetic) eddy turnover time
$\tau_{\rm eddy} = l/v$.
%
%
Accordingly, for very low kinetic Reynolds numbers, we can
neglect the left-hand side of Navier-Stokes equation~(\ref{Eq1bis}).
This means that the system will approach asymptotically a state
where the Lorentz force equilibrates the damping force~\cite{Banerjee1,Banerjee2},
\begin{equation}
\label{Eq15} {\textbf F}_L = {\textbf F}_d \, .
\end{equation}
In this ``equilibrium state'', the autoinduction equation~(\ref{Eq2bis}) reduces to
an equation which depends only on the magnetic field,
\begin{equation}
\label{Eq16} \frac{\partial {\textbf B}}{\partial t} = \tau_d \, \nabla \times
\left[ \, {\textbf B} \times \left({\textbf B} \times \nabla \times {\textbf B} \right)\right]
+ \eta \nabla^2 {\textbf B},
\end{equation}
and will be basis for the study of the evolution of the magnetic field itself.

Multiplying both sides of Eq.~(\ref{Eq15}) by ${\textbf B}$, we find that
the cross helicity is zero in the equilibrium state,
\begin{equation}
\label{Eq14bis} H_c(t) = 0.
\end{equation}
Also, there is no equipartition between kinetic and
magnetic energy in this state. Indeed, defining the $\Gamma$-ratio
\begin{equation}
\label{Gamma} \Gamma = \frac{E_v}{E_B} \, ,
\end{equation}
we get
\begin{equation}
\label{GammaRe} \Gamma \sim \text{Re} \ll 1,
\end{equation}
since $\Gamma \sim v^2/B^2 \sim {\text{Re}} \: |{\textbf F}_L|/|{\textbf F}_d| = {\text{Re}}$,
where $B$ is the typical intensity of the magnetic field.
Moreover, $\tau_d$ and $\tau_{\rm eddy}$ are related by
\begin{equation}
\label{nice} \tau_d \sim \Gamma \tau_{\rm eddy},
\end{equation}
since $1 \simeq |{\textbf F}_d|/|{\textbf F}_L| \sim \Gamma \tau_{\rm eddy}/\tau_d \, $.
Let us observe that $|\dot{E}_v|/|\dot{E}_B| \sim {\text{Re}}^{3/2}$,
$|\alpha E_v|/|\dot{E}_B| \sim {\text{Re}}^{1/2}$, and
$|\eta \! \int \! d^{\,3}x \, \textbf{J}^2|/|\dot{E}_B| \sim {\text{Re}}^{1/2}
{\text{Re}}_B^{-1}$, where a dot indicates a time derivative. Consequently,
in the limit of low kinetic Reynolds number and large magnetic Reynolds number,
Eq.~(\ref{Eq7}) can be approximated with
\begin{equation}
\label{Eq12} \frac{\partial {E}_B}{\partial t} \simeq -2\alpha E_v.
\end{equation}
The above equation connects the evolutions of magnetic and kinetic energies,
and will be useful in the following.
Finally, since we are considering a very large Prandtl number, we can
neglect the second term in the right-hand side of Eq.~(\ref{Eq8}).
In fact, we have $|\eta \int \! d^{\,3} x \, {\textbf J} \cdot {\boldsymbol \omega}|/|\alpha H_c| \sim 1/{\text{Pr}}$.
Accordingly, we get $\dot{H}_c \simeq - \alpha H_c$, whose solution
\begin{equation}
\label{Eq13bis} H_c(t) = H_c(t_i) \exp \!\! \left[ -\! \int_{t_i}^t \! dt' \alpha(t') \right]
\end{equation}
gives the evolution law of the cross helicity in the equilibrium state.
In the next subsection, we show that the above equation is indeed (asymptotically) compatible
with Eq.~(\ref{Eq14bis}).

\subsection{IIc. Scaling arguments}

Let us suppose, and this is indeed the case in a cosmological context~\cite{Banerjee1,Banerjee2},
that the drag coefficient $\alpha$ evolves in time as a simple power law,
\begin{equation}
\label{s1} \alpha(t) = \alpha(t_i) \left(\frac{t}{t_i} \right)^{\!\!a},
\end{equation}
where $t_i$ is a reference time.

In numerical simulation of dragged magnetohydrodynamics~\cite{Banerjee1,Banerjee2},
it is observed that the magnetic energy scales in time as a power law
if the initial magnetic spectrum is assumed to be a simple power of the
wavenumber. Let us then parameterize the magnetic energy as
\begin{equation}
\label{s2} E_B(t) = E_B(t_i) \left(\frac{t}{t_i} \right)^{\!-\beta}.
\end{equation}
Inserting Eqs.~(\ref{s1}) and (\ref{s2}) into Eq.~(\ref{Eq12}), we find
the evolution laws for the kinetic energy and $\Gamma$-ratio,
\begin{eqnarray}
\label{s3} && E_v(t) = E_v(t_i) \left(\frac{t}{t_i} \right)^{\!-(1 + a + \beta)},  \\
\label{s4} && \Gamma(t) = \Gamma(t_i) \left(\frac{t}{t_i} \right)^{\!-(1+a)},
\end{eqnarray}
respectively, where $\Gamma(t_i) = (\beta/2) [\tau_d(t_i)/t_i]$. From this last relation we
get $\beta > 0$, which means that the magnetic energy decreases in time during a dragged MHD
phase (as in the case of turbulent MHD).

Moreover, recalling that $\Gamma \sim {\text{Re}} = v/\alpha l$ and
taking $v \sim E_v^{1/2}$, we find that the typical length scale evolves in time as
\begin{equation}
\label{s5} l(t) \propto t^{(1 - a - \beta)/2},
\end{equation}
giving, in turn, the evolution law
\begin{equation}
\label{s6} \tau_{\rm eddy}(t) \propto t
\end{equation}
for the eddy turnover time.
%
Finally, inserting Eq.~(\ref{s1}) in Eq.~(\ref{Eq13bis}), we get
\begin{equation}
\label{Eq14} H_c(t) = H_c(t_i)
\exp \! \left\{ -\frac{t_i}{\tau_d(t_i)} \, \frac{1}{1+a} \left[ \left( \frac{t}{t_i} \right)^{\! 1+a} - 1 \right] \! \right\}
\end{equation}
if $a \neq -1$, and
\begin{equation}
\label{Eq14tris} H_c(t) = H_c(t_i) \left( \frac{t}{t_i} \right)^{-t_i/\tau_d(t_i)}
\end{equation}
if $a = - 1$. In both cases, since $t_i \sim \tau_{\rm eddy}(t_i) \gg \tau_d(t_i)$ [see Eqs.~(\ref{GammaRe}) and (\ref{nice})],
we get that the cross helicity goes to zero asymptotically ($t \gg t_i$), in agreement with Eq.~(\ref{Eq14bis}).

It is amazing to observe that, even without knowing the value of the exponent $\beta$,
simple scaling arguments, when combined with the exact laws~(\ref{Eq12}) and (\ref{Eq13bis}),
give the evolution laws for the $\Gamma$-ratio, the eddy turnover time, and the cross helicity.

\subsection{IId. Expanding universe}

In the case of an expanding universe (with zero spatial curvature) described by a Friedmann-Robertson-Walker
line element, 
it has been shown that the dragged MHD equations are the same as the Eqs.~(\ref{Eq1}) and (\ref{Eq2}) provided that time,
coordinates and dynamical variables are replaced by suitable quantities. The form of these new ``tilded'' quantities
is different in radiation an matter eras.

{\it Radiation era: comoving variables}.-- In a radiation-dominated
universe the tilded quantities are~\cite{Banerjee1,Banerjee2}:
\begin{eqnarray}
\label{Ex1}  && t \rightarrow \tilde{t} = \int \! R^{-1} dt, \;\;\;\;
                {\textbf x} \rightarrow \tilde{{\textbf x}} = R {\textbf x}, \\
\label{Ex3}  && {\textbf B} \rightarrow \tilde{{\textbf B}} = R^2 {\textbf B}, \;\;\;\;\;\;\,
                {\textbf v} \rightarrow \tilde{{\textbf v}} = {\textbf v}, \\
\label{Exx1} && \rho \rightarrow \tilde{\rho} = R^4 \rho, \;\;\;\;\;\;\;\;\;\,
                P \rightarrow \tilde{P} = R^4 P, \\
\label{Ex4}  && \eta \rightarrow \tilde{\eta} = R^{-1}\eta, \;\;\;\;\;\;\;\,
                \alpha \rightarrow \tilde{\alpha} = R \alpha,
\end{eqnarray}
where $R(t)$ is the expansion parameter, $\tilde{t}$ is the so-called conformal time, and we note that ${\textbf v}$ is not scaled.

{\it Matter era: supercomoving variables}.-- In a matter-dominated universe the tilded quantities
are~\cite{Banerjee1,Banerjee2}:
\begin{eqnarray}
\label{Ex6}  && t \rightarrow \tilde{t} = \int \! R^{-3/2} dt, \;\;\;\;
                {\textbf x} \rightarrow \tilde{{\textbf x}} = R {\textbf x}, \\
\label{Ex8}  && {\textbf B} \rightarrow \tilde{{\textbf B}} = R^2 {\textbf B}, \;\;\;\;\;\;\;\;\;\,
                {\textbf v} \rightarrow \tilde{{\textbf v}} = R^{1/2} {\textbf v}, \\
\label{Exx3} && \rho \rightarrow \tilde{\rho} = R^3 \rho, \;\;\;\;\;\;\;\;\;\;\;\;\:
                P \rightarrow \tilde{P} = R^3 P, \\
\label{Ex10} && \eta \rightarrow \tilde{\eta} = R^{-1/2}\eta, \;\;\;\;\;\;\;\,
                \alpha \rightarrow \tilde{\alpha} = R^{3/2} \alpha.
\end{eqnarray}
Due to the formal coincidence of the dragged MHD equations in Minkowski and Friedmann spacetimes,
we can analyze the evolution properties of a primordial magnetic field in both cases in a similar way.
For definiteness, we consider first the case of a non-expanding universe and then translate the results
to the case of interest of an expanding universe.

We observe that, since $\rho \propto P \propto R^{-4}$ in the radiation-dominated era and
$\rho \propto P \propto R^{-3}$ in the matter-dominated era, the quantity
$\tilde{\rho} + \tilde{P}$ is a constant during the evolution of the universe.
This result was used in section IIa.


\section{III. Olesen's approach}

In this section, using a scaling approach first introduced by Olesen in~\cite{Olesen} for the case of
freely-decaying MHD turbulence, we express the unknown exponent $\beta$ introduced in section
IIc as a function of the exponents of the assumed initial power-law magnetic spectrum,
${\mathcal E}_B(k,t=0) \propto k^p$, and of the power law for the drag coefficient,
$\alpha(t) \propto t^a$.

{\it Non-helical case}.-- We are interested in the case of large magnetic Reynolds numbers,
and then we neglect the dissipative term in the
autoinduction equation~(\ref{Eq2bis}). We can then write the dragged MHD equations as
\begin{eqnarray}
\label{Olesen1} && {\textbf v} = \frac{1}{\alpha}\left(\nabla \times {\textbf B}\right) \times {\textbf B}, \\
\label{Olesen2} && \frac{\partial {\textbf B}}{\partial t}  = \nabla \times ({\textbf v} \times {\textbf B}).
\end{eqnarray}
A direct inspection shows that, under the scaling transformations
${\textbf x} \rightarrow \ell \, {\textbf x}$ and $t \rightarrow \ell^{1-u} \, t$,
%
%
equations~(\ref{Olesen1}) and (\ref{Olesen2}) admit solutions of the type
\begin{eqnarray}
\label{Eq18} && {\textbf v}(\ell \, {\textbf x},\ell^{1-u} \, t) = \ell^{\,u} \, {\textbf v}({\textbf x},t), \\
\label{Eq19} && {\textbf B}(\ell \, {\textbf x},\ell^{1-u} \, t) = \ell^{\,(1 + u + m)/2} \, {\textbf B}({\textbf x},t), \\
\label{scalingalpha} && \alpha(\ell^{1-u} \, t) = \ell^m \alpha(t),
\end{eqnarray}
where $\ell > 0$ is the ``scaling factor'', and $u$ and $m$ real parameters.
Differentiating Eq.~(\ref{scalingalpha}) with respect to $\ell$, and putting $\ell = 1$ afterwards, we get
$\alpha(t) \propto t^{m/(1-u)}$, and then $m = a (1-u)$, since we are assuming that $\alpha \propto t^a$.
Defining $p = -(2 + u + m)$ and using Eqs.~(\ref{Eq18}) and (\ref{Eq19}), we straightforwardly obtain
\begin{eqnarray}
\label{Eq20} {\mathcal E}_B(k,t) \!\!& = &\!\! \lambda_B k^p \, \psi_B(k^{3+p} \, t^{1-a}),
\\
\label{Eq21} {\mathcal E}_v(k,t) \!\!& = &\!\! \lambda_v k^r \, \psi_v(k^{3+p} \, t^{1-a}),
\end{eqnarray}
where we have defined $r = [3(1+a) + 2p \,]/(1-a)$ for notational
convenience. Here, $\lambda_B$ and $\lambda_v$ are constants, while
$\psi_B$ and $\psi_v$ are arbitrary scaling-invariant functions of
their arguments.

What is observed in dragged MHD~\cite{Banerjee1,Banerjee2} (as well as in turbulent MHD~\cite{Banerjee1,Banerjee2}) is that
the evolution of the magnetic spectrum in the non-helical case proceeds through the so-called
``selective decay'' (see section V), so that the initial magnetic spectrum retains its form
(for lengths below the characteristic dissipation scale) for all times. This means that if
we assume that the initial magnetic spectrum is a simple power of the wavenumber $k$, then
the coefficient $p$ is the index of that power. In fact, we have
${\mathcal E}_B(k,0) = \lambda_B k^p \, \psi_B(0)$,
%
%
a result which is true also in turbulent MHD (see~\cite{Olesen}).
Inserting Eq.~(\ref{Eq20}) into Eq.~(\ref{Eq3}) and
Eq.~(\ref{Eq21}) into Eq.~(\ref{Eq4}), we obtain Eqs.~(\ref{s2}) and (\ref{s3}), respectively,
where now $\beta =(1-a)(1+p)/(3+p)$ is expressed as a function of $p$ and $a$,
\begin{eqnarray}
\label{scalingEB} &&  E_B \propto t^{-\beta} = t^{-(1-a)(1+p)/(3+p)}, \\
\label{scalingEv} &&  E_v \propto t^{-(1 + a + \beta)} = t^{-2(2+a+p)/(3+p)}.
\end{eqnarray}
Equations~(\ref{s4}) and (\ref{s6}) are accordingly confirmed. Also, we find that
\begin{equation}
\label{correlation} \xi_B \propto \xi_v \propto t^{(1 - a - \beta)/2} = t^{(1-a)/(3+p)},
\end{equation}
confirming the scaling relation~(\ref{s5}).

{\it Helical case}.-- In the case when the magnetic helicity is different from zero, we can still apply the
Olesen's approach to find the scaling behavior of the magnetic helicity spectrum.
We find, straightforwardly,
\begin{equation}
\label{Eq23} {\mathcal H}_B(k,t) = \mu_B k^{p-1} \, \phi_B(k^{3+p}
\, t^{1-a}),
\end{equation}
where $\mu_B$ is a constant and $\phi_B$ an arbitrary
scaling-invariant function.
In dragged MHD~\cite{Banerjee1,Banerjee2} (as well as in turbulent MHD~\cite{Biskamp,Banerjee1,Banerjee2}),
the magnetic spectrum evolves through a phase of so-called ``inverse cascade'' (see section V),
during which magnetic modes with small wavelengths get modified by the longer ones and viceversa.
During this process, the information enclosed in the initial magnetic spectrum is lost,
so that the coefficient $p$ in Eq.~(\ref{Eq20}) does not represent, necessarily, the index
of the initial power-law spectrum. Indeed,
since magnetic helicity turns out to be (quasi-)conserved in
dragged MHD~\cite{Banerjee1,Banerjee2} (as well as in turbulent MHD~\cite{Biskamp,Banerjee1,Banerjee2}),
we must have $p = 0$. In fact, equations~(\ref{Eq5}) and (\ref{Eq23}) imply
$H_B(t) = H_B(t_i) (t/t_i)^{-(1-a)p/(3+p)}$.
%
%
Accordingly, the evolution laws for the magnetic and kinetic energies and correlation lengths
are obtained in the helical case taking $p=0$ in the evolution laws previously found for the non-helical case:
\begin{eqnarray}
\label{scalingEB-helical}   &&  E_B \propto t^{-(1-a)/3}, \\
\label{scalingEv-helical}   &&  E_v \propto t^{-2(2+a)/3}, \\
\label{correlation-helical} && \xi_B \propto \xi_v \propto t^{(1-a)/3} \propto E_B^{-1}.
\end{eqnarray}
Equations~(\ref{scalingEB}), (\ref{scalingEv}), (\ref{correlation}),
and equations~(\ref{scalingEB-helical}), (\ref{scalingEv-helical}), (\ref{correlation-helical}),
are in agreement with the results of~\cite{Banerjee1,Banerjee2}.


\section{IV. Mean-field approximation}

In this section, working in mean-field-theory approximation,
we derive the evolution integro-differential equations for the magnetic energy and
helicity spectra, which will be solved later on in section V.
We start from the ``simplest'' mean-field approximation first introduced in~\cite{Cornwall},
we then study the dragged MHD equations in the so-called ``one-point-closure''~\cite{Biskamp}
and ``quasi-normal''~\cite{Campanelli5} approximations.

\subsection{IVa. Cornwall's approximation}

It is useful to re-write the autoinduction equation~(\ref{Eq16}) as
\begin{equation}
\label{Eq16bis} \frac{\partial {\textbf B}}{\partial t} =
\tau_d \nabla \times [ ({\textbf J} \cdot {\textbf B}) {\textbf B} -
{\textbf B}^2 {\textbf J}] + \eta \nabla^2 {\textbf B}.
\end{equation}
We now proceed as in~\cite{Cornwall} by replacing the
quadratic terms ${\textbf J} \cdot {\textbf B}$ and
${\textbf B}^2$ in the above equation with
${\textbf J} \cdot {\textbf B} \rightarrow \frac{1}{3} \, \langle {\textbf J} \cdot {\textbf B} \rangle$
and ${\textbf B}^2 \rightarrow \frac{1}{3} \, \langle {\textbf B}^2 \rangle$,
%
%
where the brackets $\langle ... \rangle$ indicate a suitable
average to be defined later. This operation allow us to
linearize the autoinduction equation,
\begin{equation}
\label{MF2} \frac{\partial {\textbf B}}{\partial t} =
\frac{\tau_d}{3} \, \nabla \times [ \langle {\textbf J} \cdot
{\textbf B} \rangle {\textbf B} - \langle {\textbf B}^2 \rangle
{\textbf J}] + \eta \nabla^2 {\textbf B}.
\end{equation}
Whatever is the averaging operation, the averaged
quantities in the above equation must satisfy a consistency
relation coming from conservation of magnetic helicity in the case
of null dissipation. Indeed, observing that Eq.~(\ref{MF2}) gives
\begin{equation}
\label{MF2a} \frac{\partial H_B}{\partial t} = \frac{2 \tau_d}{3} \int \! d^{\,3} x
\left[ \langle {\textbf J} \cdot {\textbf B} \rangle {\textbf B}^2
- \langle {\textbf B}^2 \rangle {\textbf J} \cdot {\textbf B}
\right] - 2\eta \! \int \! d^{\,3} x \, {\textbf J} \cdot {\textbf B},
\end{equation}
we get
\begin{equation}
\label{MF2b} \int \! d^{\,3} x \, \langle {\textbf J} \cdot
{\textbf B} \rangle {\textbf B}^2 = \int \! d^{\,3} x \, \langle
{\textbf B}^2 \rangle {\textbf J} \cdot {\textbf B}.
\end{equation}
Assuming now, as in~\cite{Cornwall}, that the operation of
averaging is just a volume average,
$\langle {\textbf J} \cdot {\textbf B} \rangle = \int \! d^{\,3} x \, {\textbf J} \cdot {\textbf B}$
and $\langle {\textbf B}^2 \rangle = \int \! d^{\,3} x \, {\textbf B}^2$,
%
%
we find that Eq.~(\ref{MF2b}) is satisfied for all magnetic field
configurations and, taking into account Eqs.~(\ref{Eq9}) and (\ref{Eq3}), that
$\langle {\textbf J} \cdot {\textbf B} \rangle = -\dot{H}_B/(2\eta)$ and
$\langle {\textbf B}^2 \rangle = 2 E_B$.
%
%
Finally, we can re-write Eq.~(\ref{MF2}) as
\begin{equation}
\label{MF3} \frac{\partial {\textbf B}}{\partial t} = \alpha_B
\nabla \times {\textbf B} + \eta_{\rm eff} \nabla^2 {\textbf B},
\end{equation}
where we have introduced the quantities
$\alpha_B(t) = -\dot{H}_B \, \tau_d/(6\eta)$,
$\eta_{\rm eff}(t) = \eta + \eta_{\rm T}$, and
$\eta_{\rm T}(t) = 2E_B \tau_d/3$.
%
%
Equation (\ref{MF3}) describes the well-known $\alpha-$dynamo
effect~\cite{Biskamp}, in which the dynamo, the turbulent
diffusion, and the total effective diffusion coefficients, are
given by $\alpha_B$, $\eta_{\rm T}$, and $\eta_{\rm eff}$, respectively.

\subsection{IVb. One-point-closure approximation}

Cornwall's arguments find a full justification in the framework of
one-point-closure approximation, originally introduced in turbulent MHD~\cite{Biskamp},
and that we apply now to dragged MHD.

{\it Equation}.-- Let us assume that ${\textbf v}$ and ${\textbf B}$ can be
decomposed into an average part varying only on large scales
and a weak, small-scale fluctuating part,
\begin{equation}
\label{MFT1} {\textbf v} = {\textbf v}_0 + \widetilde{\textbf v}, \;\;\;\;
{\textbf B} = {\textbf B}_0 + \widetilde{\textbf B},
\end{equation}
with $\langle \widetilde{\textbf v} \rangle = \langle\widetilde{\textbf B} \rangle = 0$,
and $|\widetilde{\textbf v}| \ll |{\textbf v}_0|$, $|\widetilde{\textbf B}| \ll |{\textbf B}_0|$.
In a moving coordinate system such that ${\textbf v}_0 = 0$, the
equation governing the evolution of the mean magnetic field is a
dynamo equation
\begin{equation}
\label{MFT3} \frac{\partial {\textbf B}_0}{\partial t} = \alpha'_B
\nabla \times {\textbf B}_0 + (\eta'_{\rm T} + \eta) \nabla^2
{\textbf B}_0,
\end{equation}
with dynamo and turbulent diffusion coefficients given by
\cite{Biskamp}
\begin{eqnarray}
\label{MFT4} \alpha'_B(t) \!\!& = &\!\! - \frac{\tau_{\rm
eddy}}{3} \, \langle \widetilde{\textbf v} \cdot \nabla \times
\widetilde{\textbf v} \rangle \simeq - \frac{\tau_{\rm eddy}}{3}
\, H_v,
\\
\label{MFT5} \eta'_{\rm T}(t) \!\!& = &\!\! \frac{\tau_{\rm
eddy}}{3} \, \langle \widetilde{\textbf v}^2 \rangle \simeq
\frac{2\tau_{\rm eddy}}{3} \, E_v = \frac{2}{3} \,
E_B \, \Gamma \tau_{\rm eddy}.
\end{eqnarray}
In Eq.~(\ref{MFT4}), we have introduced the the so-called kinetic helicity density~\cite{Biskamp},
$H_v(t) = \int\! d^{\,3}x \, \widetilde{\textbf v} \cdot \widetilde{\boldsymbol \omega}$,
with $\widetilde{\boldsymbol \omega} = \nabla \times \widetilde{\textbf v}$, while in the last
equality of Eq.~(\ref{MFT5}), we used Eq.~(\ref{Gamma}).

Starting from Eq.~(\ref{MFT3}) and imposing the conservation of the mean
magnetic helicity $\int \! d^{\,3}x \, {\textbf A}_0 \cdot
{\textbf B}_0 \simeq \int \! d^{\,3}x \, {\textbf A} \cdot
{\textbf B}$ for vanishing resistivity, we get
$\alpha'_B \! \int \! d^{\,3}x \, {\textbf B}_0^2 = \eta'_{\rm T} \! \int \! d^{\,3}x \, {\textbf J}_0 \cdot {\textbf B}_0$.
Since $\int \! d^{\,3}x \, {\textbf B}_0^2 \simeq \int \! d^{\,3}x \, {\textbf B}^2 = 2 E_B$ and
$\int \! d^{\,3}x \, {\textbf J}_0 \cdot {\textbf B}_0 \simeq
\int \! d^{\,3}x \, {\textbf J} \cdot {\textbf B} = -\dot{H}_B/(2\eta)$ [see Eq.~(\ref{Eq9})],
we have $2 E_B \alpha'_B = -[\dot{H}_B/(2\eta)] \, \eta'_{\rm T}$ and in turn, using
Eqs.~(\ref{MFT4}) and (\ref{MFT5}), we find $H_v = (\Gamma/2\eta) \, \dot{H}_B$.
Therefore, we can re-write the dynamo coefficient as
$\alpha'_B(t) \simeq -\dot{H}_B \, \Gamma \tau_{\rm eddy}/(6\eta)$.
%
%
Both the dynamo and turbulent diffusion coefficients are then consistent with the
expressions for $\alpha_B$ and $\eta_{\rm T}$ because of Eq.~(\ref{nice}).

{\it Solution}.-- In order to solve the dynamo equation~(\ref{MF3}) [or, which is the same, Eq.~(\ref{MFT3})]
it is useful to introduce the orthonormal helicity base $\{{\textbf e}_{+}, {\textbf e}_{-}, {\textbf e}_{3} \}$,
with ${\textbf e}_{\pm} = ({\textbf e}_{1} \pm i {\textbf e}_{2})/\sqrt{2}$
and ${\textbf e}_{3} = {\textbf k}/k$, where
$\{{\textbf e}_{1}, {\textbf e}_{2}, {\textbf e}_{3} \}$ is
a right-handed orthonormal base. In the helicity base, the magnetic field
in Fourier space can be decomposed as
${\textbf B}(k,t) = B_{+}(k,t) \, {\textbf e}_{+} + B_{-}(k,t) \, {\textbf e}_{-}$,
%
%
where $B_{+}(k,t)$ and $B_{-}(k,t)$ represent the positive and
negative helicity components of ${\textbf B}(k,t)$, respectively.
Equation~(\ref{MF3}) becomes
$\dot{B}_{\pm} =  \pm \alpha_B k B_{\pm} - \eta_{\rm eff} k^2 B_{\pm}$
%
%
in Fourier space. The solution of the above equation is easily found,
\begin{equation}
\label{MF7} B_{\pm}(k,t) = B_{\pm}(k,0) \exp (\pm k \ell_{\alpha} - k^2 \ell_{\rm diss}^2),
\end{equation}
where we have defined the ``dynamo'' and ``dissipation'' lengths,
$\ell_{\alpha}(t) = \int_{0}^{t} dt \, \alpha_B$ and
$\ell_{\rm diss}^2(t) = \int_{0}^{t} dt \, \eta_{\rm eff}$,
%
%
respectively. In the helicity base, the energy and helicity spectra read
\begin{eqnarray}
\label{MF9} {\mathcal E}_B (k,t) \!\!& = &\!\! \left( \frac{k}{2\pi} \right)^{\!2} \left( \, |B_{+}|^2 + |B_{-}|^2 \right) \! , \\
\label{MF10} {\mathcal H}_B (k,t) \!\!& = &\!\! \frac{k}{2\pi^2} \left( \, |B_{+}|^2 - |B_{-}|^2 \right) \! .
\end{eqnarray}
To proceed further, let us suppose that
$|B_{-}(k,0)|^2 = (1-h) |B_{+}(k,0)|^2$,
%
and define ${\mathcal H}^{\mbox{\scriptsize max}}_B (k,t) = 2 k^{-1} {\mathcal E}_B (k,t)$
%
%
so that we can write the initial helicity spectrum as
${\mathcal H}_B (k,0) = h_B {\mathcal H}^{\mbox{\scriptsize max}}_B (k,0)$,
%
%
where $h_B = h/(2-h)$.
%
%
Indeed, we have made the simplifying assumption that the initial
helicity is just a fraction of the initial maximal helicity. In
other words, we restrict our analysis only to so-called magnetic
fields with (initial) ``fractional helicity''.
Inserting Eq.~(\ref{MF7}) in Eqs.~(\ref{MF9}) and (\ref{MF10}) we
get, respectively,
\begin{eqnarray}
\label{MF13} && {\mathcal E}_B (k,t) = {\mathcal E}_B (k,0) \left[ \cosh(2k \ell_{\alpha}) + h_B \sinh(2k \ell_{\alpha}) \right]
\exp(-2k^2 \ell_{\rm diss}^2),
\\
\label{MF14} && {\mathcal H}_B (k,t) = {\mathcal H}^{\mbox{\scriptsize max}}_B (k,0) \left[ \sinh(2k \ell_{\alpha})
+ h_B \cosh(2k \ell_{\alpha}) \right] \exp (-2k^2 \ell_{\rm diss}^2).
\end{eqnarray}
The above equations are integro-differential equations for the magnetic energy and helicity, and will be solved in section V.

\subsection{IVc. Quasi-normal approximation}

We now show that the results in the above two subsections, can be also derived using a two-point-closure
approximation, also known as ``quasi-normal'' approximation.

We start by observing that the autoinduction equation~(\ref{Eq16bis}), in Fourier space, reads:
\begin{equation}
\label{Eq16tris} \frac{\partial B_i({\textbf k})}{\partial t} = \tau_d \! \int \!\! \frac{d^3k'}{(2\pi)^3} \!
\int \!\! \frac{d^3q}{(2\pi)^3} \, \varepsilon_{ijk} k_j q_r B_s({\textbf q}) \left[ \varepsilon_{krs} B_n({\textbf k}' -
{\textbf q}) B_n({\textbf k}-{\textbf k}') - \varepsilon_{rsm} B_k({\textbf k}-{\textbf k}') B_m({\textbf k}'-{\textbf q}) \right]
- \eta k^2 B_i({\textbf k}),
\end{equation}
where $\varepsilon_{ijk}$ is the totally antisymmetric tensor and
summation over repeated indexes is understood. In quasi-normal approximation,
the four-point magnetic correlator is decomposed, in terms of two-point correlator, as
\cite{Biskamp}:
\begin{equation}
\label{Wick}
\langle B_i({\textbf k}) B_j({\textbf k}') B_k({\textbf q}) B_l({\textbf q}') \rangle  =
\langle B_i({\textbf k}) B_j({\textbf k}') \rangle
\langle B_k({\textbf q}) B_l({\textbf q}') \rangle +
\langle B_i({\textbf k}) B_k({\textbf q}) \rangle
\langle B_j({\textbf k}') B_l({\textbf q}') \rangle +
\langle B_i({\textbf k}) B_l({\textbf q}') \rangle
\langle B_j({\textbf k}') B_k({\textbf q}) \rangle,
\end{equation}
where $\langle ... \rangle$ denotes an ensemble average.
Multiplying Eq.~(\ref{Eq16tris}) respectively by $B_i^*({\textbf k})$
and $A_i^*({\textbf k})$, and then averaging out, we get
\begin{eqnarray}
\label{Eqspectrum1} && \frac{\partial{\mathcal E}_B}{\partial t} = -2 \eta_{\rm eff} k^2 {\mathcal E}_B + \alpha_B k^2 {\mathcal H}_B, \\
\label{Eqspectrum2} && \frac{\partial{\mathcal H}_B}{\partial t} = -2 \eta_{\rm eff} k^2 {\mathcal H}_B + 4 \alpha_B {\mathcal E}_B.
\end{eqnarray}
As it is easy to check, the solution of the above equations in the case of magnetic
fields with initial fractional helicity is given by Eqs.~(\ref{MF13})-(\ref{MF14}),
so that the one-point- and two-point-closure approximations give exactly,
and perhaps surprisingly, the same identical results.


\section{V. Results}

In this section, we find the evolution laws of the magnetic intensity and correlation length
in a radiation-dominated universe for the case of free-steaming neutrinos and photons, by
solving Eqs.~(\ref{MF13})-(\ref{MF14}) for some specific initial conditions. When
working in a Minkowski spacetime, we take $t=0$ as the initial time for the sake of simplicity.

\subsection{Va. Initial conditions}

{\it Magnetic spectrum}.-- For the sake of simplicity, we assume that the initial magnetic energy
spectrum is represented by the simple function
\begin{equation}
\label{MF15} {\mathcal E}_B (k,0) = \lambda_B k^p \exp (-2k^2 \ell_{B}^2),
\end{equation}
where $\lambda_B$ and $\ell_B$ are constants. For $k \ll \ell_{B}^{-1}$,
the magnetic energy spectrum possesses a power-law behavior with exponent
$p$ which, to avoid infinities in the magnetic energy, we assume to be
greater than 1. For the same reason (finiteness of energy), an exponential
cut-off has been introduced. The quantity $\ell_B$ can be related to the initial
correlation length, $\xi_B(0)$, by
$\ell_B = \Gamma[(1+p)/2] \: \xi_B(0)/[\sqrt{2} \, \Gamma(p/2)]$,
%
%
where $\Gamma(x)$ is the Euler gamma function~\cite{Gradshteyn}.

It is worth noting that the power-law behavior for the initial magnetic spectrum
is predicted by many models of generation of cosmic magnetic fields in the early
universe. In particular, inflation-produced magnetic fields~\cite{Generation2} usually
exhibit a power-law spectrum when they re-enter the horizon. Also, most of
the generating mechanisms after inflation~\cite{Generation4}
repose on microphysical processes acting only on small scales.
Therefore, the structure of the resulting magnetic field appears as a set of
uncorrelated (Gaussian-distributed) eddies, which implies a $k^2$-like spectrum.
To see this, suppose that the magnetic field at the initial time
$t=0$ is a stochastic variable with gaussian distribution,
$\mathcal{P} \! \left[ \, {\textbf B}({\textbf x},0) \, \right] \propto
\exp \! \left[ -(3\pi/\lambda_B) \int\! d^{\,3}x \, {\textbf B}^2({\textbf x},0) \right]$,
%
%
where $\lambda_B$ is a constant. The magnetic energy spectrum,
\begin{equation}
\label{Gaussian2} {\mathcal E}_B(k,0) = 2\pi k^2 \langle |{\textbf B}({\textbf k},0)|^2 \rangle =
2 \pi k^2 \!\! \int\! d^{\,3}x \! \int\! d^{\,3}y \: e^{i {\textbf k} \cdot ({\textbf x}-{\textbf y})}
\sum_{k=1}^{3}\langle B_k({\textbf x},0) \, B_k({\textbf y},0) \rangle,
\end{equation}
is then proportional to $k^2$. In fact, observing that
\begin{equation}
\label{Gaussian3} \langle B_i({\textbf x},0) B_j({\textbf y},0) \rangle  =
\frac{\int \! \mathcal{D} [B_k({\textbf z},0)] \mathcal{P} \! \left[ \, {\textbf B}({\textbf z},0) \, \right]
B_i({\textbf x},0) B_j({\textbf y},0)}{\int \! \mathcal{D} [B_k({\textbf z},0)] \mathcal{P} \!
\left[ \, {\textbf B}({\textbf z},0) \, \right]} =
\frac{\lambda_B}{6\pi} \, \delta_{ij} \, \delta({\textbf x}-{\textbf y}),
\end{equation}
where in the last equality we used the properties of Gaussian integral~\cite{Gaussian},
we get ${\mathcal E}_B(k,0) = \lambda_B k^2$.

Recently enough~\cite{Caprini}, however, it has been shown that the analyticity of the
magnetic field correlator $B_i({\textbf x}) B_j({\textbf y})$
defined on a compact support, together with the divergenceless of ${\textbf B}$,
implies that the spectral index $p$ has to be
even and equal or larger than 4. Moreover, in~\cite{Tevzadze}, it has been shown
that, starting from suitable initial conditions which should reflect the production
mechanism of a magnetic field from bubble collision in a first-order phase transition,
a Batchelor spectrum ${\mathcal E}_B \propto k^4$ is established
after a short time interval and at small wavenumbers.
Nevertheless, we can always assume that on very large scales (very small wavenumbers
$k$) the initial spectrum satisfies the above requirements, but that on
smaller scales, where the gross of the magnetic energy is stored,
it behaves like $k^2$. In any case, in the following we leave the index $p$
as a free parameter, privileging the cases $p=2,3,4$
when we show graphically our results (for the choice $p=3$, see below).

{\it Drag coefficient}.-- We assume that the drag coefficient
$\alpha$ scales in time following the simple law
\begin{equation}
\label{alpha} \alpha(t) = \alpha(0) [1 + \tau/\gamma(0)]^a,
\end{equation}
where $\tau$ is the normalized time
\begin{equation}
\label{normalized} \tau = \frac{t}{\tau_{\rm eddy}(0)} \, ,
\end{equation}
and $\gamma(0)$ and $a$ are constants.
The above parametrization is useful since $\alpha(t) \simeq \alpha(0)$ for
$t \ll \tau_{\rm eddy}(0)$, when the system is, as we will see below, in a
quiescent phase [i.e. the integral quantities like $E_B(t)$ and $\xi_B(t)$
remain almost constant in time], and $\alpha(t)$ follows asymptotically a
power-law behavior $\alpha(0) [\tau/\gamma(0)]^a$ for large times, when
magnetohydrodynamic effects operate in changing the state of the system.
Using the results in~\cite{Banerjee1,Banerjee2}, we find that in comoving variables $a=-3$ for the
case of free-steaming photons, while $a=-4$ for the case of free-steaming neutrinos.
The constant $\gamma(0)$ is generally different from unity, and its
explicit expression will be derived in section Vh.

{\it Setup}.-- Numerical integration of dragged MHD equations has been performed only in~\cite{Banerjee1,Banerjee2}.
There, the simple case of constant dragged coefficient was analyzed.
Also, one of the cases discussed in~\cite{Banerjee1,Banerjee2}, was that with $p=3$. Hence,
in the light of the above discussions, we consider the three cases
\begin{equation}
\label{setup1}  (p,a) =
\left\{ \begin{array}{lll}
        (2,-3), &  \;\; \mbox{Gaussian, photon case},  \\
        (3,0),  &  \;\; \mbox{Banerjee-Jedamzik case}, \\
        (4,-4), &  \;\; \mbox{causal, neutrino case},
    \end{array}
    \right.
\end{equation}
for the non-helical case, and the three cases
\begin{equation}
\label{setup2}  (p,a,h) =
\left\{ \begin{array}{lll}
        (2,-3,10^{-5}),  &  \;\; \mbox{Gaussian, photon case},  \\
        (3,0,1),         &  \;\; \mbox{Banerjee-Jedamzik case}, \\
        (4,-4,10^{-15}), &  \;\; \mbox{causal, neutrino case},
    \end{array}
    \right.
\end{equation}
for the helical case. Note that in the helical, Banerjee-Jedamzik case,
the magnetic field is taken to be maximally helical, as it is in the numerical
analysis of~\cite{Banerjee1,Banerjee2}. Also, the very small value $h=10^{-15}$ is taken
to clearly show the transition from the selective decay phase to the inverse cascade phase
to be discussed in section Vd.

\subsection{Vb. Master equations}

Inserting Eqs.~(\ref{MF13}) and (\ref{MF14}) in
Eqs.~(\ref{Eq3}) and (\ref{Eq5}) we find, respectively,
\begin{eqnarray}
\label{MF17} && E_B(t) =  E_B(0) (1 + \zeta_{\rm diss}^2)^{-\frac{1+p}{2}} \left( \Phi_1 + p \, h_B \frac{\zeta_B}{2} \,
\frac{\zeta_\alpha}{\sqrt{1 + \zeta_{\rm diss}^2}} \, \Phi_2 \right) \! , \\
\label{MF18} && H_B(t) = H_B(0) (1 + \zeta_{\rm diss}^2)^{-\frac{p}{2}} \left( \Phi_3 + \frac{1}{h_B} \, \frac{2}{\zeta_B} \,
\frac{\zeta_\alpha}{\sqrt{1 + \zeta_{\rm diss}^2}} \, \Phi_4 \right) \! ,
\end{eqnarray}
with
\begin{eqnarray}
\label{phi12} \Phi_1 \!\!& = &\!\!  _1 F_1 \! \left( \frac{1+p}{2} ,
\frac{1}{2} ; \frac{1}{2} \, \frac{\zeta_\alpha^2}{1 + \zeta_{\rm diss}^2} \right), \;\;\;\;
\Phi_2 = _1 F_1 \!
\left( \frac{2+p}{2} , \frac{3}{2} ; \frac{1}{2} \,
\frac{\zeta_\alpha^2}{1 + \zeta_{\rm diss}^2} \right), \\
\label{phi34} \Phi_3 \!\!& = &\!\! _1 F_1 \! \left( \frac{p}{2} ,
\frac{1}{2} ; \frac{1}{2} \, \frac{\zeta_\alpha^2}{1 + \zeta_{\rm diss}^2} \right), \;\;\;\;\;\;\;\;\;\;\,
\Phi_4 = _1 F_1 \!
\left( \frac{1+p}{2} , \frac{3}{2} ; \frac{1}{2} \,
\frac{\zeta_\alpha^2}{1 + \zeta_{\rm diss}^2} \right),
\end{eqnarray}
where $_1 F_1 (a,b;z)$ is the Kummer confluent hypergeometric
function~\cite{Gradshteyn}, and we have introduced the quantities
$\zeta_\alpha(t) = \ell_{\alpha}(t)/\ell_B$,
$\zeta_{\rm diss}(t) = \ell_{\rm diss}(t)/\ell_B$, and
$\zeta_B = \xi_B(0)/\ell_B$.
%
%
Also, inserting Eq.~(\ref{MF13}) in Eq.~(\ref{Eq10}) we get
\begin{equation}
\label{xi} \xi_B(t) = \xi_B(0) \, (1 + \zeta_{\rm diss}^2)^{1/2}  \; \frac{\Phi_3 + h_B \, \frac{2}{\zeta_B} \,
\frac{\zeta_\alpha}{\sqrt{1 + \zeta_{\rm diss}^2}} \, \Phi_4}{\Phi_1 + p \, h_B
\frac{\zeta_B}{2} \, \frac{\zeta_\alpha}{\sqrt{1 + \zeta_{\rm diss}^2}} \, \Phi_2} \, .
\end{equation}
It is useful, for the following discussion, to define accurately
the magnetic Reynolds number and the eddy turnover time as
\begin{equation}
\label{Rey-eddy2}
{\text{Re}}_B = \frac{v_{\rm rms} \xi_B}{\eta} \, , \;\;\;\;
\tau_{\rm eddy} = \frac{\xi_B}{v_{\rm rms}} \, ,
\end{equation}
where we used the magnetic correlation length and the root-mean-square value of the velocity
field,
$v_{\textmd{rms}}^2 = \int \! d^{\,3} x \, {\textbf v}^2({\textbf x},t) = 2E_v$,
%
%
as typical length scale and velocity, respectively.
With the aid of the above definitions, the integro-differential equations~(\ref{MF17}) and
(\ref{MF18}) for $E_B$ and $H_B$ can be transformed into ordinary differential equations for the quantities
$\zeta_{\rm diss}$ and $\zeta_{\alpha}$. We have
\begin{eqnarray}
\label{differential1} &&  \frac{d\zeta_{\rm diss}^2}{d\tau} = \frac{\zeta_B^2}{\text{Re}_B(0)} +
                          \frac{\zeta_B^2}{3} \, \delta(0) [1 + \tau/\gamma(0)]^{-a} \, \frac{E_B(\tau)}{E_B(0)} \, , \\
\label{differential2} &&  \frac{d\zeta_{\alpha}}{d\tau} = -\frac{\zeta_B}{6}  \delta(0) h_B
                          \text{Re}_B(0) [1 + \tau/\gamma(0)]^{-a} \, \frac{d}{d\tau} \frac{H_B(\tau)}{H_B(0)} \, ,
\end{eqnarray}
with $E_B$ and $H_B$ as a function of $\zeta_{\rm diss}$ and
$\zeta_{\alpha}$ given by Eqs.~(\ref{MF17}) and (\ref{MF18}),
respectively. The constant
\begin{equation}
\label{delta} \delta(0) = \frac{\tau_d(0)}{\tau_{\rm eddy}(0) \Gamma(0)} \,
\end{equation}
is of order unity [see Eq.~(\ref{nice})]. In section Vf, we will found that
this order-unity constant is indeed a parameter which weakly depends on $p$.

Equations~(\ref{MF17})-(\ref{MF18}) and (\ref{differential1})-(\ref{differential2})
are our master equations in the study of the evolution of the magnetic field.

\subsection{Vc. Solutions: non-helical case}

In the non-helical case, $h_B=0$, the solution of
Eqs.~(\ref{differential1})-(\ref{differential2}) is such that
$\zeta_\alpha(t) = 0$, that is $H_B(t) = 0$ for all times. Then, from
Eqs.~(\ref{MF17}), (\ref{xi}), and (\ref{MF13}) we get
\begin{eqnarray}
\label{Es0} && E_B(t) = E_B(0) (1 + \zeta_{\rm diss}^2)^{-\frac{1+p}{2}}, \\
\label{xis0} && \xi_B(t) = \xi_B(0) \, (1 + \zeta_{\rm diss}^2)^{1/2}, \\
\label{spectrum0} && {\mathcal E}_B (k,t) = {\mathcal E}_B (k,0) \exp(-2k^2 \ell_B^2 \zeta_{\rm diss}^2).
\end{eqnarray}
For large magnetic Reynolds numbers, the first term in the
right-hand-side of Eq.~(\ref{differential1}) can be neglected with
respect to the second one. In this case, the solution of Eq.~(\ref{differential1}) is
\begin{equation}
\label{zetadiss} \zeta_{\rm diss}^2(\tau) =
\left\{ c_0 \! \left\{[1 + \tau/\gamma(0)]^{1-a} - 1 \right\} + 1 \right\}^{\!\frac{2}{3+p}} - 1,
\end{equation}
where $c_0$ is given in Appendix A. For $\tau \gg 1$, we have
\begin{equation}
\label{zetadissapprox} \zeta_{\rm diss}^2(\tau) \simeq c_1 \, \tau^{\frac{2(1-a)}{3+p}} ,
\end{equation}
where $c_1$ is given in Appendix A,
so that
\begin{eqnarray}
\label{Es} && E_B(\tau) \simeq c_2 E_B(0) \, \tau^{-\frac{(1-a)(1+p)}{3+p}} ,
\\
\label{xis} && \xi_B(\tau) \simeq c_3 \xi_B(0) \, \tau^{\frac{1-a}{3+p}} ,
\\
\label{spectrum2} && {\mathcal E}_B (k,t) \simeq {\mathcal E}_B (k,0)
\exp \! \left [ -c_4 k^2 t^{\frac{2(1-a)}{3+p}} \right ] \! ,
\end{eqnarray}
where $c_2,c_3,c_4$ are given in Appendix A.
Equations~(\ref{Es}), (\ref{xis}), and (\ref{spectrum2}), are in agreement with
Eqs.~(\ref{scalingEB}), (\ref{correlation}), and
(\ref{Eq20}), respectively. We deduce that the mean-field-theory approach and
the simple scaling arguments applied to MHD equations are
compatible.
The mean-field-theory approach gives us two more
pieces of information with respect to scaling arguments. Firstly,
the time when the system enters into the scaling regime is, looking at Eqs.~(\ref{Es0}),
(\ref{xis0}), (\ref{spectrum0}), and (\ref{zetadiss}), approximatively given by the
initial eddy turnover time times $\gamma(0)$. Secondly, comparing
Eq.~(\ref{spectrum2}) with Eq.~(\ref{Eq20}) we get the expression
for the scaling function $\psi_B(x)$:
\begin{equation}
\label{psi} \psi_B (x) = e^{-c_4 x^2}.
\end{equation}
In Fig.~1, we plot the spectrum of the magnetic energy at different times
for the three cases in Eq.~(\ref{setup1}) [in all figures of this paper, we take $\gamma(0) = 1$].
It is clear that the decay of the magnetic energy and the growth
of the magnetic correlation length (which is, roughly speaking, the inverse of the wavenumber
corresponding to the peak in the magnetic spectrum)
advance through {\it selective decay}, a phenomenon well described in~\cite{Son}
in the case of turbulent MHD (see also~\cite{Biskamp}). There is no direct transfer of
magnetic energy from small scales  to large scales but modes with larger wavenumbers
just decay faster than those whose wavenumbers are small. Accordingly, as the time goes on,
the magnetic field survives only on larger and larger scales, as it is evident in Fig.~1.

The dotted lines in Fig~1 represent the magnetic energy spectra at a certain time
$t/\tau_{\rm eddy}(0) = \tau_{\mathcal{E}}$ (to be defined in section Vd) when, if
the field were helical, the magnetic field would enter in a phase of {\it inverse cascade}
(see below). In the case at hand, the magnetic field has no helicity and, indeed, nothing special
happens when $\tau$ reaches $\tau_{\mathcal{E}}$.


\begin{figure}[t!]
\begin{center}
\includegraphics[clip,width=0.45\textwidth]{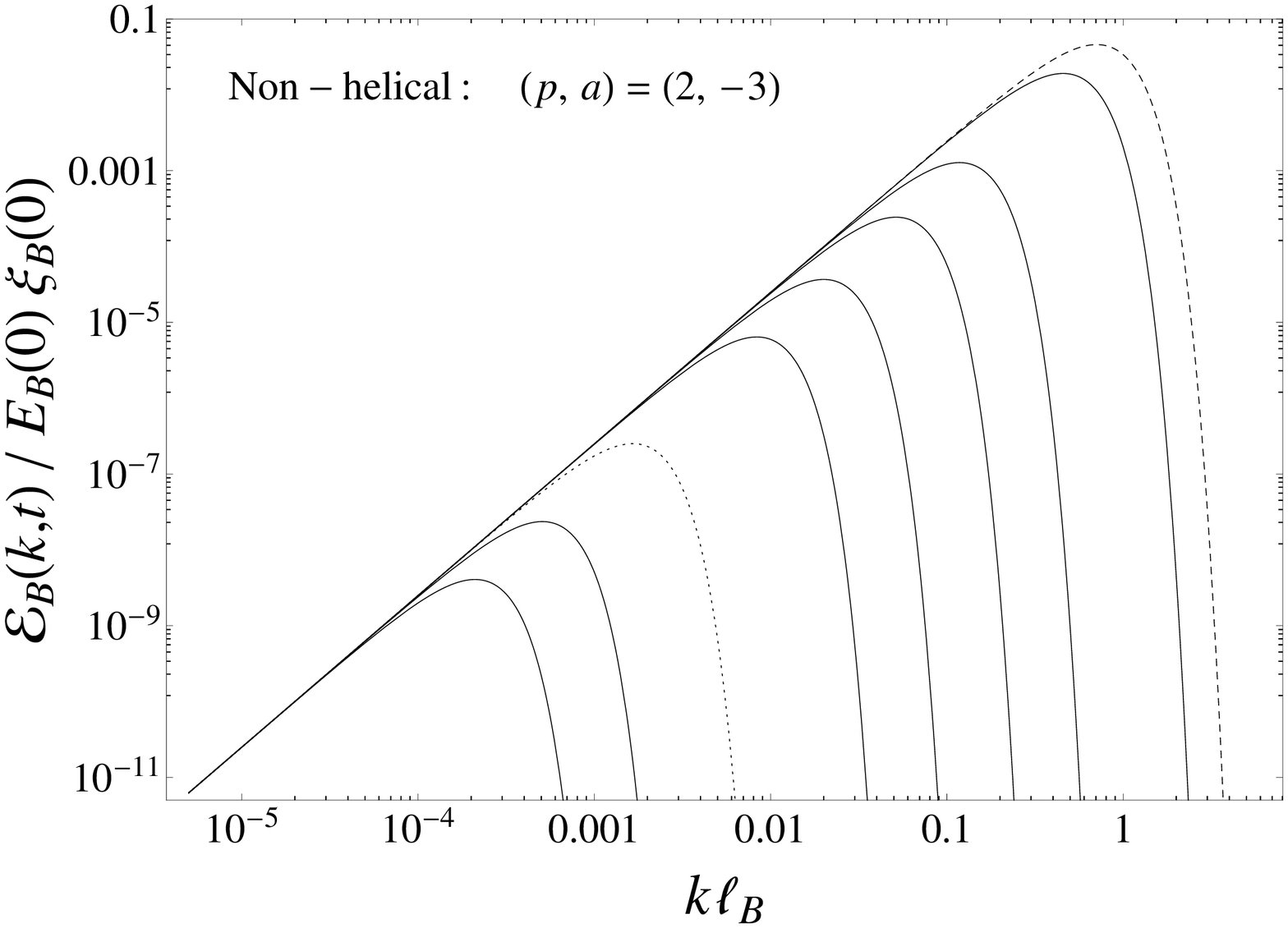}
\hspace{0.4cm}
\includegraphics[clip,width=0.45\textwidth]{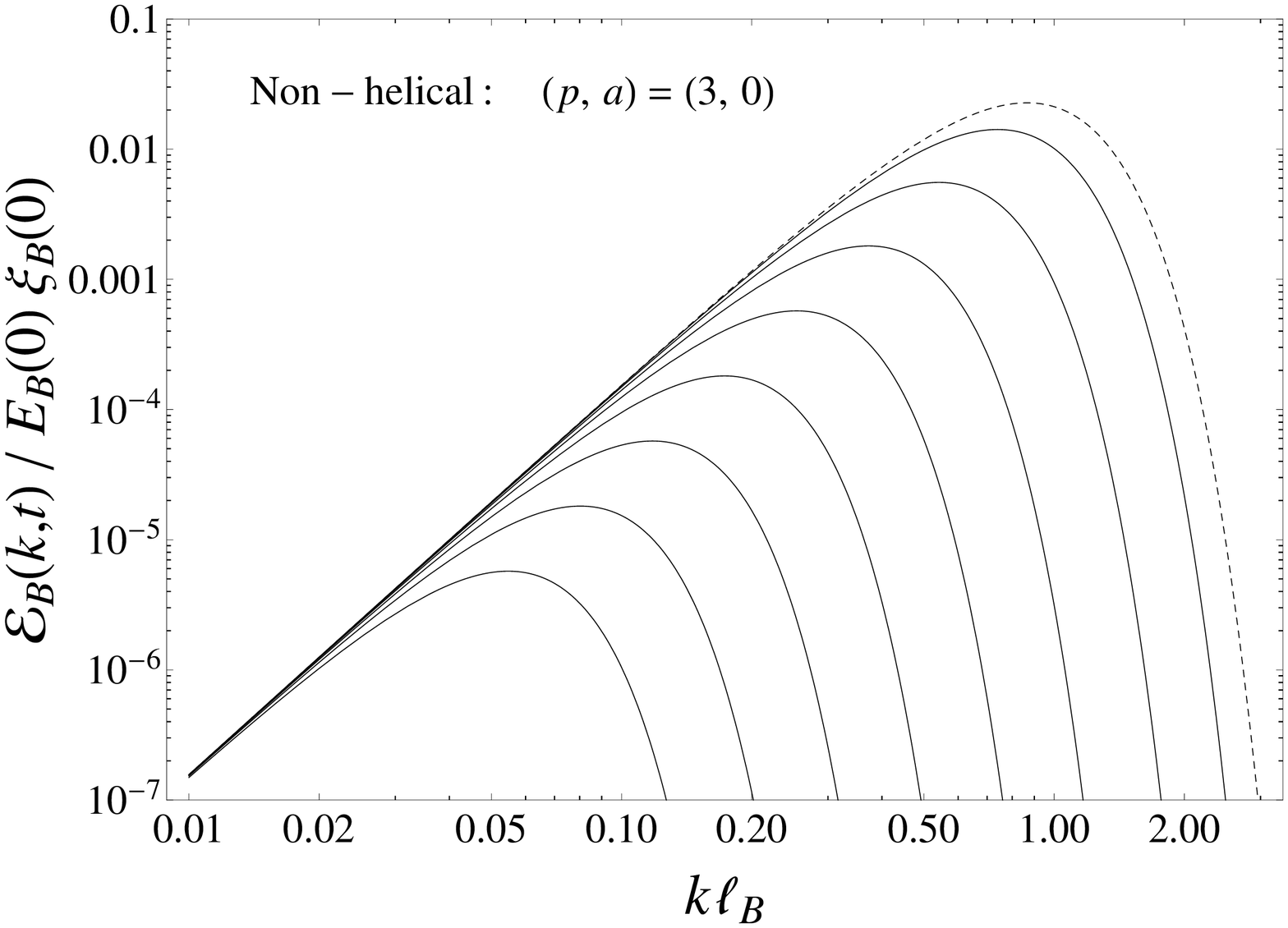}
\vspace{0.4cm}

\includegraphics[clip,width=0.45\textwidth]{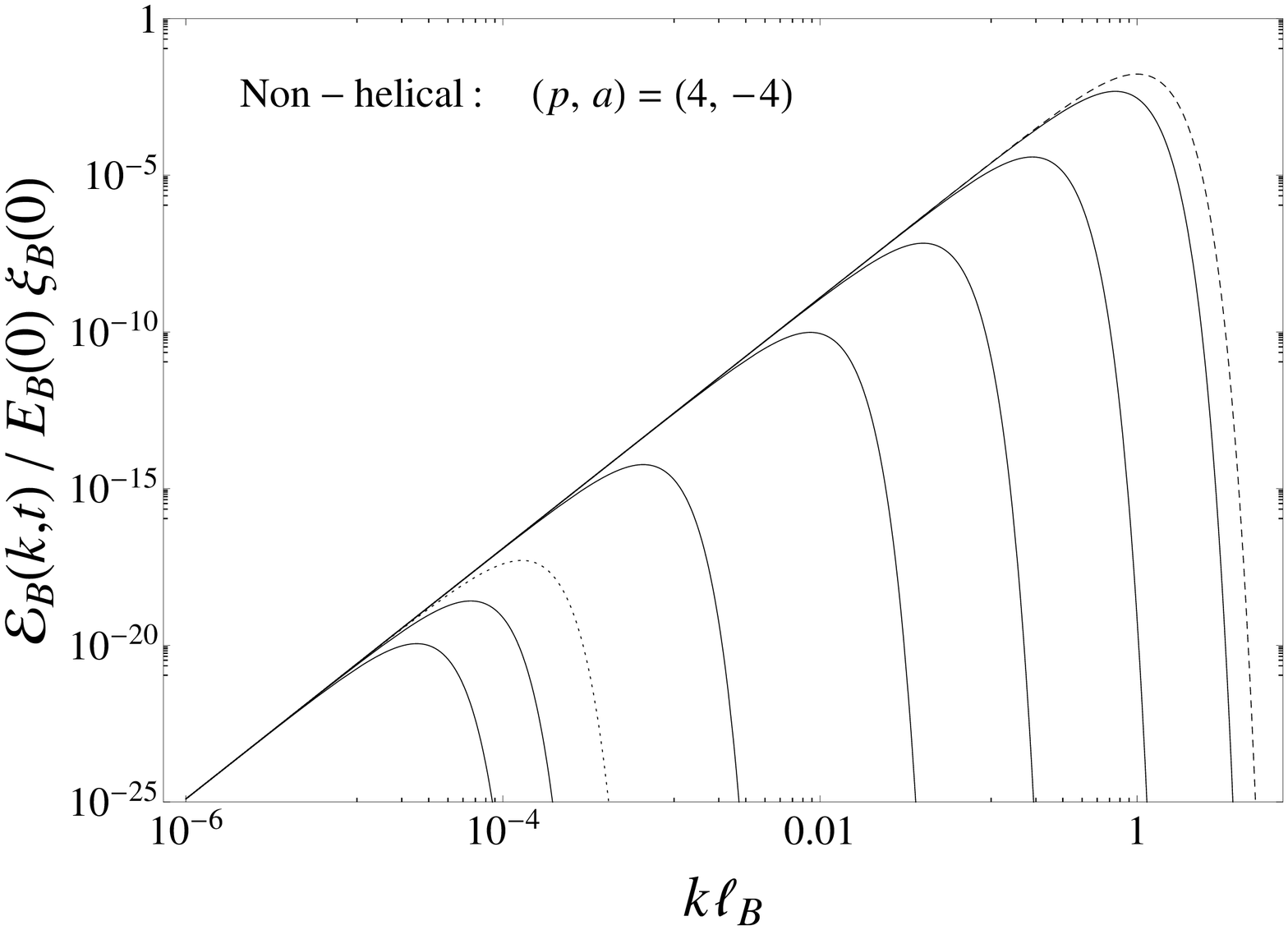}
\caption{Energy spectra in the non-helical case for $(p,a)=(2,-3)$ (upper left
panel), $(p,a)=(3,0)$ (upper right panel) and $(p,a)=(4,-4)$ (lower panel). Dashed lines correspond to the
initial spectra. Continuous lines correspond, from left to
right, to $t/\tau_{\rm eddy}(0) = 1,10,30,10^2,3 \times 10^2,10^4,3 \times 10^4$ (upper left panel),
$t/\tau_{\rm eddy}(0) = 1,10,10^2,10^3,10^4,10^5,10^6,10^7$ (upper right panel), and
$t/\tau_{\rm eddy}(0) = 1,10,10^2,10^3,3 \times 10^4,10^6,3 \times 10^6$ (lower panel).
Dotted lines correspond to $t/\tau_{\rm eddy}(0) \simeq 2.3 \times 10^3$ (upper left panel),
and $t/\tau_{\rm eddy}(0) \simeq 3.5 \times 10^5$ (lower panel).}
\end{center}
\end{figure}


\subsection{Vd. Solutions: helical case}

In the helical case, the dynamo and dissipation lengths, as well
as the ratio $\zeta_\alpha/\zeta_{\rm diss}$, are increasing
functions of time. This is inferred by the results of numerical
integration of Eqs.~(\ref{differential1})-(\ref{differential2}) as
shown, for example, in the left panel of Fig.~2. [Here, and in the following, we consider
the case of a very high magnetic Reynolds number, ${\text{Re}}_B = 10^{15}$].
In the Appendix B, we show that there are two different regimes depending on
the value of $\zeta_\alpha/\zeta_{\rm diss}$. For
$\zeta_\alpha/\zeta_{\rm diss} \ll 1$ the system behaves as if the
magnetic helicity were zero, so that the system evolves by
selective decay. The asymptotic solutions are then the
same as obtained in section Vc. For
$\zeta_\alpha/\zeta_{\rm diss} \gg 1$, instead, the mechanism operating in
the evolution of the system is such that energy is transferred
from small to large scales. In turbulent MHD, this mechanism is known as
{\it inverse cascade} and will be discussed, in more detail, later on in this subsection.

The following asymptotic ($\tau \rightarrow \infty$) solutions of
Eqs.~(\ref{differential1}) and (\ref{differential2}) are derived
in the Appendix B in the case of very large magnetic Reynolds numbers:
\begin{eqnarray}
\label{expansion-diss} \zeta_{\rm diss}(\tau) \!\!& \simeq &\!\!
c_5 \, (\ln \tau)^{1/6} \, \tau^{(1-a)/3},
\\
\label{expansion-alpha} \zeta_{\alpha}(\tau) \!\!& \simeq &\!\!
c_6 \, (\ln \tau)^{2/3} \, \tau^{(1-a)/3},
\end{eqnarray}
and
\begin{eqnarray}
\label{expansion-Energy} E_B(\tau) \!\!& \simeq &\!\! c_7 \,
E_B(0)
 \, (\ln \tau)^{1/3} \, \tau^{-(1-a)/3},
\\
\label{expansion-correlation} \xi_B(\tau) \!\!& \simeq &\!\! c_8
\, \xi_B(0) \, (\ln \tau)^{-1/3} \, \tau^{(1-a)/3},
\end{eqnarray}
where $c_5,c_6,c_7,c_8$ are given in Appendix A.
The above relations have been obtained supposing that the magnetic
helicity is almost constant during the evolution of the system.
This assumption is well justified by the results of
the numerical integration of Eqs.~(\ref{differential1}) and (\ref{differential2}),
as shown, for example, in the right panel of Fig.~2.


\begin{figure}[t!]
\begin{center}
\includegraphics[clip,width=0.45\textwidth]{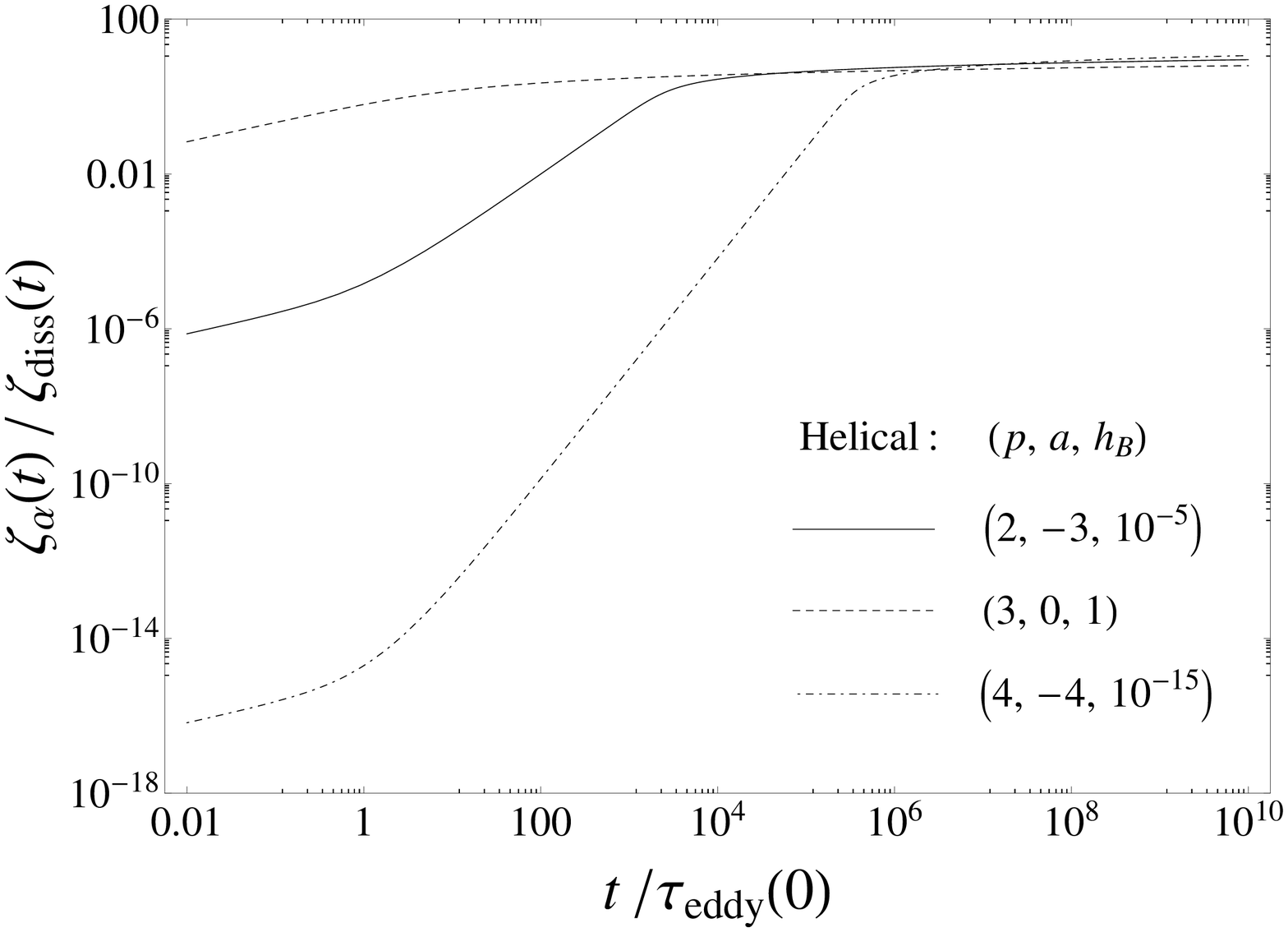}
\hspace{0.4cm}
\includegraphics[clip,width=0.45\textwidth]{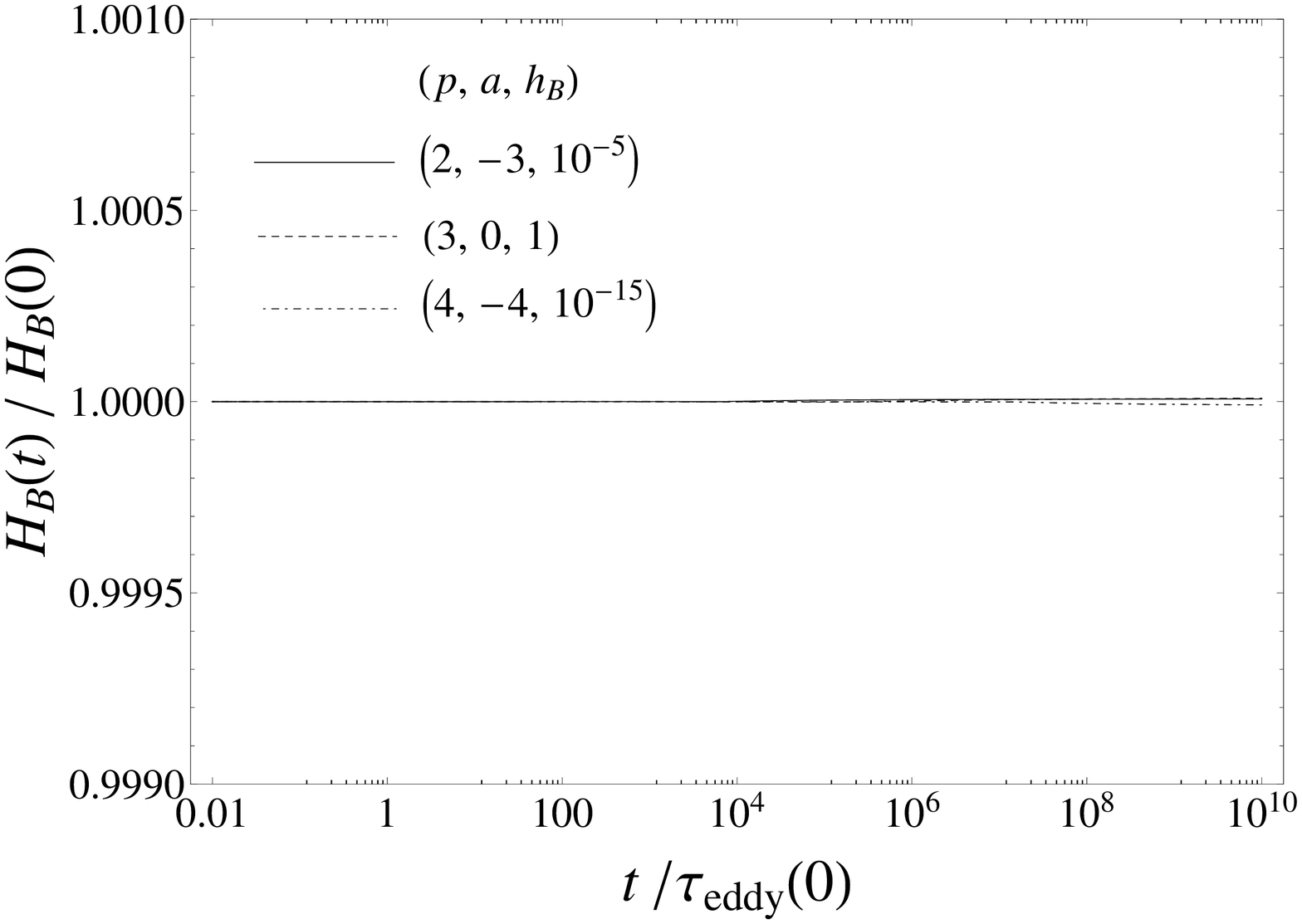}
\caption{Evolution of the $\alpha$-length normalized to the dissipation length (left panel),
and magnetic helicity (right panel), for the tree cases in Eq.~(\ref{setup2}).}
\end{center}
\end{figure}


In the upper panels of Fig.~3, we show the magnetic energy and correlation length as a function
of time for the three cases in Eq.~(\ref{setup2}). Dotted lines are
the asymptotic expansions~(\ref{expansion-Energy}) and (\ref{expansion-correlation})
for $\tau \geq \tau_E$ and $\tau \geq \tau_\xi$, respectively,
and the analytical results~(\ref{Es0}) and (\ref{xis0}) [with $\zeta_{\rm diss}$
given by Eq.~(\ref{zetadiss})] for $0 \leq \tau \leq \tau_E$ and $0 \leq \tau \leq \tau_\xi$,
respectively. Here, $\tau_E$ and $\tau_\xi$ are the
times when, approximatively, the integral quantities $E_B$ and $\xi_B$
enter into the inverse cascade regime.
We can find them by simply matching the
asymptotic solutions in the two different regimes, namely equating
Eqs.~(\ref{Es}) and (\ref{xis}) to Eqs.~(\ref{expansion-Energy}) and (\ref{expansion-correlation}),
respectively. We find,
$\tau_E \simeq (c_7/c_2)^{-q} \left[ \ln (c_7/c_2)^{-q} \right]^{-q/3}$
and $\tau_\xi \simeq (c_8/c_3)^{-2q} \left[ \ln (c_8/c_3)^{-2q} \right]^{2q/3}$,
%
%
where $q = 3(3+p)/[2p(1-a)]$.
%
%
In the case $h_B \ll 1$, the above results can be approximated as
\begin{eqnarray}
\label{tauEapprox} \tau_E \!\!& \simeq &\!\! c_9 \left(\ln h_B^{-1} \right)^{-q/3} h_B^{-2q/3} \! ,
\\
\label{tauxiapprox} \tau_\xi \!\!& \simeq &\!\! c_{10} \left(\ln h_B^{-1} \right)^{2q/3} h_B^{-2q/3} \! ,
\end{eqnarray}
where $c_9,c_{10}$ are given in Appendix A.


\begin{figure}[t!]
\begin{center}
\includegraphics[clip,width=0.45\textwidth]{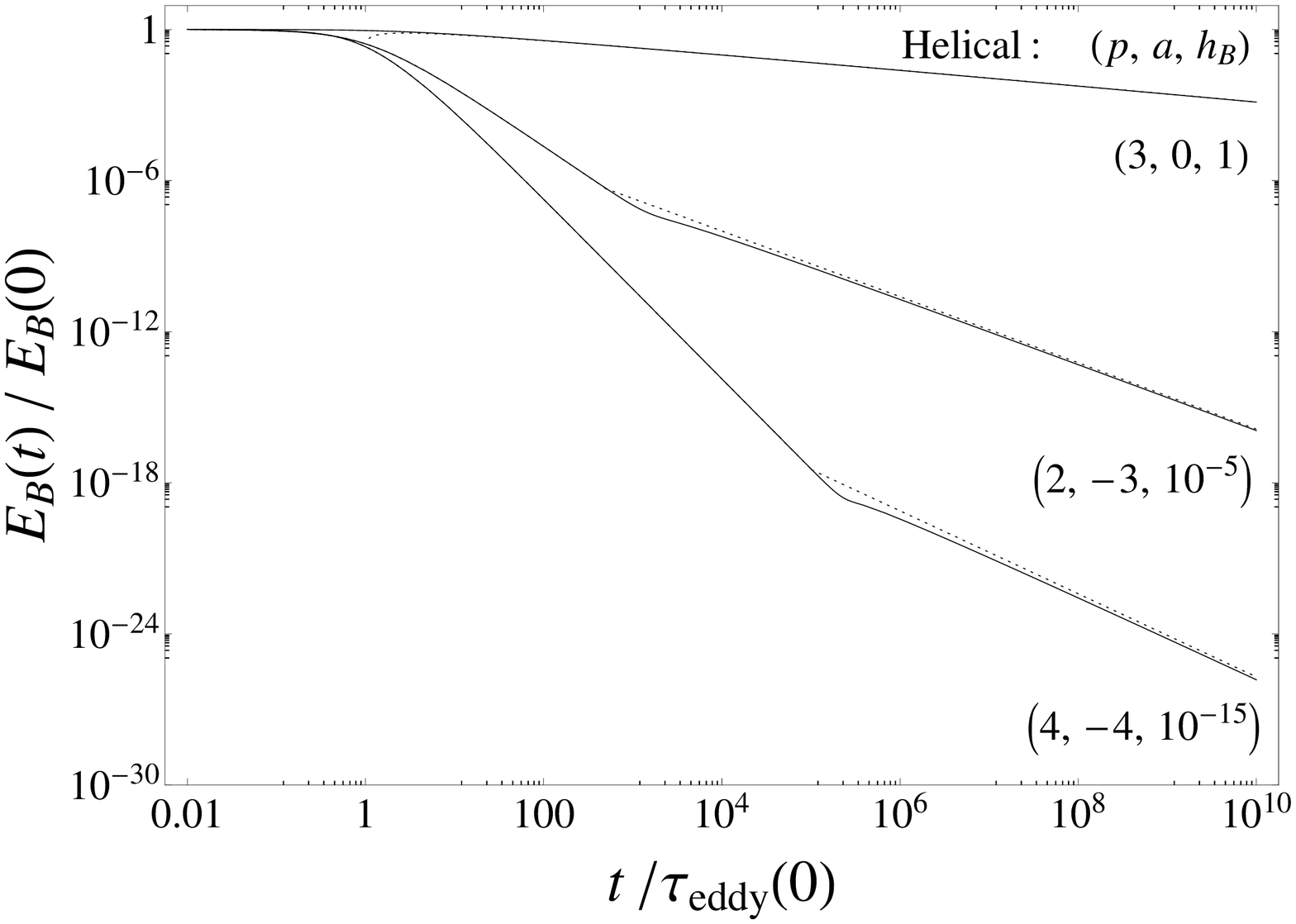}
\hspace{0.4cm}
\includegraphics[clip,width=0.45\textwidth]{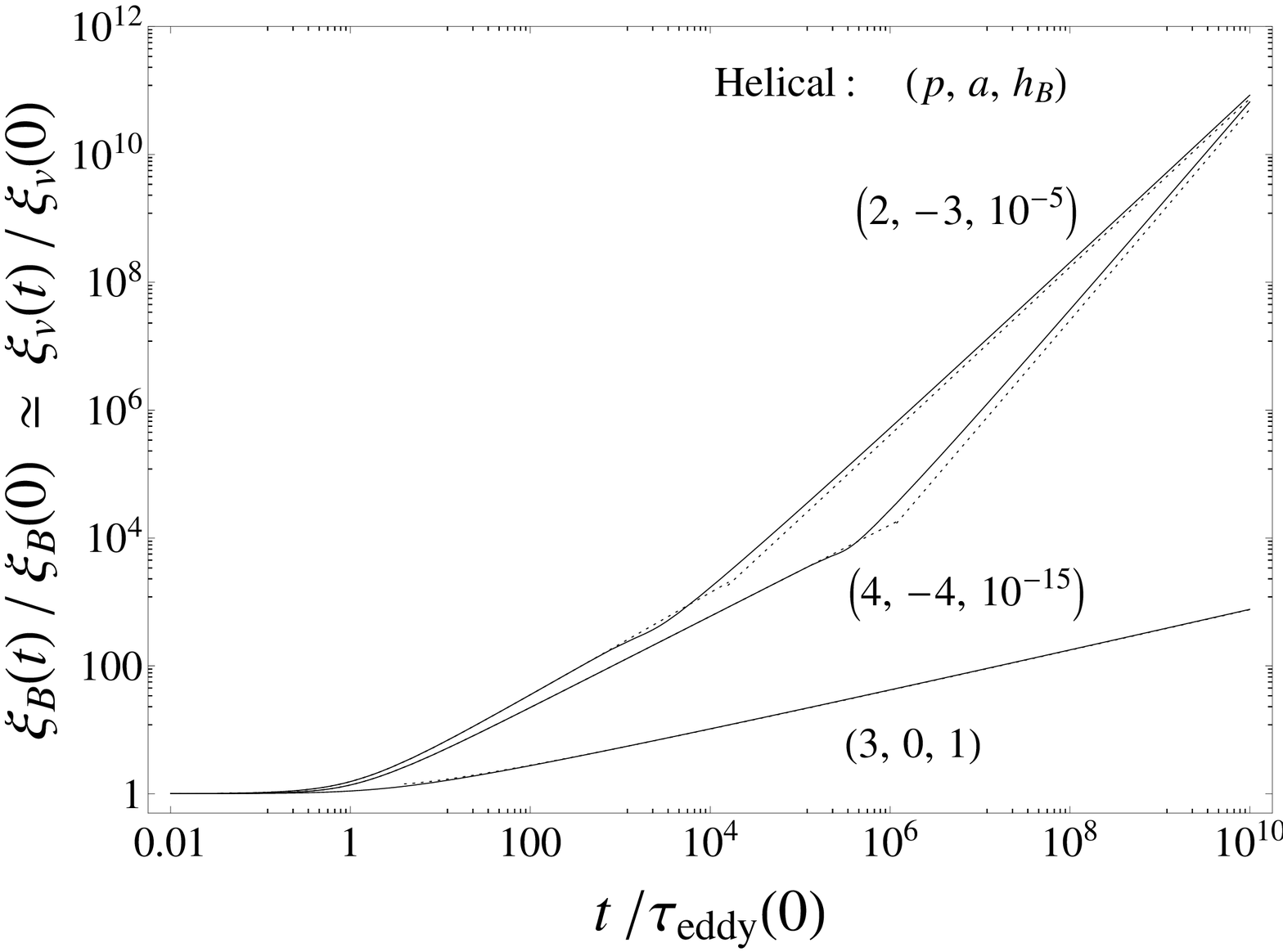}
\vspace{0.4cm}

\includegraphics[clip,width=0.45\textwidth]{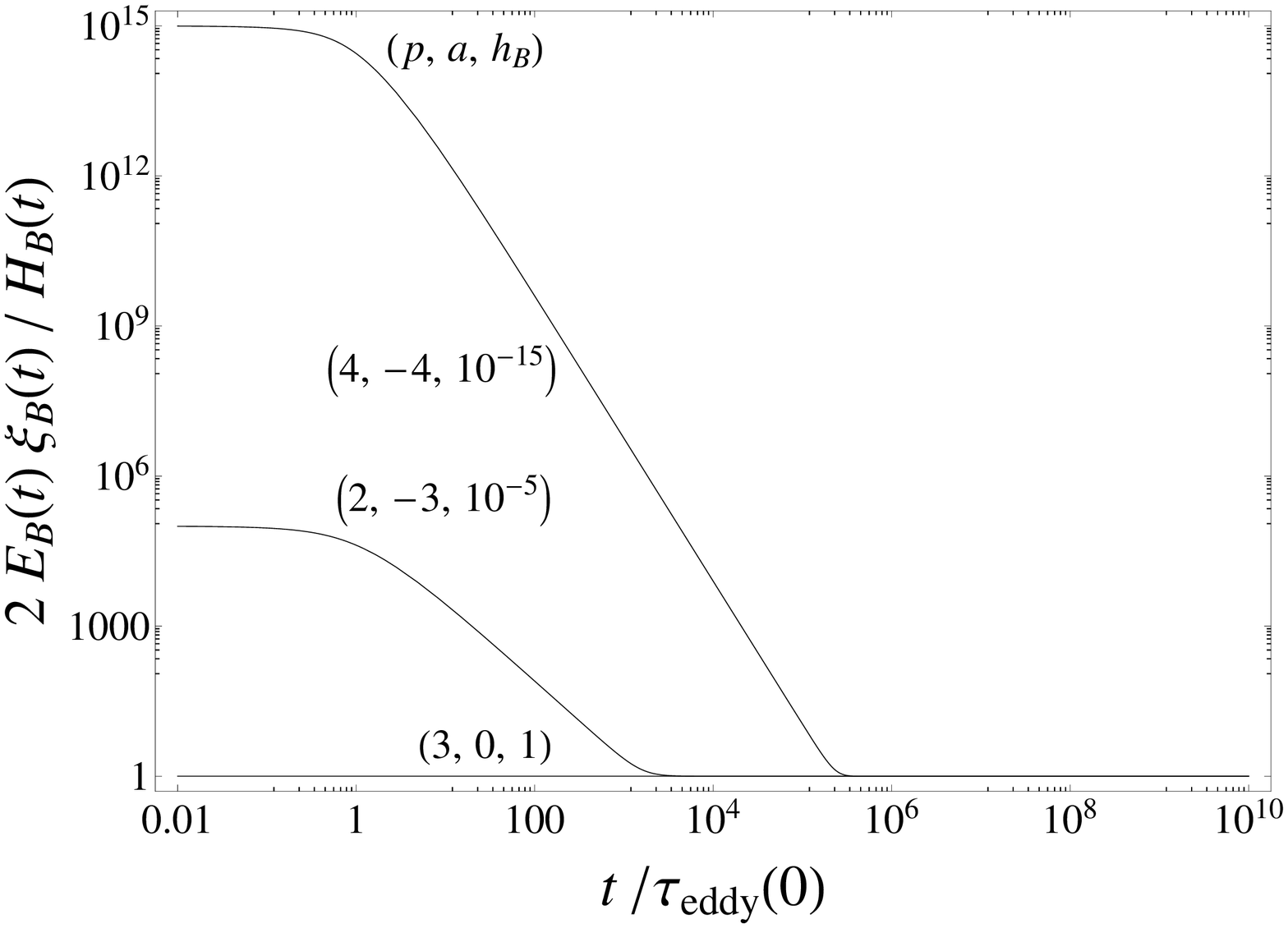}
\caption{The magnetic energy (upper left panel), the magnetic correlation
length (upper right panel), and the quantity $2E_B \xi_B / H_B$ (lower panel), as a function of time
for the tree cases in Eq.~(\ref{setup2}).
Dotted lines are the asymptotic expansions (indistinguishable from the continuous lines in the lower panel).}
\end{center}
\end{figure}


We also note that, in the inverse cascade regime, the product of
magnetic energy and correlation length is a quasi-conserved quantity. In
fact, as we show in the Appendix B, we have
\begin{equation}
\label{Exi0} E_B(\tau) \, \xi_B(\tau) \simeq \frac{H_B(\tau)}{2} \, .
\end{equation}
From the above equation, we deduce that a magnetic field with
fractional helicity becomes maximally helical approximatively
after the system enters into the inverse cascade phase. The time when this happens, $\tau_H$,
can be found by matching the product of asymptotic
solutions Eqs.~(\ref{Es})-(\ref{xis}) and (\ref{Exi0}). We get
\begin{equation}
\label{tauH} \tau_{H} \simeq c_{11} h_B^{-2q/3},
\end{equation}
where $c_{11}$ is given in Appendix A.
In the lower panel of Fig.~3, we show the quantity $2E_B \xi_B / H_B$ as a function of time
for the tree cases in Eq.~(\ref{setup2}). The asymptotic expansions for $\tau \leq \tau_H$
and $\tau \geq \tau_H$ are practically indistinguishable from the numerical solutions.

In the helical case, the evolution laws of magnetic energy and
correlation length do not depend on the index of the initial
magnetic energy spectrum. Moreover, apart logarithmic corrections,
and for the case of constant drag coefficient ($a=0$),
we find the $t^{-1/3}$ and $t^{1/3}$ laws which are indeed
observed in numerical simulations of dragged MHD~\cite{Banerjee1,Banerjee2}.

It is evident from all the above figures that the (approximate) analytical
expansions fit very well the numerical solutions. Moreover, the
magnetic helicity is, as expected, an almost constant function of
time. Because of the conservation of the magnetic helicity,
small-scale modes are not dissipated during the decay but their
energy is transferred to larger scales. This process of
inverse cascade is manifest in Fig.~4
(see, in particular, the magnetic spectra).
It is worth noting that, as explained above, this mechanism begins to operates
only after a certain time, that is when the selective decay
mechanism ends. We can suitably define the time when the inverse
cascade begins, $\tau_{\mathcal{E}}$, as the time when the maximum
of the magnetic spectrum meets the initial spectrum (see the magnetic
spectra in Fig.~4). Assuming for
simplicity $h_B \ll 1$, we get (see Appendix C)
\begin{equation}
\label{tauSpectrum} \tau_{\mathcal{E}} \simeq c_{12} h_B^{-2q/3} \! ,
\end{equation}
where $c_{12}$ is given in Appendix A.
Looking at Eqs.~(\ref{normalized}), (\ref{tauEapprox}), (\ref{tauxiapprox}),
(\ref{tauH}), (\ref{tauSpectrum}), and roughly
speaking, we can say that the time when system enters into the
inverse cascade regime is
$t_{\rm inv \, cas} \sim h_B^{-(3+p)/[p\,(1-a)]} \, \tau_{\rm eddy}(0)$.
%

Finally, we note that magnetohydrodynamic effects cannot change the characteristics of the
magnetic field on scales well above the integral scale (the magnetic correlation length);
in particular, in the limit $k \rightarrow 0$, the initial spectrum must retain its form for all times.
This is due to the fact that
the transfer of magnetic energy from small to larger scales (inverse cascade) is not instantaneous and needs
some time to develop. As a result, there will be always, at a given time,
a tail in the small wavenumber part of the magnetic spectrum which is leaved unprocessed by
the inverse cascade.
This phenomenon of inefficiency of inverse cascade for $k \rightarrow 0$
is visible in the case shown in the upper left panel of Fig.~4, where it is evident that
the magnetic energy spectrum retains its initial power law form at very large scales.
Such a phenomenon is however present in all the cases we have analyzed, although it
is not evident in the other panels of Fig.~4
since the magnetic energy and helicity spectra are not shown for very small wavenumbers.


\begin{figure*}[t!]
\begin{center}
\includegraphics[clip,width=0.45\textwidth]{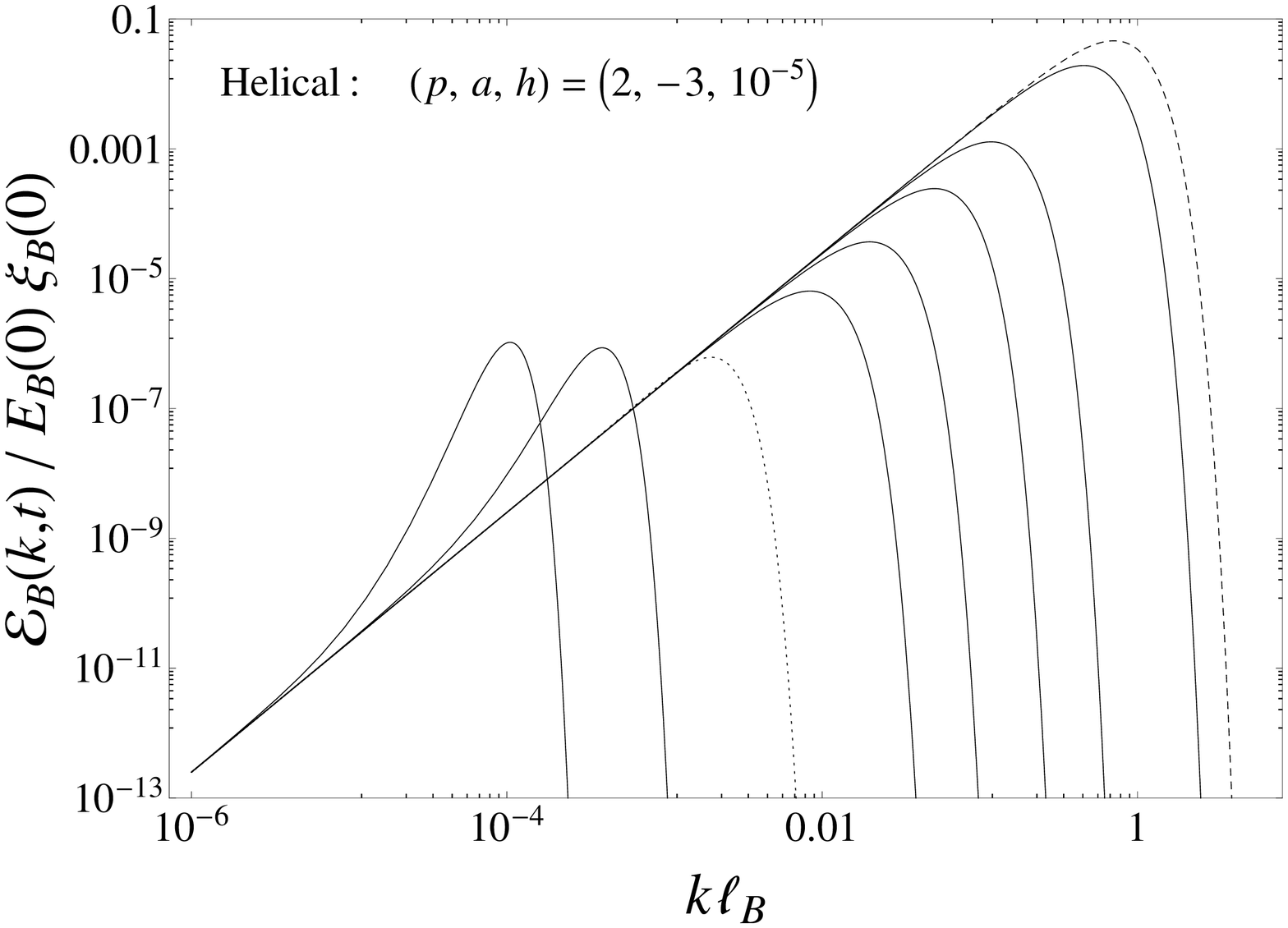}
\hspace{0.4cm}
\includegraphics[clip,width=0.45\textwidth]{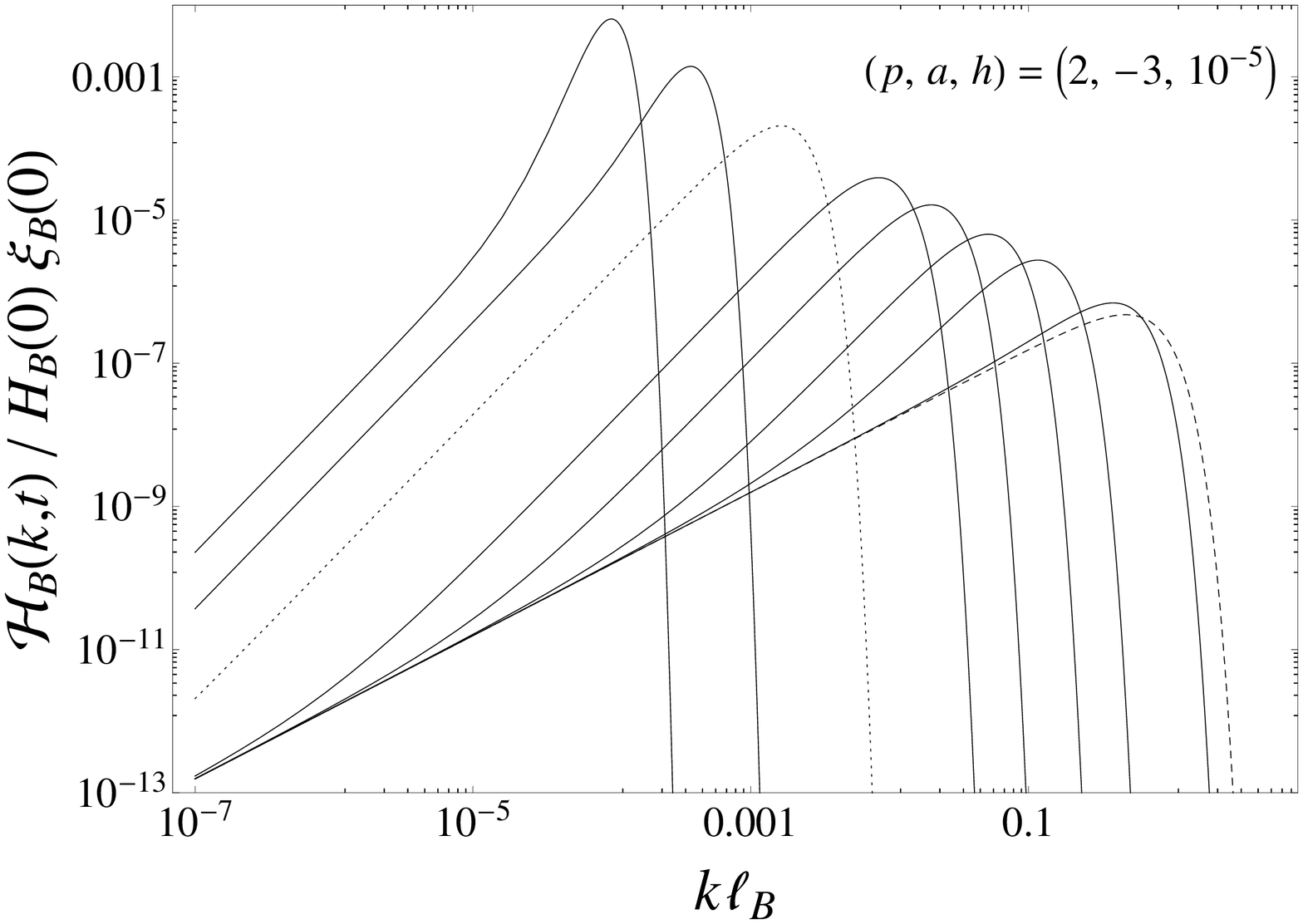}

\includegraphics[clip,width=0.45\textwidth]{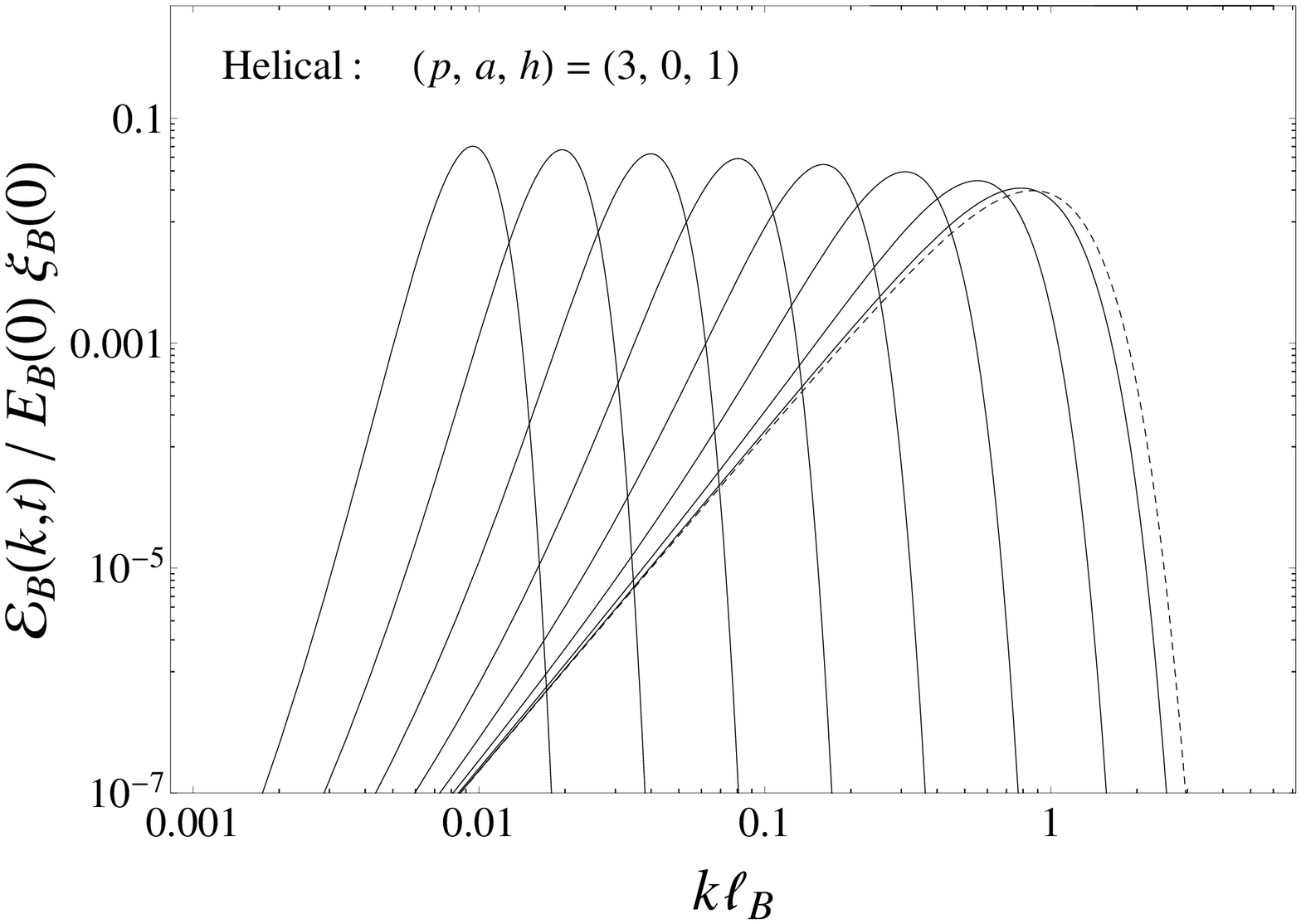}
\hspace{0.4cm}
\includegraphics[clip,width=0.45\textwidth]{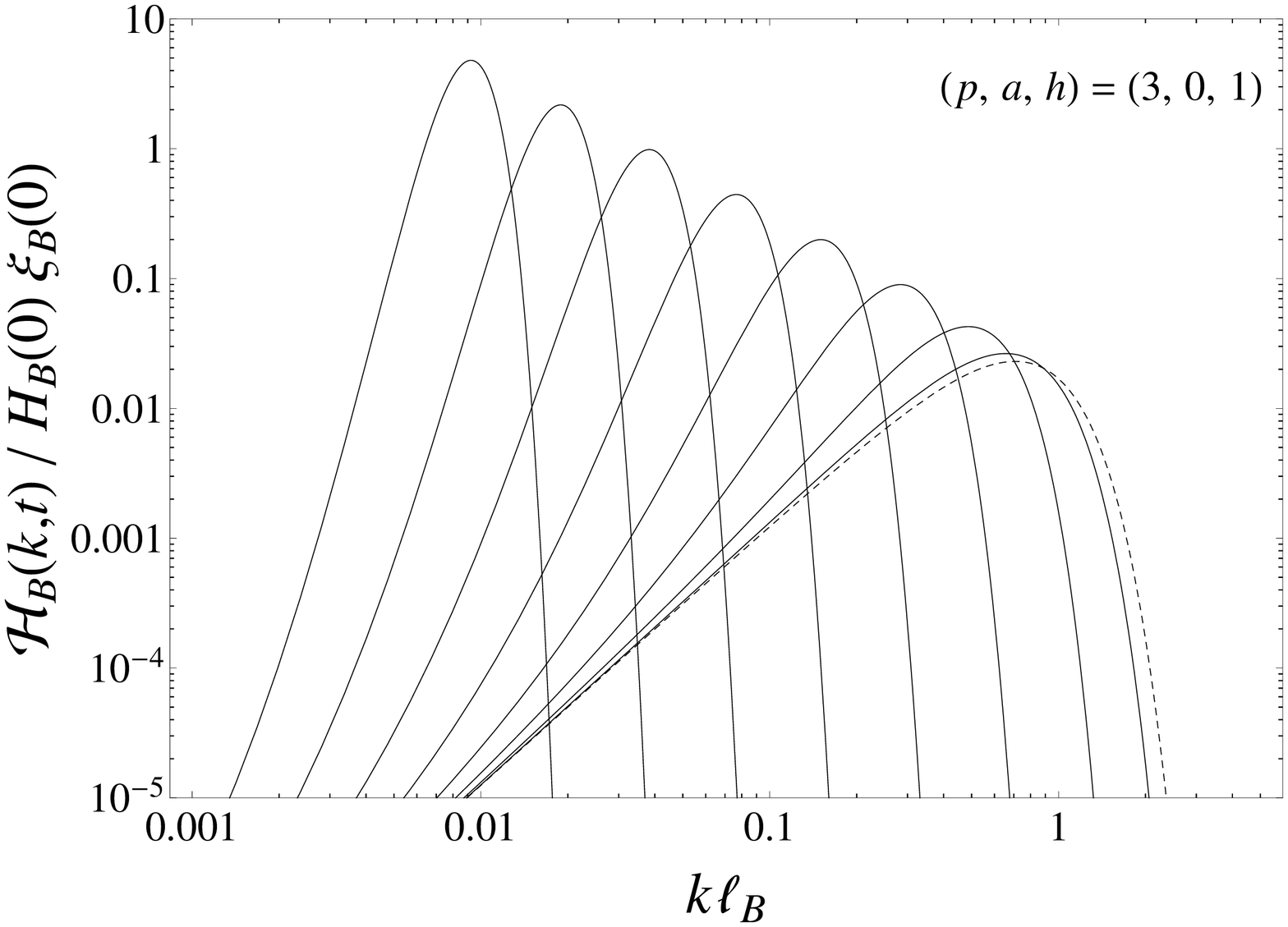}

\includegraphics[clip,width=0.45\textwidth]{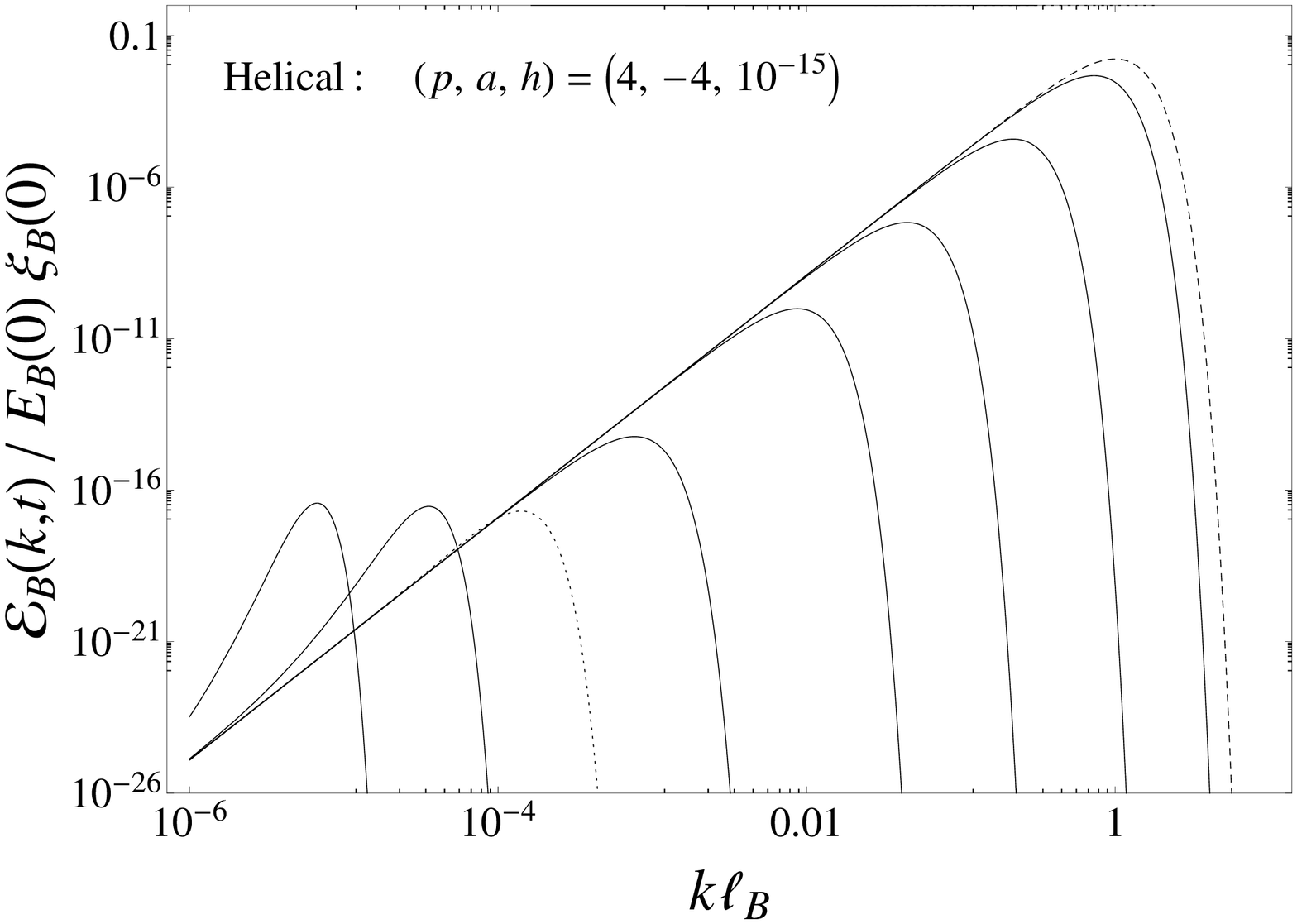}
\hspace{0.4cm}
\includegraphics[clip,width=0.45\textwidth]{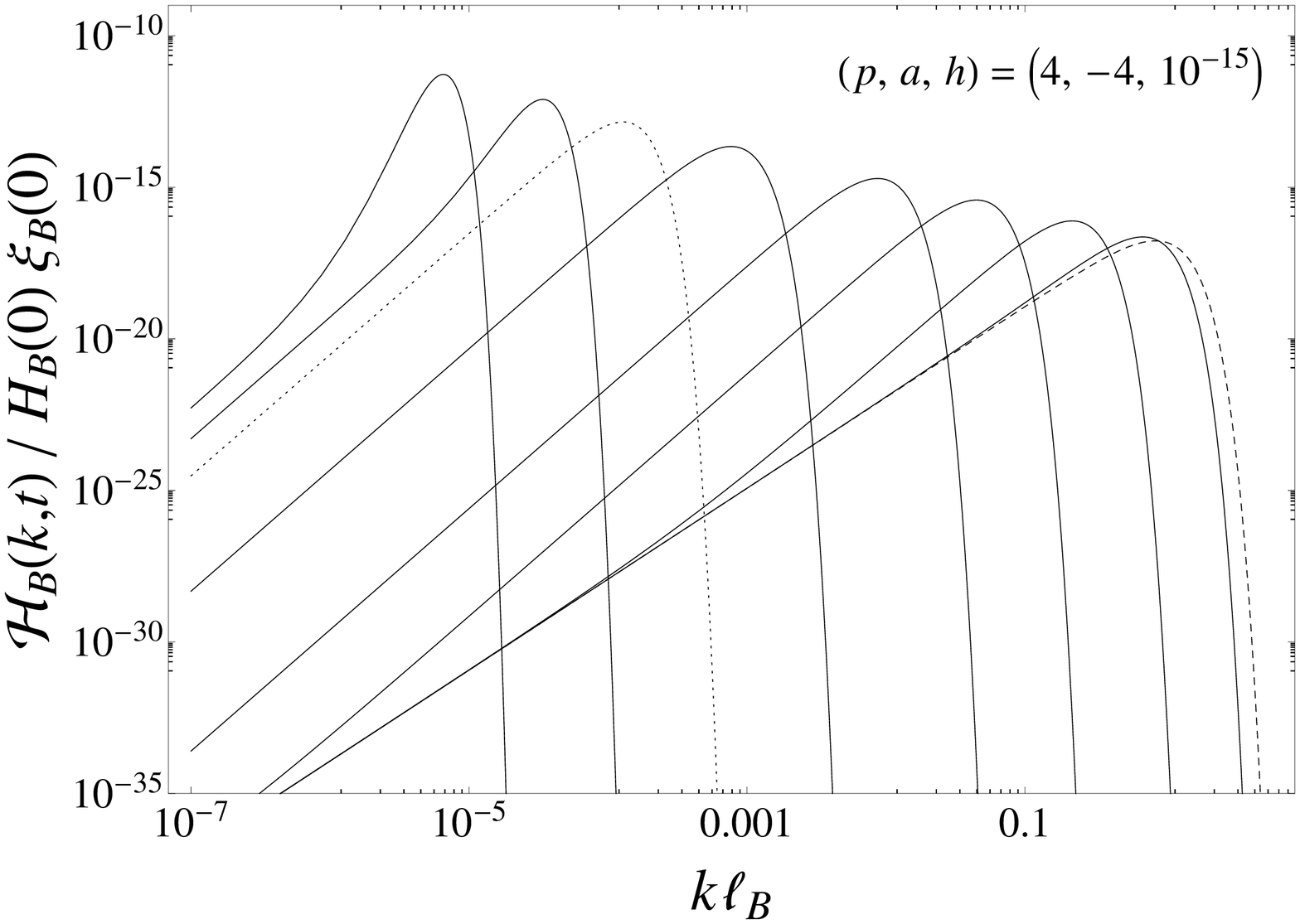}
\caption{{\it Upper panels}. Magnetic energy (left panel) and helicity (right panel)
spectra in the helical case for $(p,a,h)=(2,-3,10^{-5})$.
Dashed lines correspond to the initial spectra, while continuous lines correspond,
from left to right, to $t/\tau_{\rm eddy}(0) = 1,10,30,10^2,3 \times 10^2,10^4,3 \times 10^4$. Dotted lines
correspond to $t/\tau_{\rm eddy}(0) = \tau_{\mathcal{E}} \simeq 2.3 \times 10^3$.
{\it Middle panels}. As in the upper panels but for $(p,a,h)=(3,0,1)$ and
$t/\tau_{\rm eddy}(0) = 1,10,10^2,10^3,10^4,10^5,10^6,10^7$.
{\it Lower panels}. As in the upper panels but for $(p,a,h)=(4,-4,10^{-15})$,
$t/\tau_{\rm eddy}(0) = 1,10,10^2,10^3,3 \times 10^4,10^6,3 \times 10^6$, and
$\tau_{\mathcal{E}} \simeq 3.5 \times 10^5$.}
\end{center}
\end{figure*}


\subsection{Ve. Solutions: maximally helical case}

In the case where the initial magnetic field is maximally
helical, $h=1$, we get two exact analytical results. Firstly, from
Eqs.~(\ref{MF13}) and (\ref{MF14}) we find that the magnetic field
remains maximally helical for all times,
${\mathcal H}_B (k,t) = 2 k^{-1} {\mathcal E}_B(k,t)$.
%
%
Secondly, integrating the above expression in $k$ we have
$E_B(t) \, \xi_B(t) = H_B(t)/2$,
%
%
a result usually used in the literature for the evolution of a (maximally) helical magnetic field,
but never fully justified.

\subsection{Vf. Kinetic energy}

Starting from Eq.~(\ref{Eq12}) and taking into account Eq.~(\ref{alpha}), it is straightforward to find
the asymptotic expansions of $E_v$ for $\tau \gg 1$:
\begin{equation}
\label{v4} E_v(\tau) \propto
\left\{ \begin{array}{ll}
        \tau^{-\frac{2(2+a+p)}{3+p}},                &  \;\; \mbox{non-helical}, \\
        (\ln\tau)^{1/3} \, \tau^{-\frac{2(2+a)}{3}}, &  \;\; \mbox{helical},
    \end{array}
    \right.
\end{equation}
where we used Eq.~(\ref{Es}) for the non-helical case, and
Eq.~(\ref{expansion-Energy}) for the helical case.
Equation~(\ref{v4}) is in agreement with Eq.~(\ref{scalingEv}).

Moreover, evaluating Eq.~(\ref{Eq12}) at the time $\tau = 0$, and taking into account Eqs.~(\ref{Gamma}),
(\ref{delta}), (\ref{Es0}), and Eq.~(\ref{differential1}) in the limit of large magnetic Reynolds number,
%
%
we straightforwardly find that
$E_v(0) = [(1+p)/12] \, \xi_B(0)^2 \delta(0)^2 E_v(0)$, from which we get
\begin{equation}
\label{delta0} \delta(0) = \left( \frac{12}{1+p} \right)^{\!\!1/2} \! \frac{1}{\zeta_B} \, .
\end{equation}
Finally, starting from the definition of $\delta(0)$ in Eq.~(\ref{delta}),
and using Eqs.~(\ref{Gamma}) and (\ref{Rey-eddy2}) evaluated at $\tau = 0$, we find
\begin{equation}
\label{Ev0} E_v(0) = 2 \left[\frac{E_B(0)}{\delta(0) \, \alpha(0) \, \xi_{B}(0)} \right]^2 ,
\end{equation}
which relates the kinetic energy at the onset of the dragged phase to the initial values
of the drag coefficient, the magnetic field energy, and the correlation length.

\subsection{Vg. Exiting the dragged phase}

{\it Neutrino and photon mean free paths}.-- The dragged phase is
characterized by the condition that the comoving mean free path of
neutrinos, $\ell_{\rm mfp}^{(\nu)}$, and/or photons, $\ell_{\rm mfp}^{(\gamma)}$,
is much greater than the comoving magnetic
correlation length. Therefore, it is not obvious that this condition
in maintained during the evolution of the system, since
$\xi_B$ and both $\ell_{\rm mfp}^{(\nu)}$ and $\ell_{\rm mfp}^{(\gamma)}$
evolve in time. Observing that
$\ell_{\rm mfp}^{(\nu)} \propto T^{-4} \propto R^4 \propto \tilde{t}^4$ and
$\ell_{\rm mfp}^{(\gamma)} \propto T^{-2} \propto R^2 \propto \tilde{t}^2$
in radiation-dominated era~\cite{Banerjee1,Banerjee2}, we have
$\ell_{\rm mfp}^{(\nu)} \propto \tau^4$
and $\ell_{\rm mfp}^{(\gamma)} \propto \tau^2$.
%
%
Here, we used the results in section IId, the fact that $R \propto t^{1/2}$ in radiation era
(with $t$ being the cosmic time), and the fact that the temperature $T$ and the scale factor $R$
are related by $R \propto g_{*,S}^{-1/3} \, T^{-1}$~\cite{Kolb},
%
%
with $g_{*,S}$ being the effective number of entropy degrees of freedom at the temperature $T$.
On the other hand, the maximum growth
for the correlation length happens for the helical case and for the neutrino free-streaming case ($a=-4$)
[see Eqs.~(\ref{xis}) and (\ref{expansion-correlation})]. Namely, $\xi_B$ increases at most as
$\xi_B \propto (\ln \tau)^{-1/3} \, \tau^{5/3}$.
%
%
As a consequence, if the system is in a dragged phase such that
$\xi_B \ll \ell_{\rm mfp}^{(\nu)}$, $\ell_{\rm mfp}^{(\gamma)}$, it will
remain in it for all times.

However, the evolution laws for the magnetic energy and correlation length has been
obtained in the equilibrium state, where the kinetic Reynolds number
is much smaller than unity, $\text{Re} \ll 1$. This quantity generally evolves in time as the typical
velocity and correlation length of the fluid motion, and the drag parameter
change in time. Consequently, there could exist a time when the
condition $\text{Re} \ll 1$ ceases to be satisfied and the system exits the dragged phase.

{\it Kinetic Reynolds number}.-- Defining the kinetic Reynolds number accurately as
\begin{equation}
\label{R0} {\text{Re}} = \frac{v_{\rm rms}}{\xi_v \alpha} \, ,
\end{equation}
we find ${\text{Re}} = (-\dot{E}_B)^{1/2} \, \xi_v^{-1} \, \alpha^{-3/2}$,
%
%
where we used Eq.~(\ref{Eq12}).
The asymptotic expansions (for $\tau \gg 1$) of the kinetic Reynolds number
in the non-helical and helical cases, and for large magnetic Reynolds numbers, are
\begin{equation}
\label{R2}  \text{Re}(\tau) \propto
\left\{ \begin{array}{ll}
        \tau^{-(1+a)},         &  \;\; \mbox{non-helical}, \\
        (\ln\tau)^{1/2} \: \tau^{-(1+a)}, &  \;\; \mbox{helical},
    \end{array}
    \right.
\end{equation}
where we used the fact that $\xi_v(t) \propto \xi_B(t)$
[see the first relation in Eq.~(\ref{correlation})],
Eq.~(\ref{alpha}), Eqs.~(\ref{Es})-(\ref{xis}) for the non-helical case, and
Eqs.~(\ref{expansion-Energy})-(\ref{expansion-correlation}) for the helical case.
From Eq.~(\ref{R2}) we see that in the physical cases of interest, $a=-3,-4$, the
kinetic Reynolds number is an increasing function of time. Therefore, there will be
a time $t_{\rm exit}$ when the system leaves the dragged phase characterized by $\text{Re} \ll 1$.
This time can be operatively defined by the condition $\text{Re}(t_{\rm exit}) = 1$.
%

Finally, let us remember that the kinetic Reynolds number is approximately equal to
the $\Gamma$ ratio, $\text{Re}(t) \sim \Gamma(t)$ [see Eq.~(\ref{GammaRe})], so that
Eq.~(\ref{R2}) is in agreement with Eq.~(\ref{s4}).

\subsection{Vh. Evolution laws in expanding universe}

Returning to the ``tilde'' notation and indicating the time when the dragged phase begins as $\tilde{t}_i$,
equation~(\ref{s1}) reads
\begin{equation}
\label{zzz1} \alpha(\tilde{t}) = \alpha(\tilde{t}_i) \left( \frac{\tilde{t}}{\tilde{t}_i} \right)^{\!\!a}.
\end{equation}
Introducing the new variable $\tilde{\tau}$ as
\begin{equation}
\label{tautilde} \tilde{\tau} = \frac{\tilde{t}}{\tau_{\rm eddy}(\tilde{t}_i)} \, ,
\end{equation}
we can write Eq.~(\ref{zzz1}) as
$\alpha(\tilde{t}) = \alpha(\tilde{t}_i) [ \tau/\gamma(\tilde{t}_i) ]^{a}$
%
%
with $\gamma(\tilde{t}_i) = \tilde{t}_i/\tau_{\rm eddy}(\tilde{t}_i)$.
%
%
In radiation-dominated era, it is easy to see that $\gamma(\tilde{t}_i)$ can be expressed as
$\gamma(\tilde{t}_i) = [d_H(t_i)/\xi_{B,\rm phys}(t_i)]\,  v_{\rm rms}(t_i)$,
%
%
where $d_H = 2t$ is the length of the Hubble horizon at the time $t$, with $t$ being now the cosmic time.
Therefore, in the notation of section Va, we have $\gamma(0) = N_i v_i$,
%
%
where we have introduced the initial number of magnetic domains per horizon length
\begin{equation}
\label{zz1} N_i = \frac{d_H(0)}{\xi_{B,\rm phys}(0)} \, ,
\end{equation}
and we have defined $v_i = v_{\rm rms}(0)$
%
%
for the sake of simplicity.
Moreover, the normalized time $\tilde{\tau}$ in the radiation-dominated era is easily found to be
\begin{equation}
\label{X2} \tilde{\tau} = N_i v_i \frac{R}{R_i} \, .
\end{equation}
Finally, using Eq.~(\ref{Ev0}), we can relate $v_i$ to the initial values of the magnetic field strength,
$B^2 = B_{\textmd{rms}}^2 = \int \! d^{\,3} x \, {\textbf B}^2({\textbf x},t) = 2E_B$,
%
%
the physical magnetic correlation length,
$\xi_{B,\rm phys} = R \xi_B$,
%
%
and drag coefficient, as
\begin{equation}
\label{vi} v_i = \kappa_0 \, \frac{B(R_i)^2}{\alpha(R_i) \, \xi_{B,\rm phys}(R_i)} \, ,
\end{equation}
where $\kappa_0$ is given in Appendix A.
[We note that in the notation in which the quantity $\rho + P$ is not taken to be $1$,
the magnetic field strength $B(R_i)$ in Eq.~(\ref{vi}) should be replaced by $B(R_i)/\sqrt{\rho(R_i) + P(R_i)}$.]

{\it Non-helical case}.-- Translating the results previously found in the case of a Minkowski spacetime to the
case of a flat Friedmann universe, we get
\begin{equation}
\label{X4} \frac{R^2 B(R)}{R^2_i B(R_i)} \simeq
   \left\{ \begin{array}{ll}
        1, \;\;\;\;\;\;\;\;\;\;\;\;\;\;\;\;\;\;\;\;\;\;\;\;\;\;\;\;\;\;\;\;\;\;\;\;\;\;\;\;\;\;\;\;\;\;\;\;\: R_i \lesssim R \lesssim R_1, \\
        \kappa_{1} \, (N_i v_i)^{-\frac{1+p}{2(3+p)}} \left( \frac{R}{R_i} \right)^{\!\! -\frac{(1-a)(1+p)}{2(3+p)}},  \; R \gtrsim R_1,
    \end{array}
    \right.
\end{equation}
for the comoving magnetic field strength, and
\begin{equation}
\label{X5} \frac{\xi_B(R)}{\xi_B(R_i)} \simeq
\left\{ \begin{array}{ll}
        1, \;\;\;\;\;\;\;\;\;\;\;\;\;\;\;\;\;\;\;\;\;\;\;\;\;\;\;\;\;\;\;\;\;\;\;\;\;\;\, R_i \lesssim R \lesssim R_1, \\
        \kappa_{2} \, (N_i v_i)^{1/(3+p)} \left( \frac{R}{R_i} \right)^{\! \frac{1-a}{3+p}},  \; R \gtrsim R_1,
    \end{array}
    \right.
\end{equation}
for the comoving magnetic correlation length. Here,
\begin{equation}
\label{X6}
\frac{R_1}{R_i} \simeq \kappa_{3} (N_i v_i)^{-1/(1-a)} ,
\end{equation}
and $\kappa_1,\kappa_2,\kappa_3$ are given in Appendix A. In Eq.~(\ref{X6}), we have assumed that
$\kappa_{3} (N_i v_i)^{-1/(1-a)} > 1$. If this is not the case, $R_1$ is (approximatively) equal to $R_i$.

{\it Helical case}.-- In the case of helical magnetic fields, we find
\begin{equation}
\label{X7} \frac{R^2 B(R)}{R^2_i B(R_i)} \simeq
    \left\{ \begin{array}{lll}
        1, \;\;\;\;\;\;\;\;\;\;\;\;\;\;\;\;\;\;\;\;\;\;\;\;\;\;\;\;\;\;\;\;\;\;\;\;\;\;\;\;\;\;\;\;\;\;\;\;\;\;\;\;\;\;\;\;\;\;\;\;\;
        \;\;\;\;\;\;\;\;\;\;\;\,
        R_i \lesssim R \lesssim R_1, \\
        \kappa_{1} \, (N_i v_i)^{-\frac{1+p}{2(3+p)}} \left( \frac{R}{R_i} \right)^{\!\! -\frac{(1-a)(1+p)}{2(3+p)}},
        \;\;\;\;\;\;\;\;\;\;\;\;\;\;\;\;\;\;\;\;\;\;\;\;\, R_1 \lesssim R \lesssim R_2, \\
        \kappa_{4} \, h_B^{1/3} (N_i v_i)^{-1/6} \left(\ln N_i v_i \frac{R}{R_i}\right)^{\!1/6} \left( \frac{R}{R_i} \right)^{-(1-a)/6},
        \; R \gtrsim R_2,
    \end{array}
    \right.
\end{equation}
for the comoving magnetic field strength, and
\begin{equation}
\label{X8} \frac{\xi_B(R)}{\xi_B(R_i)} \simeq
\left\{ \begin{array}{lll}
        1, \;\;\;\;\;\;\;\;\;\;\;\;\;\;\;\;\;\;\;\;\;\;\;\;\;\;\;\;\;\;\;\;\;\;\;\;\;\;\;\;\;\;\;\;\;\;\;\;\;\;\;\;\;\;\;\;\;\;\;\;\;
        \;\;\;\;\;\;\;\;
        \; R_i \lesssim R \lesssim R_1, \\
        \kappa_{2} \, (N_i v_i)^{1/(3+p)} \left( \frac{R}{R_i} \right)^{\! \frac{1-a}{3+p}},
        \;\;\;\;\;\;\;\;\;\;\;\;\;\;\;\;\;\;\;\;\;\;\;\;\;\;\;\;\;\;\;\;\, R_1 \lesssim R \lesssim R'_2, \\
        \kappa_{5} \, h_B^{1/3} (N_i v_i)^{1/3} \left(\ln N_i v_i \frac{R}{R_i}\right)^{\!-1/3} \left( \frac{R}{R_i} \right)^{\! (1-a)/3}, \;  R \gtrsim R'_2,
    \end{array}
    \right.
\end{equation}
for the comoving magnetic correlation length. Here,
\begin{equation}
\label{X9} \frac{R_2}{R_i} \simeq \frac{\kappa_{6}}{N_i v_i} \: \omega \, (\ln \omega)^{-q/3} , \;\;\;\;
\frac{R'_2}{R_i} \simeq \frac{\kappa_{7}}{N_i v_i} \: \omega \, (\ln \omega)^{2q/3} ,
\end{equation}
with $\omega = \left( N_i v_i \right)^{-a/(1-a)} h_B^{-2q/3}$,
%
%
and $\kappa_4,\kappa_5,\kappa_6,\kappa_7$ are given in Appendix A. In the above equations,
we have assumed that $\omega$ is much greater than unity.

\section{VI. Dragged MHD in matter-dominated universe: free-streaming photons and hydrogen atoms}

We now discuss the evolution of a magnetic field in matter-dominated universe,
when the drag term in the MHD equations is ruled by the free-streaming properties
of photons or hydrogen atoms.

\subsection{VIa. Evolution laws in supercomoving variables}

The drag coefficients for photons and hydrogen atoms evolve as a function of the
temperature as~\cite{Banerjee1,Banerjee2} $\alpha_\gamma(T) \propto T^4$ and $\alpha_H(T) \propto T^3$, respectively.
In supercomoving variables (namely in Minkowski spacetime), their evolution with time is then
straightforwardly given by
\begin{equation}
\label{matter1} \alpha(\tilde{\tau}) = \alpha(0) \, e^{a \tilde{\tau}/\gamma(0)} ,
\end{equation}
where, taking into account the results of section IId, we used the fact that the normalized time $\tilde{\tau}$
in Eq.~(\ref{tautilde}) is
\begin{equation}
\label{F12}
\tilde{\tau} = \frac{1}{2} \, N_i v_i \ln \frac{R}{R_i}
\end{equation}
in matter-dominated era, when the scale factor evolves as $R \propto t^{2/3}$.
In Eq.~(\ref{matter1}),
\begin{equation}
\label{matter2} a =
    \left\{ \begin{array}{ll}
        -3, & \;\; \mbox{photons}, \\
        -5, & \;\; \mbox{hydrogens},
    \end{array}
    \right.
\end{equation}
and $\gamma(0) = N_i v_i/3$.
%
%
The initial number of magnetic domains per horizon length, $N_i$, is the same as in Eq.~(\ref{zz1}),
the length of the Hubble horizon in matter-dominated era being $d_H = 3t$ (with $t$ the cosmic time),
and $v_i = v_{\rm rms}(0)$ is given, as in the case of radiation-dominated universe, by Eq.~(\ref{vi}).

In supercomoving variables, the master equation~(\ref{differential1}) reads, for the case at hand,

\begin{equation}
\label{MD1} \frac{d\zeta_{\rm diss}^2}{d\tilde{\tau}} = \frac{\zeta_B^2}{\text{Re}_B(0)} +
\frac{\zeta_B^2}{3} \, \delta(0) \, e^{-a \tilde{\tau}/\gamma(0)} \: \frac{E_B(\tilde{\tau})}{E_B(0)} \, .
\end{equation}
Defining the new variable
$\gamma(0) [e^{-a \tilde{\tau}/\gamma(0)} - 1] = N_i v_i (R/R_i)^{3/2}/3$,
%
%
the above equation becomes
\begin{equation}
\label{MD3} \frac{d\zeta_{\rm diss}^2}{d\bar{\tau}} = \frac{\zeta_B^2}{\text{Re}_B(0)} +
\frac{\zeta_B^2}{3} \, \delta(0) [1 + \bar{\tau}/\gamma(0)]^{-(a+1)} \, \frac{E_B(\bar{\tau})}{E_B(0)} \, .
\end{equation}
Comparing Eq.~(\ref{MD3}) with Eq.~(\ref{differential1}), we see that the evolution
of the magnetic energy and correlation length can be derived by applying the substitutions
$\tau \rightarrow \bar{\tau}$ and $a \rightarrow a + 1$
%
%
in the solutions found in the case of a Minkowski spacetime in sections Vc, Vd, and Ve.

\subsection{VIb. Evolution laws in expanding universe}

{\it Non-helical case}.-- Using the results in the above subsection, we straightforwardly find the
evolution laws for the the comoving magnetic field strength,
\begin{equation}
\label{Y4} \frac{R^2 B(R)}{R^2_i B(R_i)} \simeq
\left\{ \begin{array}{ll}
        1, \;\;\;\;\;\;\;\;\;\;\;\;\;\;\;\;\;\;\;\;\;\;\;\;\;\;\;\;\;\;\;\;\;\;\;\;\;\;\;\;\;\;\;\,
        R_i \lesssim R \lesssim R_1, \\
        \kappa_{8} \, (N_i v_i)^{-\frac{1+p}{2(3+p)}} \left( \frac{R}{R_i} \right)^{\! \frac{3a(1+p)}{4(3+p)}},  \; R \gtrsim R_1,
    \end{array}
    \right.
\end{equation}
and for the comoving magnetic correlation length,
\begin{equation}
\label{Y5} \frac{\xi_B(R)}{\xi_B(R_i)} \simeq
\left\{ \begin{array}{ll}
        1, \;\;\;\;\;\;\;\;\;\;\;\;\;\;\;\;\;\;\;\;\;\;\;\;\;\;\;\;\;\;\;\;\;\;\;\;\;\;\;\;\;\;\;\, R_i \lesssim R \lesssim R_1, \\
        \kappa_{9} \, (N_i v_i)^{1/(3+p)} \left( \frac{R}{R_i} \right)^{\! -\frac{3a}{2(3+p)}},  \; R \gtrsim R_1,
    \end{array}
    \right.
\end{equation}
where
\begin{equation}
\label{Y6}
\frac{R_1}{R_i} \simeq \kappa_{10}(N_i v_i)^{\frac{2}{3a}} ,
\end{equation}
and $\kappa_8,\kappa_9,\kappa_{10}$ are given in Appendix A. In Eq.~(\ref{Y6}), we have assumed that
$\kappa_{10}(N_i v_i)^{2/(3a)} > 1$. If this is not the case, then $R_1 \simeq R_i$.

{\it Helical case}.-- In the helical case, we get
\begin{equation}
\label{Y7} \frac{R^2 B(R)}{R^2_i B(R_i)} \simeq
\left\{ \begin{array}{lll}
        1, \;\;\;\;\;\;\;\;\;\;\;\;\;\;\;\;\;\;\;\;\;\;\;\;\;\;\;\;\;\;\;\;\;\;\;\;\;\;\;\;\;\;\;\;\;\;\;\;\;\;\;\;\;\;\;\;\;
        \;\;\;\;\;\;\;\;\;\;\;\;\;\;\;\;\;\;\;\,
        R_i \lesssim R \lesssim R_1, \\
        \kappa_{8} \, (N_i v_i)^{-\frac{1+p}{2(3+p)}} \left( \frac{R}{R_i} \right)^{\!\! \frac{3a(1+p)}{4(3+p)}},
        \;\;\;\;\;\;\;\;\;\;\;\;\;\;\;\;\;\;\;\;\;\;\;\;\;\;\;\;\;\;\;\;\;\;\,
        R_1 \lesssim R \lesssim R_2, \\
        \kappa_{11} \, h_B^{1/3} (N_i v_i)^{-1/6} \left[\ln \frac13 N_i v_i \! \left(\frac{R}{R_i}\right)^{3/2}\right]^{\!1/6} \left( \frac{R}{R_i} \right)^{\! a/4},                            \; R \gtrsim R_2,
    \end{array}
    \right.
\end{equation}
for the comoving magnetic field strength, and
\begin{equation}
\label{Y8}  \frac{\xi_B(R)}{\xi_B(R_i)} \simeq
\left\{ \begin{array}{lll}
        1, \;\;\;\;\;\;\;\;\;\;\;\;\;\;\;\;\;\;\;\;\;\;\;\;\;\;\;\;\;\;\;\;\;\;\;\;\;\;\;\;\;\;\;\;\;\;\;\;\;\;\;\;\;\;\;\;\;
        \;\;\;\;\;\;\;\;\;\;\;\;\;\;\;\;\;\;\;\;\;\,
        R_i \lesssim R \lesssim R_1, \\
        \kappa_{9} \, (N_i v_i)^{1/(3+p)} \left( \frac{R}{R_i} \right)^{\! -\frac{3a}{2(3+p)}},
        \;\;\;\;\;\;\;\;\;\;\;\;\;\;\;\;\;\;\;\;\;\;\;\;\;\;\;\;\;\;\;\;\;\;\;\,
        R_1 \lesssim R \lesssim R'_2, \\
        \kappa_{12} \, h_B^{1/3} (N_i v_i)^{1/3} \left[\ln \frac13 N_i v_i \! \left(\frac{R}{R_i}\right)^{3/2}\right]^{\!-1/3} \left( \frac{R}{R_i} \right)^{\! -a/2},          \;  R \gtrsim R'_2,
    \end{array}
    \right.
\end{equation}
for the comoving magnetic correlation length, where
\begin{equation}
\label{Y9} \frac{R_2}{R_i}  \simeq \left[\frac{\kappa_{13}}{\frac13 N_i v_i} \: \bar{\omega} \, (\ln \bar{\omega})^{-\bar{q}/3}\right]^{2/3}, \;\;\;\;\frac{R'_2}{R_i} \simeq \left[\frac{\kappa_{14}}{\frac13 N_i v_i} \: \bar{\omega} \, (\ln \bar{\omega})^{2\bar{q}/3}\right]^{2/3},
\end{equation}
with $\bar{\omega} = (N_i v_i)^{(1+a)/a} h_B^{-2\bar{q}/3}/3$
and $\bar{q} = -3(3+p)/(2pa)$,
%
%
and $\kappa_{11},\kappa_{12},\kappa_{13},\kappa_{14}$ are given in Appendix A. In the above equations, we have assumed that
$\bar{\omega} \gg 1$.


\subsection{VIc. Exiting the dragged phase}

{\it Photon mean free path}.-- In the matter-dominated era,
the comoving mean free path of photons scales in time as
$\ell_{\rm mfp}^{(\gamma)} \propto T^{-2} \propto R^2$.
%
%
On the other hand, the maximum growth
for the correlation length happens for helical fields in the case of free-streaming photons.
Accordingly, $\xi_B$ increases at most as
$\xi_B \propto (\ln R^{3/2})^{-1/3} R^{2}$.
%
%
As a consequence, if the system is in a dragged phase such that
$\xi_B \ll \ell_{\rm mfp}^{(\gamma)}$, it will remain in it for all times.

{\it Hydrogen atoms mean free path}.-- The comoving mean free path of
hydrogen atoms evolves as
$\ell_{\rm mfp}^{(H)} \propto T^{-2} \propto R^2$.
%
%
The correlation length in the hydrogen atoms free-streaming case, instead,
increases as $\xi_B \propto R^{15/[2(3+p)]}$ in the non-helical case, and
$\xi_B \propto (\ln R^{3/2})^{-1/3} R^{5/2}$ in the helical case.
%
%
As a consequence, the system will eventually exit the dragged phase in the helical case, and in the
non-helical case for the case $p \leq 4/3$ (we remember that we are assuming $p > 1$).
Let us call $t_{1, \rm exit}$ the time when this happens, which can be defined by the condition
$\xi_B(t_{1, \rm exit}) = \ell_{\rm mfp}^{(H)}(t_{1, \rm exit})$.
%

{\it Kinetic Reynolds number}.-- The asymptotic expansions of the kinetic
Reynolds number in the non-helical and helical cases, and for large magnetic Reynolds numbers, are
\begin{equation}
\label{RR2}  \text{Re}(R) \propto
\left\{ \begin{array}{ll}
        R^{-3a/2},         &  \;\; \mbox{non-helical}, \\
        \left( \ln R^{3/2} \right)^{1/2} \: R^{-3a/2}, &  \;\; \mbox{helical}.
    \end{array}
    \right.
\end{equation}
From the above equation, we see that the kinetic Reynolds number is an increasing function of time
in the physical cases of interest, $a=-3,-5$. Therefore, there will be
a time $t_{2, \rm exit}$ when the system will leave the dragged phase characterized by $\text{Re} \ll 1$.
This time can be operatively defined by the condition
$\text{Re}(t_{2, \rm exit}) = 1$.
%
%
Accordingly, and in the cases of physical interest $p=2,4$, we have that the dragged phase terminates at
the time $t_{\rm exit}$ given by
\begin{equation}
\label{exit4}  t_{\rm exit} =
\left\{ \begin{array}{lll}
        t_{2, \rm exit} \: ,                           &  \;\; \mbox{photon case}, \\
        t_{2, \rm exit} \: ,                           &  \;\; \mbox{hydrogen, non-helical case}, \\
        \min\{t_{1, \rm exit} \, ,t_{2, \rm exit}\},   &  \;\; \mbox{hydrogen, helical case}.
    \end{array}
    \right.
\end{equation}
Finally, we observe that, as in the radiation-dominated era, we find that $\text{Re} \sim \Gamma$,
in agreement with Eq.~(\ref{s4}).


\section{VII. Justifying the Banerjee and Jedamzik scaling arguments}

In this section, we reconsider some scaling arguments used by Banerjee and Jedamzik
in~\cite{Banerjee1,Banerjee2} to explain their results obtained by direct numerical integration
of dragged MHD equations. Our goal here, it to show that some working hypotheses used
in~\cite{Banerjee1,Banerjee2} find indeed a justification in the light of our analytical results
obtained in sections II, V and VI.

{\it Master equation}.-- In order to discuss the Banerjee and Jedamzik's scaling arguments,
let us introduce the notations used in~\cite{Banerjee1,Banerjee2}. We define the magnetic energy in
the mode $k$, $E_B(k,t)$, by $E_B(t) = \int_0^\infty \! dk k^{-1} E_B(k,t)$.
%
%
This is related to the magnetic energy spectrum by $E_B(k,t) = k \mathcal{E}_B(k,t)$.
Accordingly, we can define the magnetic energy on the scale $l \sim 1/k$ as
$E_B(l,t) = (1/l) \, \mathcal{E}_B(1/l,t)$.
%
%
At the initial time $t=0$, the magnetic energy on the scale $l$ is proportional to a power of $l$,
$E_B(l,0) \propto l^{-n}$, where $n = 1+p$ in the notation of~\cite{Banerjee1,Banerjee2}.

In~\cite{Banerjee1,Banerjee2}, it is assumed (without a full justification) that the evolution of the magnetic energy
on the integral scale $l_B(t)$ (namely the scale under which the magnetic energy is dissipated) is
ruled by the simple scaling equation
\begin{equation}
\label{Z1} \frac{dE_B(l_B)}{dt} \sim \frac{E_B(l_B)}{\tau_{\rm eddy}} \, ,
\end{equation}
where $E_B(l_B) = E_B(l_B(t),t)$. The above equation can be justified in the light of our results
in section IIb. In fact, from Eq.~(\ref{Eq12}), it follows that
\begin{equation}
\label{Z2} \frac{dE_B(t)}{dt} \sim  -2 \, \frac{E_B(t)}{\tau_{\rm eddy}} \, ,
\end{equation}
where $E_B(t)$ is the magnetic energy. In order to obtain the above equation,
we eliminated $E_v$ in Eq.~(\ref{Eq12}) in favor of $\Gamma$ and then used the fact that
$\alpha \Gamma \sim 1/\tau_{\rm eddy}$ [see Eq.~(\ref{nice})].
We now use the fact that the magnetic spectrum, in both the helical and non-helical cases,
is peaked at a wavenumber $k_{\rm max}(t) \sim 1/\xi_B(t) \sim 1/l_B(t)$,
as it is not hard to see using the expressions for the magnetic energy spectrum and correlation
length in section V,
so that $E_B(t) = \int_0^\infty \! dk \, \mathcal{E}_B(k,t) \sim k_{\rm max} \mathcal{E}_B(k_{\rm max},t) \sim E_B(l_B)$.
This last observation, when combined with Eq.~(\ref{Z2}), gives Eq.~(\ref{Z1}).

Now, we can transform Eq.~(\ref{Z1}) into an equation which does not involve the bulk field $v$
(which enters in the equation through the quantity $\tau_{\rm eddy}$). To this end, we use Eq.~(\ref{Olesen1})
and approximate the nabla operator by $\nabla \sim 1/l_B$. We find that the typical velocity $v$ and
the typical magnetic field strength $B$ are related through
$v \sim B^2/(\alpha l_B) \sim E_B/(\alpha l_B)$.
In turn, this gives for the eddy turnover time
$\tau_{\rm eddy} \sim l_B^2/(\alpha B)$.
%
%
Inserting the above equation in Eq.~(\ref{Z1}) we get
\begin{equation}
\label{Z3} \frac{dE_B(l_B)}{dt} \sim \frac{E_B^2(l_B)}{\alpha l_B^2} \, .
\end{equation}
This is the master equation used in~\cite{Banerjee1,Banerjee2} to derive the evolution laws
for the magnetic energy and correlation length. As we have just shown, this equation
is a direct consequence of Eqs.~(\ref{Eq15}) and (\ref{Eq12}).

{\it Non-helical case}.-- In~\cite{Banerjee1,Banerjee2}, it is assumed that the magnetic energy on the integral scale
$l_B$ is related to the integral scale itself by a simple power law
\begin{equation}
\label{Z4} E_B(l_B) \propto l_B^{-n}.
\end{equation}
This is indeed legitimated by the fact that, as we have shown in sections V and VI, the
evolution of the magnetic spectrum in the non-helical case proceeds through selective decay,
so that the initial magnetic spectrum retains its form (for length greater than the integral scale)
for all times.

From Eq.~(\ref{Z4}), we have $l_B \propto E_B(l_B)^{-1/n}$ that, when inserted in
Eq.~(\ref{Z3}), gives $E_B(t) \propto \left[ \int dt/\alpha(t) \right]^{-n/(2+n)}$
%
%
and, accordingly, $l_B(t) \propto \left[ \int dt/\alpha(t) \right]^{1/(2+n)}$.
%
%
Remembering that the index $n$ is related to our index $p$ through $n = 1+p$, and
taking into account the expressions for $\alpha$ as a function of time, we get
$E_B(t) \propto t^{-(1-a)(1+p)/(3+p)}$ and $\xi_B(t) \propto t^{(1-a)/(3+p)}$
%
%
in the case of free-streaming neutrinos and photons in radiation-dominated universe, and
$E_B(t) \propto e^{[a(1+p)/(3+p)]\, t}$ and $\xi_B(t) \propto e^{-[a/(3+p)] \, t}$
%
%
in the case of free-streaming photons and hydrogen atoms in matter-dominated universe.

The above evolution laws are in agreement with those derived in mean-field approximation
(see sections V and VI).

{\it Helical case}.-- In the helical case, the magnetic helicity is related to the magnetic
energy on the integral scale $l_B$ through
$E_B(l_B) \, l_B \simeq H_B(l_B)$.
%
%
Since the magnetic helicity on the integral scale $l_B$ is a quasi-conserved
quantity, $H_B(l_B) \simeq \mbox{constant}$, we have $l_B \propto E_B(l_B)^{-1}$.
Inserting the previous equation in Eq.~(\ref{Z3}), we obtain
$E_B(t) \propto \left[ \int dt/\alpha(t) \right]^{-1/3}$
%
%
and, consequently, $l_B(t) \propto \left[ \int dt/\alpha(t) \right]^{1/3}$.
%
%
Writing $\alpha$ as a function of time, we get $E_B(t) \propto t^{-(1-a)/3}$ and
$\xi_B(t) \propto t^{(1-a)/3}$
%
%
for free-streaming neutrinos and photons in radiation-dominated era, and
$E_B(t) \propto e^{(a/3) \, t}$ and $\xi_B(t) \propto e^{-(a/3) \, t}$
%
%
for free-streaming photons and hydrogen atoms in matter-dominated era.

Also in the helical case, the above evolution laws are in agreement with those derived in mean-field approximation
(see sections V and VI).

\section{VIII. Freely-decaying turbulent magnetic fields}

The problem of determining the evolution properties of freely-decaying
magnetic fields in MHD turbulence, in contrast to the case of dragged MHD,
has been deeply and widely discussed in the literature. Essentially, there are
been five ways to tackle it:
$i$) Applying scaling arguments to turbulent MHD equations~\cite{Olesen,Shiromizu,Kalelkar,Campanelli3};
$ii$) Using simple toy models to ``mimic'' the full turbulent MHD equations~\cite{Brandenburg,Brandenburg1,Kalelkar};
$iii$) Solving the turbulent MHD equations in Eddy-Damped Quasi-Normal Markovian (EDQNM) approximation~\cite{Son};
$iv$) Studying MHD turbulence equations in mean-field-theory approximation~\cite{Campanelli5};
$v$) Solving numerically the turbulent MHD equations~\cite{Biskamp1,Christensson1,Kalelkar,Banerjee1,Banerjee2,Christensson2,
Kahniashvili3,Tevzadze,Kahniashvili4}.

The invariance of turbulent MHD equations under scaling transformations has
been firstly used by Olesen~\cite{Olesen} to study the evolution
properties of the magnetic spectrum and correlation length. Although
this is a very simple and elegant way to study MHD turbulence,
this approach lives some unresolved questions, such as when the
system enters into the scaling regime, what is exactly the shape
of magnetic spectrum, etc. 

The most famous toy model for hydrodynamic turbulence is the
so-called {\it shell model} based on the original idea by Desnyanski
and Novikov~\cite{DN}, developed later by Gledzer~\cite{G},
Ohkitani and Yamada~\cite{OY}, and now know as GOY-model.
Subsequently, it was generalized by Gloaguen {\it et al.} to
include the case of magnetic turbulence~\cite{GOY_B}. The basic
idea on which repose the GOY-model is that the interaction due to
non-linear terms in the MHD equations are local in $k$-space. Then,
convolutions in MHD equations are approximated by sums over the
nearest and next nearest neighbors in a discretized wavenumber
space (the shell space). Moreover, the shell-model equations mimic the
full MHD equations retaining their main characteristic, such as
energy conservation in the case of null dissipation, phase space
conservation, etc.
Also, there exists a continuous version of discretized shell model in
which the distance between shells goes to zero. It was firstly
introduced by Parisi~\cite{Parisi} in a hydrodynamical context and
later extended to the case of MHD turbulence by Brandenburg,
Enqvist, and Olesen~\cite{Brandenburg1}.
Although many properties of turbulence, such as energy transfer,
general spectral properties, etc., have been studied successfully
using shell models, the physical validity of such toy models
remains an open question (for a review on shell models
of magnetohydrodynamic turbulence, see~\cite{Frick et al}).

The EDQNM approximation is the most widely two-point-closure
approximation used for studying hydrodynamical turbulence. It was first
proposed by Millionshtchikov~\cite{EDQNM1}, developed by Proudman
and Reid~\cite{EDQNM2}, and extended to MHD turbulence by Pouquet
{\it et al.}~\cite{EDQNM3}. It consists in replacing correlation
functions by suitable statistical averages~\cite{Verma},
neglecting (conveniently) high-order correlators, and using a
phenomenological expressions for the so-called ``eddy damping rate''.
Even if the resulting equations are more simple than the full MHD
equations, they can be solved only numerically and, moreover, some
unresolved questions concerning EDQNM approximation still exist~\cite{Biskamp}.

In~\cite{Campanelli5}, analytical results for the evolution of the magnetic
energy density and correlation length were obtained in mean-field-theory approximation.
Below, we briefly review such an approach to the study of turbulent MHD equations, and
extend the results obtained in non-expanding universe to the case of a flat Friedmann spacetime.

A direct integration of the full set of turbulent MHD equations is
certainly the best way to study the dynamics of freely-decaying MHD
turbulence. However, due to the complexity of such highly non-linear
equations, the numerical results in the literature are often contrasting

In Tab.~1, indeed, we show some results found in the literature concerning
the evolution laws of magnetic energy and correlation length in
freely-decaying MHD turbulence in Minkowski spacetime.
In all cases, the assumed initial
magnetic energy spectrum has a simple power-law behavior, and the
dissipation parameters are constants. Obviously, one should keep
in mind that these results were obtained by using different
initial conditions, such as initial magnetic Reynolds number,
initial $\Gamma$-ratio, an so on.
As it is evident from the Table, the situation is still unclear.
Nevertheless, we see that convergence in the results, at least for the most recent works
and for the case of helical magnetic fields seems achieved
(see, in particular, Refs.~\cite{Banerjee1,Banerjee2,Campanelli5,Tevzadze,Kahniashvili4}).


\begin{table}[t!]
\caption{Evolution laws for the magnetic energy, $E_B(t) \propto t^{-\sigma}$, and magnetic
correlation length, $\xi_B(t) \propto t^{\varsigma}$, in helical and non-helical freely-decaying
magnetohydrodynamic turbulence. All results were obtained assuming an initial power-law for the
magnetic energy spectrum, ${\mathcal E}_B(k,0) \propto k^p$, and constant dissipation parameters.
The asterisk refers to analytical solutions of continuous shell model.}
\vspace{0.5cm}
\begin{tabular}{ccccccc}
\hline \hline
& Reference                     & Method     ~~& $p$ ~~~& $\sigma$       ~~~& $\varsigma$ ~& Helical \\
\hline
& \cite{Olesen}                 & Scaling    ~~& $p$ ~~~& --             ~~~& $2/(3+p)$   ~& No  \\
& \cite{Shiromizu}              & Scaling    ~~& $p$ ~~~& --             ~~~& $2/(3+p)$   ~& No  \\
& \cite{Kalelkar}               & Scaling    ~~& $1$ ~~~& $1$            ~~~& $1/2$       ~& No  \\
&                               & Shell      ~~& $1$ ~~~& $1$            ~~~& $0.5$       ~& No  \\
&                               & Numerical  ~~& $1$ ~~~& $\sim 1$       ~~~& $0.4$       ~& No  \\
& \cite{Campanelli3}            & Scaling    ~~& $p$ ~~~& $1$            ~~~& $1/2$       ~& No  \\
&                               & Scaling    ~~& $p$ ~~~& $1/2$          ~~~& $1/2$       ~& Yes \\
& \cite{Brandenburg}            & Shell      ~~& $2$ ~~~& --             ~~~& $0.25$      ~& No  \\
& \cite{Brandenburg1}           & Shell$^*$  ~~& $p$ ~~~& --             ~~~& $1/(3+p)$   ~& No  \\
& \cite{Son}                    & EDQNM      ~~& $2$ ~~~& $\sim 6/5$     ~~~& $\sim 2/5$  ~& No  \\
& \cite{Biskamp1}               & Numerical  ~~& $2$ ~~~& $0.5$          ~~~& --          ~& Yes \\
& \cite{Christensson1}          & Numerical  ~~& $4$ ~~~& $0.7$          ~~~& $0.4$       ~& No  \\
&                               & Numerical  ~~& $4$ ~~~& $1.1$          ~~~& $0.5$       ~& Yes \\
& \cite{Banerjee1,Banerjee2}    & Numerical  ~~& $2$ ~~~& $1.05$         ~~~& --          ~& No  \\
&                               & Numerical  ~~& $4$ ~~~& $0.5$          ~~~& --          ~& Yes \\
& \cite{Christensson2}          & Numerical  ~~& $4$ ~~~& $0.5$          ~~~& $0.5$       ~& Yes \\
& \cite{Campanelli5}            & Mean-field ~~& $p$ ~~~& $2(1+p)/(3+p)$ ~~~& $2/(3+p)$   ~& No  \\
&                               & Mean-field ~~& $p$ ~~~& $2/3$          ~~~& $2/3$       ~& Yes \\
& \cite{Kahniashvili3}          & Numerical  ~~& $4$ ~~~& $\sim 1$       ~~~& $\sim 1/2$  ~& No  \\
& \cite{Tevzadze}               & Numerical  ~~& $4$ ~~~& $\sim 1$       ~~~& $\sim 1/2$  ~& No  \\
&                               & Numerical  ~~& $4$ ~~~& $\sim 2/3$     ~~~& $\sim 2/3$  ~& Yes \\
& \cite{Kahniashvili4}          & Numerical  ~~& $4$ ~~~& $0.9$          ~~~& $\sim 1/2$  ~& No  \\
&                               & Numerical  ~~& $4$ ~~~& $\sim 2/3$     ~~~& $\sim 2/3$  ~& Yes \\
\hline \hline
\end{tabular}
\end{table}


\subsection{VIIIa. Evolution laws in non-expanding universe}

Prior to recombination, namely for temperatures
$T \gtrsim T_{\rm rec} \simeq 0.3 \, \mbox{eV}$~\cite{Kolb},
the equations of turbulent MHD are given by Eqs.~(\ref{Eq1}) and (\ref{Eq2}) by the replacement
\begin{equation}
\label{freely1} - \alpha {\textbf v} \rightarrow \nu \nabla^2 {\textbf v},
\end{equation}
where $\nu$ is the kinematic viscosity.
After recombination, instead, and in the tight-coupling regime between ions and neutral particle species,
the turbulent MHD equations are given by Eqs.~(\ref{Eq1}) and (\ref{Eq2}) by the
replacement~\cite{Banerjee1,Banerjee2}
\begin{equation}
\label{freely2} - \alpha {\textbf v} \rightarrow
\frac{1}{\rho_{i} \alpha_{\rm in}} \nabla \times ({\textbf F}_L \times {\textbf B}),
\end{equation}
where $\rho_{i}$ is the matter density of ions and $\alpha_{\rm in}$ is the momentum transfer rate
due to ion-neutral collisions~\cite{Banerjee1,Banerjee2}.

Although the form of the dissipation term is different before and after recombination,
it is inessential to our discussion, since we work in the hypothesis of large
kinetic Reynolds numbers, namely assuming that dissipation is negligible.
[After recombination, the condition of tight-coupling is equivalent to assuming that the
dissipative term in Eq.~(\ref{freely2}) is negligible in the Navier-Stokes
equation~\cite{Banerjee1,Banerjee2}.]

Proceeding as in section IVb, we can decompose the magnetic field according to Eq.~(\ref{MFT1}).
In this case, the quadratic term ${\textbf v} \cdot \nabla {\textbf v}$ can be approximated by
${\textbf v} \cdot \nabla {\textbf v} \simeq {\textbf v}_0 \cdot \nabla \widetilde{{\textbf v}}$.
This term in the Navier-Stokes equation can be neglected if the condition
$\Gamma |\widetilde{{\textbf v}}|/|{\textbf v}_0| \ll 1$ is satisfied. In fact,
comparing it with the Lorentz force, we have
$|{\textbf v} \cdot \nabla {\textbf v}|/ |{\textbf F}_L| \sim
\Gamma |\widetilde{{\textbf v}}|/|{\textbf v}_0|$. Since we are assuming
$|\widetilde{{\textbf v}}| \ll |{\textbf v}_0|$, the above condition is certainly valid
if the $\Gamma$ ratio does not assume very large values
(and this is the case studied in~\cite{Campanelli5} and in this paper).
The approximation just discussed corresponds to assuming that the small-scale components
of the velocity field do not play any role in the development of MHD turbulence, at least
on large scales. This means, in turn, that the Lorentz force acting on the charged particles
of the fluid is responsible for the development of turbulence on large scales.
In the limit of vanishing dissipation or high Reynolds numbers (see below),
we have from Eq.~(\ref{Eq1bis}) that $\partial_t {\textbf v} \simeq {\textbf F}_L$.

To further simplify the Navier-Stokes equation, we write $\partial_t \sim 1/\tau_d$,
giving ${\textbf v} \simeq \tau_d {\textbf F}_L$,
where $\tau_d$ is the fluid-response time to the
Lorentz force. This sort of ``drag time' was first introduced in~\cite{Sigl}, and its explicit expression
as a function of time was derived in~\cite{Campanelli5},
$\tau_d = \tau_d(0) ( 1 +  \tau/ \gamma_\infty)$,
%
%
where $\tau_d(0) = \Gamma(0) \, \tau_{\rm eddy}(0)$,
%
%
and
\begin{equation}
\label{F3} \gamma_\infty = \lim_{t \rightarrow \infty} \frac{\Gamma(0)}{\Gamma(t)} \frac{t}{\tau_{\rm eddy}(t)} \, .
\end{equation}
The quantity $\gamma_\infty$ is a constant since the results of numerical integration
of turbulent MHD equations give, asymptotically, $\Gamma(t) \sim 1$ in the non-helical
case and $\Gamma(t) \sim \mbox{constant}$ (eventually different from unity) in the
helical case~\cite{Banerjee1,Banerjee2}, while scaling arguments indicate that
$\tau_{\rm eddy}(t) \propto t$~\cite{Campanelli5}.

Consequently, the turbulent MHD equations are, formally, the same as the dragged MHD equations and then
the master equation for the quantity $\zeta_{\rm diss}$, which rules then the evolution
of the magnetic energy and correlation length is
\begin{equation}
\label{FF1} \frac{d\zeta_{\rm diss}^2}{d\tau} = \frac{\zeta_B^2}{\text{Re}_B(0)} +
\frac{\zeta_B^2}{3} (1 + \tau/\gamma_\infty) \frac{E_B(\tau)}{E_B(0)} \, .
\end{equation}
Looking at Eqs.~(\ref{FF1}) and (\ref{differential1}), we see that the substitutions
$\delta(0) = 1$, $\gamma(0) \rightarrow \gamma_\infty$, and $a = -1$
%
%
in the solutions found for the dragged case (see sections Vc, Vd, and Ve),
give directly the solutions for the magnetic energy and correlation length
in freely-decaying MHD turbulence.

\subsection{VIIIb. Evolution laws in radiation-dominated universe}

{\it Non-helical case}.-- Using the results in the above subsection and taking into account
Eq.~(\ref{X2}), we easily find the evolution
laws of the magnetic field in radiation-dominated universe:
\begin{equation}
\label{F6} \frac{R^2 B(R)}{R^2_i B(R_i)} \simeq
\left\{ \begin{array}{ll}
        1,  \;\;\;\;\;\;\;\;\;\;\;\;\;\;\;\;\;\;\;\;\;\;\;\;\;\;\;\;\;\;\;\;\;\;\;\;\;\;\: R_i \lesssim R \lesssim R_1, \\
        \kappa_{15} \, (N_i v_i)^{-\frac{1+p}{3+p}} \left( \frac{R}{R_i} \right)^{\!\! -\frac{1+p}{3+p}},  \; R \gtrsim R_1,
    \end{array}
    \right.
\end{equation}
for the comoving magnetic field strength, and
\begin{equation}
\label{F7} \frac{\xi_B(R)}{\xi_B(R_i)} \simeq
\left\{ \begin{array}{ll}
        1,  \;\;\;\;\;\;\;\;\;\;\;\;\;\;\;\;\;\;\;\;\;\;\;\;\;\;\;\;\;\;\;\;\;\;\;\;\;\;\;\, R_i \lesssim R \lesssim R_1, \\
        \kappa_{16} \, (N_i v_i)^{2/(3+p)} \left( \frac{R}{R_i} \right)^{\! \frac{2}{3+p}},  \; R \gtrsim R_1,
    \end{array}
    \right.
\end{equation}
for the comoving magnetic correlation length, where
$v_i = \Gamma_i^{1/2} B(R_i)$
%
%
[$B(R_i)$ should be replaced by $B(R_i)/\sqrt{\rho(R_i) + P(R_i)}$ in the notation in
which the quantity $\rho + P$ is not taken to be $1$, as already noted in section Vh].
In the above equations,
\begin{equation}
\label{F8}
\frac{R_1}{R_i} \simeq \kappa_{17} (N_i v_i)^{-1} ,
\end{equation}
and $\kappa_{15},\kappa_{16},\kappa_{17}$ are given in Appendix A. In Eq.~(\ref{F8}), we have assumed that
$\kappa_{17} (N_i v_i)^{-1} > 1$. In the opposite case, we have approximatively $R_1 \simeq R_i$.

{\it Helical case}.-- In the helical case, we find
\begin{equation}
\label{F9} \frac{R^2 B(R)}{R^2_i B(R_i)} \simeq
    \left\{ \begin{array}{lll}
        1, \;\;\;\;\;\;\;\;\;\;\;\;\;\;\;\;\;\;\;\;\;\;\;\;\;\;\;\;\;\;\;\;\;\;\;\;\;\;\;\;\;\;\;\;\;\;\;\;\;\;\;\;
        \;\;\;\;\;\;\;\;\;\;\;\;\;\;\;\;\;
        R_i \lesssim R \lesssim R_1, \\
        \kappa_{15} \, (N_i v_i)^{-\frac{1+p}{3+p}} \left( \frac{R}{R_i} \right)^{\!\! -\frac{1+p}{3+p}},
        \;\;\;\;\;\;\;\;\;\;\;\;\;\;\;\;\;\;\;\;\;\;\;\;\;\;\;\;\;\;\,
        R_1 \lesssim R \lesssim R_2, \\
        \kappa_{18} \, h_B^{1/3} (N_i v_i)^{-1/3} \left(\ln N_i v_i \frac{R}{R_i}\right)^{\!1/6} \left( \frac{R}{R_i} \right)^{-1/3},
        \; R \gtrsim R_2,
    \end{array}
    \right.
\end{equation}
for the comoving magnetic field strength, and
\begin{equation}
\label{F10} \frac{\xi_B(R)}{\xi_B(R_i)} \simeq
\left\{ \begin{array}{lll}
        1, \;\;\;\;\;\;\;\;\;\;\;\;\;\;\;\;\;\;\;\;\;\;\;\;\;\;\;\;\;\;\;\;\;\;\;\;\;\;\;\;\;\;\;\;\;\;\;\;\;
        \;\;\;\;\;\;\;\;\;\;\;\;\;\;\;\;\; R_i \lesssim R \lesssim R_1, \\
        \kappa_{16} \, (N_i v_i)^{2/(3+p)} \left( \frac{R}{R_i} \right)^{\! \frac{2}{3+p}},
        \;\;\;\;\;\;\;\;\;\;\;\;\;\;\;\;\;\;\;\;\;\;\;\;\;\;\:
        R_1 \lesssim R \lesssim R'_2, \\
        \kappa_{19} \, h_B^{1/3} (N_i v_i)^{2/3} \left(\ln N_i v_i \frac{R}{R_i}\right)^{\!-1/3} \left( \frac{R}{R_i} \right)^{\! 2/3}, \;  R \gtrsim R'_2,
    \end{array}
    \right.
\end{equation}
for the comoving magnetic correlation length, where
\begin{equation}
\label{F11} \frac{R_2}{R_i}  \simeq \frac{\kappa_{20}}{N_i v_i} \: h_B^{-2q/3} \left[\ln h_B^{-2q/3}\right]^{-q/3} , \;\;\;\;
\frac{R'_2}{R_i} \simeq \frac{\kappa_{21}}{N_i v_i} \: h_B^{-2q/3} \left[\ln h_B^{-2q/3}\right]^{2q/3} ,
\end{equation}
with $q=3(3+p)/(4p)$, and $\kappa_{18},\kappa_{19},\kappa_{20},\kappa_{21}$ are given in Appendix A.
In the above equations, we have assumed that $(N_i v_i)^{1/2} \, h_B^{-2q/3}$ is a quantity
much greater than unity.

\subsection{VIIIc. Evolution laws in matter-dominated universe}

{\it Non-helical case}.-- In the case of a matter-dominated universe, we find
\begin{equation}
\label{F13} \frac{R^2 B(R)}{R^2_i B(R_i)} \simeq
\left\{ \begin{array}{ll}
        1, \;\;\;\;\;\;\;\;\;\;\;\;\;\;\;\;\;\;\;\;\;\;\;\;\;\;\;\;\;\;\;\;\;\;\;\;\;\;\;\;\;\;   R_i \lesssim R \lesssim R_1, \\
        \kappa_{22} \, (N_i v_i)^{-\frac{1+p}{3+p}} \left( \ln \frac{R}{R_i} \right)^{\!\! -\frac{1+p}{3+p}},  \; R \gtrsim R_1,
    \end{array}
    \right.
\end{equation}
for the comoving magnetic field strength, and
\begin{equation}
\label{F14} \frac{\xi_B(R)}{\xi_B(R_i)} \simeq
\left\{ \begin{array}{ll}
        1, \;\;\;\;\;\;\;\;\;\;\;\;\;\;\;\;\;\;\;\;\;\;\;\;\;\;\;\;\;\;\;\;\;\;\;\;\;\;\;\;\;\;\;\, R_i \lesssim R \lesssim R_1, \\
        \kappa_{23} \, (N_i v_i)^{2/(3+p)} \left( \ln \frac{R}{R_i} \right)^{\! \frac{2}{3+p}},  \; R \gtrsim R_1,
    \end{array}
    \right.
\end{equation}
for the comoving magnetic correlation length, where we have taken into account Eq.~(\ref{F12}).
In the above equations,
\begin{equation}
\label{F15}
\frac{R_1}{R_i} \simeq \exp\!\left(\frac{\kappa_{24}}{N_i v_i}\right) \! ,
\end{equation}
and $\kappa_{22},\kappa_{23},\kappa_{24}$ are given in Appendix A.

{\it Helical case}.-- In the helical case, and neglecting loglog factors, we find
\begin{equation}
\label{F16} \frac{R^2 B(R)}{R^2_i B(R_i)} \simeq
\left\{ \begin{array}{lll}
        1, \;\;\;\;\;\;\;\;\;\;\;\;\;\;\;\;\;\;\;\;\;\;\;\;\;\;\;\;\;\;\;\;\;\;\;\;\;\;\;\;\;\;\;\;\;\;\;\, R_i \lesssim R \lesssim R_1, \\
        \kappa_{22} \, (N_i v_i)^{-\frac{1+p}{3+p}} \left( \ln \frac{R}{R_i} \right)^{\!\! -\frac{1+p}{3+p}},
        \;\;\;\;\;\; R_1 \lesssim R \lesssim R_2, \\
        \kappa_{25} \, h_B^{1/3} (N_i v_i)^{-1/3} \left(\ln \frac{R}{R_i}\right)^{\!-1/3} ,                              \; R \gtrsim R_2,
    \end{array}
    \right.
\end{equation}
for the comoving magnetic field strength, and
\begin{equation}
\label{F17} \frac{\xi_B(R)}{\xi_B(R_i)} \simeq
\left\{ \begin{array}{lll}
        1, \;\;\;\;\;\;\;\;\;\;\;\;\;\;\;\;\;\;\;\;\;\;\;\;\;\;\;\;\;\;\;\;\;\;\;\;\;\;\;\;\;\;\;\, R_i \lesssim R \lesssim R_1, \\
        \kappa_{23} \, (N_i v_i)^{2/(3+p)} \left( \ln \frac{R}{R_i} \right)^{\! \frac{2}{3+p}},  \;
        R_1 \lesssim R \lesssim R'_2, \\
        \kappa_{26} \, h_B^{1/3} (N_i v_i)^{2/3} \left(\ln \frac{R}{R_i}\right)^{\!2/3} , \;  R \gtrsim R'_2,
    \end{array}
    \right.
\end{equation}
for the comoving magnetic correlation length. Here,
\begin{equation}
\label{F18} \frac{R_2}{R_i} \simeq \exp\!\left(\frac{\kappa_{27}}{N_i v_i} \, h_B^{-2q/3}\right) \! , \;\;\;\;
\frac{R'_2}{R_i} \simeq \exp\!\left(\frac{\kappa_{28}}{N_i v_i} \, h_B^{-2q/3}\right) \! ,
\end{equation}
and $\kappa_{25},\kappa_{26},\kappa_{27},\kappa_{28}$ are given in Appendix A.
The above equations are valid in the limit $h_B \ll 1$.

\subsection{VIIId. Exiting the turbulence phase}

The turbulence phase prior to recombination is characterized by high values of the kinetic Reynolds number,
${\text{Re}} = vl/\nu \gg 1$ where, as usual,
$v$ and $l$ are the typical velocity and length scale of the fluid
motion. Since $v$, $l$, and $\nu$ are functions of time in an expanding universe, there
could be a time when the system, initially in a turbulent phase, undergoes a transition to
a viscous phase characterized by ${\text{Re}} < 1$.

After recombination, the kinetic Reynolds number, known in this case as ``ambipolar''
Reynolds number, has the form ${\text{Re}_{\rm amb}} = v l \alpha_{\rm in} X_e/B^2$,
where $X_e$ is the (constant) ionization fraction after recombination, and $B$ is the typical
magnetic field strength. Turbulence operates if ${\text{Re}_{\rm amb}} \gg 1$, a condition that,
although initially satisfied, can be (and indeed is) successively violated.

{\it Kinetic and ambipolar Reynolds numbers}.-- In order to study the evolution of the kinetic and ambipolar
Reynolds numbers, let as define them accurately as
\begin{equation}
\label{FD1} {\text{Re}} = \frac{v_{\rm rms} \xi_v}{\nu} \, , \;\;\;\;
{\text{Re}_{\rm amb}} = \frac{v_{\rm rms} \xi_v \alpha_{\rm in} X_e}{B_{\rm rms}^2} \, .
\end{equation}
In a flat Friedmann universe, the kinematic viscosity re-scales as~\cite{Brandenburg,Banerjee1,Banerjee2}
\begin{equation}
\label{nuscaled}  \nu \rightarrow \tilde{\nu} =
\left\{ \begin{array}{ll}
        R^{-1} \nu,    &  \;\; \mbox{radiation-dominated era}, \\
        R^{-1/2} \nu , &  \;\; \mbox{matter-dominated era},
    \end{array}
    \right.
\end{equation}
where $\nu$ as a function of the temperature is given by~\cite{Son}
$\nu(T) \simeq [\alpha_{\rm em} \log(1/\alpha_{\rm em}) \, T]^{-1}$ for $T \gg m_e$
and $\nu(T) \simeq (n_e \sigma_T)^{-1}$ for $T \ll m_e$,
%
%
with $\alpha_{\rm em}$ being the fine structure constant, $n_e \propto T^{3}$ the electron density,
and $\sigma_T$ the Thompson cross section.
The momentum transfer rate $\alpha_{\rm in}$ evolves with the expansion parameter as
$\alpha_{\rm in} \propto R^{-3}$~\cite{Banerjee1,Banerjee2}, so that in ``tilde'' variables
we have $\tilde{\alpha}_{\rm in} = R^{3/2} \alpha_{\rm in} \propto R^{-3/2}$.

{\it Radiation-dominated universe}.-- In radiation-dominated universe, the asymptotic expansion
of the kinetic Reynolds number is
\begin{equation}
\label{FD3}  \text{Re}(R) \propto
\left\{ \begin{array}{ll}
        R^{\frac{1-p}{3+p}},         &  \;\; T \gg m_e, \\
        R^{-\frac{5+3p}{3+p}}, &  \;\; T \ll m_e,
    \end{array}
    \right.
\end{equation}
in the non-helical case, and
\begin{equation}
\label{FD4}  \text{Re}(R) \propto
\left\{ \begin{array}{ll}
        (\ln R)^{1/3} \: R^{1/3},         &  \;\; T \gg m_e, \\
        (\ln R)^{1/3} \: R^{-5/3}, &  \;\; T \ll m_e,
    \end{array}
    \right.
\end{equation}
in the helical case.

{\it Matter-dominated universe}.-- In a matter-dominated universe, instead, we find
\begin{equation}
\label{FD3}  \text{Re}(R) \propto
\left\{ \begin{array}{ll}
        (\ln R)^{\frac{1-p}{3+p}} \: R^{-1/2},         &  \;\; T \gg m_e, \\
        (\ln R)^{\frac{1-p}{3+p}} \: R^{-5/2},   &  \;\; T \ll m_e,
    \end{array}
    \right.
\end{equation}
and
\begin{equation}
\label{FD4}  \text{Re}(R) \propto
\left\{ \begin{array}{ll}
        (\ln R)^{1/3} \: R^{-1/2},         &  \;\; T \gg m_e, \\
        (\ln R)^{1/3} \: R^{-5/2}, &  \;\; T \ll m_e,
    \end{array}
    \right.
\end{equation}
in the non-helical and helical cases, respectively.

Regarding the ambipolar Reynolds number, we have
$\text{Re}_{\rm amb} = \tilde{v}_{\rm rms} \xi_v \tilde{\alpha}_{\rm in} X_e/\tilde{B}_{\rm rms}^2$
in expanding universe. Using the fact that in a turbulent phase
$\tilde{v}_{\rm rms} \simeq \tilde{\tau} \tilde{B}_{\rm rms}^2/\tilde{\xi}_B$ (see section VIIIa),
we get $\text{Re}_{\rm amb} \simeq \tilde{\tau} \tilde{\alpha}_{\rm in} X_e$.
Since $\tilde{\tau} \propto \ln R$ in a matter-dominated universe, we finally get
\begin{equation}
\label{ambipolar1} \text{Re}_{\rm amb}(R) \propto (\ln R) \: R^{-3/2},
\end{equation}
in both non-helical and helical cases.

Excepted the case of helical magnetic fields in radiation-dominated universe, we see that
the kinetic and ambipolar Reynolds numbers are decreasing functions of time (we remember that we are assuming $p > 1$).
Hence, there will exist two times $t_{{\rm exit},1}$ and $t_{{\rm exit},2}$ when
$\text{Re}(t_{{\rm exit},1}) = 1$ and $\text{Re}_{\rm amb}(t_{{\rm exit},2}) = 1$,
%
%
thus defining the end of turbulence.
\footnote{Also for (maximally) helical fields there will be certainly a time when turbulence terminates.
This is because, after $e^{+}e^{-}$ annihilation,
the kinematic viscosity increases by a factor of order $10^{10}$, namely of order of
the photon-to-baryon ratio~\cite{Kolb}. Accordingly, somewhere below $T_{e^{+}e^{-}}$,
the kinetic Reynolds number drops below unity.}

\section{IX. On the cut-off of the initial power spectrum}

In section Va, in order to get finite results for the magnetic energy, 
we introduced a Gaussian cut-off, $\propto e^{-k^2}$, for the initial magnetic energy spectrum. In the literature
(see, for example,~\cite{Biskamp1,Christensson1,Christensson2}),
it is sometimes used a cut-off of the form $e^{-k^4}$, sharper than the Gaussian one.
Therefore, it is of some interest to consider the initial magnetic spectrum
\begin{equation}
\label{W1}  {\mathcal E}_B (k,0) = \lambda_B k^p \exp (-2k^4 \ell_{B}^4),
\end{equation}
instead of Eq.~(\ref{MF15}).

{\it $e^{-k^4}\!$ cut-off: Non-helical case}.-- Following the same arguments in section Vb, we find for the magnetic energy and correlation length
\begin{eqnarray}
\label{W2} && E_B(t) = E_B(0) \: \frac{\Gamma\!\left(\frac{3+p}{4}\right)}{\sqrt{\pi}} \:
                              U\!\left( \frac{1+p}{4} , \frac12 , \frac{\zeta_{\rm diss}^4}{2} \right) \! , \\
\label{W3} && \xi_B(t) = \xi_B(0) \: \frac{\Gamma\!\left(\frac{2+p}{4}\right)}{\Gamma\!\left(\frac{3+p}{4}\right)} \,
                                  \frac{U\!\left( \frac{p}{4} , \frac12 , \frac{\zeta_{\rm diss}^4}{2} \right)}
                                  {U\!\left( \frac{1+p}{4} , \frac12 , \frac{\zeta_{\rm diss}^4}{2} \right)} \, ,
\end{eqnarray}
instead of Eq.~(\ref{MF17}) and (\ref{MF18}), respectively. In the limit of large $\zeta_{\rm diss}$,
which corresponds to large values of the time, we have the asymptotic expressions
\begin{eqnarray}
\label{W4} && E_B(t) = E_B(0) \: \frac{2^{(1+p)/4}}{\sqrt{\pi}} \, \Gamma\!\left(\frac{3+p}{4}\right) \zeta_{\rm diss}^{-(1+p)}, \\
\label{W5} && \xi_B(t) = \xi_B(0) \: \frac{1}{2^{1/4}} \, \frac{\Gamma\!\left(\frac{2+p}{4}\right)}{\Gamma\!\left(\frac{3+p}{4}\right)} \,
                         \zeta_{\rm diss} \, ,
\end{eqnarray}
where we used the asymptotic expansion of the confluent hypergeometric function
$U(a,b,z)$, namely $U(a,b,z) \sim z^{-a}$ for $z \gg 1$~\cite{Gradshteyn},
valid in the case $a>0$.
The master (differential) equation for $\zeta_{\rm diss}$ is given by Eq.~(\ref{differential1}),
with the formal expression of $\delta(0)$ given by Eq.~(\ref{delta}).
The solution of this master equation gives Eqs.~(\ref{Es}) and (\ref{xis}), with the constants
$c_2$ and $c_3$ replaced by
\begin{eqnarray}
\label{W7} && c_2 \rightarrow
\left[ \frac{2^{(1+p)/4}}{\sqrt{\pi}} \Gamma\!\left(\frac{3+p}{4}\right) \right]^{\!\frac{2}{3+p}}
\left[ \frac{(1+p)\Gamma\!\left(\frac{1+p}{4}\right)}{2\sqrt{2} \, \Gamma\!\left(\frac{3+p}{4}\right)} \right]^{\!-\frac{1+p}{2(3+p)}}
\! c_2, \\
\label{W8}
&& c_3 \rightarrow \frac{1}{2^{1/4}}
\frac{\Gamma\!\left(\frac{2+p}{4}\right)}{\Gamma\!\left(\frac{3+p}{4}\right)}
\left[ \frac{2^{(p-2)/2}}{\pi} \, (1+p) \, \Gamma\!\left(\frac{1+p}{4}\right) \!
\Gamma\!\left(\frac{3+p}{4}\right) \right]^{\!\frac{1}{2(3+p)}} \! c_3, \nonumber \\
\end{eqnarray}
respectively. Also, we find straightforwardly that the expression of $\delta(0)$ as a function of the
index $p$ is
\begin{equation}
\label{delta0cutoff} \delta(0) = \left[ \frac{3\sqrt{2} \, \Gamma\!\left(\frac{1+p}{4}\right)}
{\Gamma\!\left(\frac{3+p}{4}\right)} \right]^{1/2} \! \frac{1}{\zeta_B} \, ,
\end{equation}
%
instead of Eq.~(\ref{delta0}).

{\it $e^{-k^4}\!$ cut-off: Helical case}.-- Following again the analysis in section Vb, we find that in the helical case,
the expressions for the magnetic energy, helicity, and correlation length, take the form
\begin{eqnarray}
\label{W9} && \frac{E_B(\tau)}{E_B(0)}  \simeq  c_{13} (1+h_B) \,
\frac{\sqrt{\pi}}{2^{p/2} \, \Gamma[(1+p)/2]} \,
\frac{\zeta_\alpha^p}{\zeta_{\rm diss}^{1+2p}} \, \exp \! \left(
\frac{\zeta_\alpha^2}{2\zeta_{\rm diss}^2} \right) \! ,
\\
\label{W10} && \frac{H_B(\tau)}{H_B(0)}  \simeq  c_{14} \frac{1+h_B}{h_B} \,
\frac{\sqrt{\pi}}{2^{(p-1)/2} \, \Gamma(p/2)} \,
\frac{\zeta_\alpha^{p-1}}{\zeta_{\rm diss}^{-1+2p}} \, \exp \!
\left( \frac{\zeta_\alpha^2}{2\zeta_{\rm diss}^2} \right) \! ,
\\
\label{W11} && \frac{\xi_B(\tau)}{\xi_B(0)}  \simeq c_{15}
\frac{2}{\zeta_B} \, \frac{\zeta_{\rm diss}^2}{\zeta_\alpha} \, ,
\end{eqnarray}
in the limits $\zeta_{\rm diss} \gg 1$ and $\zeta_\alpha/\zeta_{\rm diss} \gg 1$.
Note that they differ from the case of a Gaussian cut-off for the introduction of the coefficients
$c_{13},c_{14},c_{15}$ (given in Appendix A), not present in Eqs.~(\ref{E1}), (\ref{H1}), and (\ref{xi1}).
Therefore, following the analysis in the Appendix B, we find that the expressions for the magnetic energy
and correlation length are given, respectively, by Eqs.~(\ref{expansion-Energy}) and (\ref{expansion-correlation}),
with the replacements
\begin{eqnarray}
\label{W12} && c_7 \rightarrow
\left[ \frac{\Gamma\!\left(\frac{1+p}{2}\right) \Gamma\!\left(\frac{p}{4}\right)}{\Gamma\!\left(\frac{1+p}{4}\right) \Gamma\!\left(\frac{p}{2}\right)} \right]^{2/3}
\left[ \frac{(1+p) \, \Gamma\!\left(\frac{1+p}{4}\right)}{\sqrt{2} \, \Gamma\!\left(\frac{3+p}{4}\right)} \right]^{-1/6}
\! c_7, \\
\label{W13} &&
c_8 \rightarrow
\left[ \frac{\Gamma\!\left(\frac{1+p}{2}\right) \Gamma\!\left(\frac{p}{4}\right)}{\Gamma\!\left(\frac{1+p}{4}\right) \Gamma\!\left(\frac{p}{2}\right)} \right]^{-2/3}
\left[ \frac{(1+p) \, \Gamma\!\left(\frac{1+p}{4}\right)}{\sqrt{2} \, \Gamma\!\left(\frac{3+p}{4}\right)} \right]^{1/6}
\! c_8.
\end{eqnarray}
{\it Other cut-offs}.-- We conclude this section by observing that very similar results are obtained if the cut-off of the
initial spectrum is even sharper than the $e^{-k^4}$ one, as for example in the case
\begin{equation}
\label{W14}  {\mathcal E}_B (k,0) =
\left\{ \begin{array}{ll}
        \lambda_B k^p,   &  \;\; k < K, \\
        0,               &  \;\; k > K.
    \end{array}
    \right.
\end{equation}
In general, what we observe is that changing the form of the cut-off, just changes the form of the
coefficients $c_2$, $c_3$, $c_7$, $c_8$, and the expression for $\delta(0)$ as a function of $p$.
This, in turn, changes the form of the $\kappa$'s coefficients which enter in the expression
for the magnetic field intensity and correlation length in the case of interest of an expanding universe.
However, the power-law behaviors, and logarithmic corrections in the helical case, do not depend
on the form of the cut-off.

Finally, we note that the authors in~\cite{Tevzadze}, as already explained in section Va,
found that soon after a first-order phase transition, the magnetic spectrum
has the form of a Batchelor spectrum ${\mathcal E}_B \propto k^4$ at small wavenumbers. Also, they found
that during a turbulent phase, a Kolmogorov spectrum of the type ${\mathcal E}_B \propto k^{-5/3}$
develops at intermediate wavenumbers, $k_B < k < k_{\rm diss}$, where $k_B = 2\pi/ \xi_B$ and
$k_{\rm diss} = 2\pi/\sqrt{\eta t}$, with $\sqrt{\eta t}$ being the dissipation scale.
This means that, if a dragged phase follows a turbulent phase
(as it could be in a cosmological context~\cite{Banerjee1,Banerjee2}),
a more realistic form of the initial magnetic spectrum in dragged phase
could be given by
\begin{equation}
\label{W15}  {\mathcal E}_B (k,t_i) =
\left\{ \begin{array}{ll}
        \lambda_1 k^4,                            &  \; k < k_B,     \;\;\;\;\;\;\;\;\;\;\;\;\;\, \mbox{(Batchelor spectrum)}, \\
        \lambda_2 k^{-5/3},                       &  \; k_B < k < k_{\rm diss},              \;\; \mbox{(Kolmogorov spectrum)}, \\
        \lambda_3 \exp (-2k^2 / k_{\rm diss}^2),  &  \; k > k_{\rm diss},  \;\;\;\;\;\;\;\;\;\;\; \mbox{(dissipative cut-off)},
    \end{array}
    \right.
\end{equation}
where $\lambda_1$, $\lambda_2$, and $\lambda_3$ are constants,
$k_B = 2\pi/ \xi_B(t_i)$,  $k_{\rm diss} = 2\pi/\sqrt{\eta t_i}$, and $t_i$ the initial time.
So, in this case the cut-off is represented
by a Kolmogorov spectrum up to the initial dissipative scale, $2\pi/k_{\rm diss}$,
followed by a Gaussian cut-off due to resistivity.

In the light of the above discussions, however, we expect that even in this more realistic and
complicated case, the main results we found, namely the power laws for the magnetic intensity and correlation length
(corrected by logarithmic factors in the helical case), do not change.

\section{X. Discussion}

In this section, we discuss our results and write down the final and simplified expressions
for the magnetic field intensity and correlation length in a Friedmann universe, which are
relevant when studying the evolution of a primordial, phase-transition-generated, cosmic magnetic field.

{\it Simplified evolution laws}.-- We saw, in the previous section,
that the $\kappa$ coefficients entering in the expressions for the magnetic field and correlation
length depend on the choice of the cut-off of the initial magnetic spectrum. In the case of
magnetic fields generated in primeval phase transitions, the exact form of this cut-off
is not precisely known, and this introduces a factor of arbitrariness in the evolution laws of $B$ and $\xi_B$.
Fortunately, since the $\kappa$ coefficients are all of order unity (see section IX and Appendix A),
this factor of arbitrariness can be safely neglected when studying the evolution
of a phase-transition-generated magnetic field.

If on the one hand the $\kappa$ coefficients depend on the initial cut-off, on the other hand
the logarithmic factors we found in the case of helical magnetic fields are a general consequence
of the fact that the magnetic helicity is a quasi-conserved quantity in magnetohydrodynamics (see Appendix B).
However, these factors introduce just a small correction in the power-law evolution of $B$ and
$\xi_B$ and then, in first approximation, can be neglected.

According to the above discussion and given an initial magnetic field with
spectrum of the form ${\mathcal E}_B (k,t_i) \propto k^p$, the
(physical) magnetic field intensity, $B$, and the comoving magnetic correlation length, $\xi_B$,
evolve as a function of the temperature $T$ approximatively as
\begin{equation}
\label{Result1} \frac{B}{B_i} \sim
    \left\{ \begin{array}{lll}
        \left( \frac{T}{T_i} \right)^{\!2}, \;\;\;\;\;\;\;\;\;\;\;\;\;\;\;\;\;\;\;\;\;\;\;\;\;\;\;\;\;\;
        T_i \lesssim T \lesssim T_1, \;\;\; \mbox{(quiescent phase)}, \\
        (N_i v_i)^{\varrho_1(p)} \left( \frac{T}{T_i} \right)^{\!\varrho_2(p)},
        \;\;\;\;\;\;\;\, T_1 \lesssim T \lesssim T_2, \;\;\, \mbox{(selective decay phase)}, \\
        h_B^{1/3} (N_i v_i)^{\varrho_1(0)} \left( \frac{T}{T_i} \right)^{\!\varrho_2(0)},
        \; T \gtrsim T_2, \;\;\;\;\;\;\;\;\;\;\; \mbox{(inverse cascade phase)},
    \end{array}
    \right.
\end{equation}
and
\begin{equation}
\label{Result2} \frac{\xi_B}{\xi_{B,i}} \sim
    \left\{ \begin{array}{lll}
        1, \;\;\;\;\;\;\;\;\;\;\;\;\;\;\;\;\;\;\;\;\;\;\;\;\;\;\;\;\;\;\;\;\;\;\;\;\;\;
        T_i \lesssim T \lesssim T_1, \;\;\; \mbox{(quiescent phase)}, \\
        (N_i v_i)^{\varrho_3(p)} \left( \frac{T}{T_i} \right)^{\!\varrho_4(p)},
        \;\;\;\;\;\;\;\, T_1 \lesssim T \lesssim T_2, \;\;\, \mbox{(selective decay phase)}, \\
        h_B^{1/3} (N_i v_i)^{\varrho_3(0)} \left( \frac{T}{T_i} \right)^{\!\varrho_3(0)},
        \; T \gtrsim T_2, \;\;\;\;\;\;\;\;\;\;\; \mbox{(inverse cascade phase)},
    \end{array}
    \right.
\end{equation}
respectively, where $B_i = B(T_i)$ and $\xi_{B,i} = \xi_B(T_i)$, and we took $R \sim T^{-1}$
[neglecting small corrections due to $g_{*,S}(T)$]. We remember that $N_i$ is the initial
number of magnetic domains per horizon length, namely $N_i = d_{H,i}/\xi_{B,{\rm phys},i}$, where $d_{H,i}$ and
$\xi_{B,{\rm phys},i}$ are the length of the Hubble horizon and the physical magnetic correlation length
at the initial time.
The value of the bulk velocity at the onset of the particular regime, $v_i$, is not an independent
parameter in both dragged phase and turbulent non-helical phase, being related to
$B_i$, $\xi_{B,{\rm phys},i}$, and $\alpha_i$ by the relations
\begin{equation}
\label{Result3} v_i \sim
    \left\{ \begin{array}{ll}
        \frac{B_i^2}{\alpha_i \xi_{B,{\rm phys},i}} \, ,  &  \mbox{(dragged phase)}, \\
        B_i,                                              &  \mbox{(turbulent phase - non-helical case)}.
    \end{array}
    \right.
\end{equation}
In equations~(\ref{Result1}) and (\ref{Result2}), we quoted the expressions for $B$ and $\xi_B$
in the helical case. In the non-helical case, those expressions are still valid, the inverse cascade phase
being absent. The transition temperatures $T_1$ and $T_2$ can be easily found by matching the expressions
of $B$ (or, which is the same, those for $\xi_B$) in two consecutive phases.

We summarize our results on the evolution laws for the magnetic field and correlation length in Table~2.


\begin{table}[t!]
\caption{The exponents $\varrho_1(p)$, $\varrho_2(p)$, $\varrho_3(p)$, and $\varrho_4(p)$
in the evolution laws~(\ref{Result1}) and (\ref{Result2}) of a primordial magnetic field
in dragged and turbulent magnetohydrodynamic phases, in radiation (RD) and matter (MD) eras.
For the dragged phase, it is indicated the free-streaming species which determines
the drag coefficient. The parameter $p$ is the index of the initial magnetic
power-law spectrum, ${\mathcal E}_B (k,t_i) \propto k^p$.}
\vspace{0.5cm}
\begin{tabular}{llccccccc}
\hline \hline
& Phase         ~~& Era ~& Streaming particle ~& $\varrho_1(p)$       ~~& $\varrho_2(p)$           ~~& $\varrho_3(p)$  ~~& $\varrho_4(p)$       \\
\hline
& Dragged MHD   ~~& RD  ~& neutrino           ~& $-\frac{1+p}{2(3+p)}$ ~~& $\frac{17+9p}{2(3+p)}$  ~~& $\frac{1}{3+p}$ ~~& $-\frac{5}{3+p}$     \\
&               ~~&     ~& photon             ~& $-\frac{1+p}{2(3+p)}$ ~~& $\frac{4(2+p)}{3+p}$    ~~& $\frac{1}{3+p}$ ~~& $-\frac{4}{3+p}$     \\
&               ~~& MD  ~& photon             ~& $-\frac{1+p}{2(3+p)}$ ~~& $\frac{33+17p}{4(3+p)}$ ~~& $\frac{1}{3+p}$ ~~& $-\frac{9}{2(3+p)}$  \\
&               ~~&     ~& hydrogen           ~& $-\frac{1+p}{2(3+p)}$ ~~& $\frac{39+23p}{4(3+p)}$ ~~& $\frac{1}{3+p}$ ~~& $-\frac{15}{2(3+p)}$ \\
& Turbulent MHD ~~& RD  ~&  --                ~& $-\frac{1+p}{3+p}$    ~~& $\frac{7+3p}{3+p}$      ~~& $\frac{2}{3+p}$ ~~& $-\frac{2}{3+p}$     \\
&               ~~& MD  ~&  --                ~& $-\frac{1+p}{3+p}$    ~~& $2$                     ~~& $\frac{2}{3+p}$ ~~& $0$                  \\
\hline \hline
\end{tabular}
\end{table}


{\it Comparison with the Banerjee and Jedamzik's results}.-- Let us now compare the above results
with those of~\cite{Banerjee1,Banerjee2}, which are, to our knowledge,
the only ones in the literature which consider other than the case of freely-decaying magnetic fields,
also the case of magnetic fields in dragged phase.
To this end, we introduce the so-called Alfv\'{e}n eddy turnover time, $\tau_{A}$, as
$\tau_{A} = \xi_{B,\rm phys}/B$.
%
%
We can then write the quantity $N_i v_i$ as
\begin{equation}
\label{Dis2}  N_i v_i \sim
\left\{ \begin{array}{ll}
        (t_i/\tau_{A,i}) (\tau_{d,i}/\tau_{A,i}),  & \;\; \mbox{dragged MHD}, \\
        \Gamma_i^{1/2} (t_i/\tau_{A,i}),           & \;\; \mbox{turbulent MHD},
    \end{array}
    \right.
\end{equation}
where we remember that in turbulent regime, $\Gamma_i$ is of order unity for the non-helical case,
but can be different from one in the helical case.
Banerjee and Jedamzik~\cite{Banerjee1,Banerjee2} assume (without a full justification) that the initial
correlation length is not a free parameter but it is determined, at the initial (cosmic) time $t_i$,
by the equality of the Hubble rate $H \simeq t^{-1}$ and
the Alfv\'{e}n eddy turnover rate $1/\tau_{A}$. Namely, in their model we have
\begin{equation}
\label{Dis3} t_i \simeq \tau_{A,i}.
\end{equation}
In this case, and taking $\Gamma_i$ of order of unity (as in~\cite{Banerjee1,Banerjee2}), we get
\begin{equation}
\label{Dis4}  N_i v_i \sim
\left\{ \begin{array}{ll}
        1/(\alpha_i t_i),  & \;\; \mbox{dragged MHD}, \\
        1,                 & \;\; \mbox{turbulent MHD}.
    \end{array}
    \right.
\end{equation}
With the choice in Eq.~(\ref{Dis3}), it is easy to check that our formulas agree
with those of~\cite{Banerjee1,Banerjee2}
[with the exception of Eqs.~(62) and (63) of~\cite{Banerjee1,Banerjee2},
in which the exponents $3/(2+n)$ and
$3n/(2+n)$ should be $4/(2+n)$ and $4n/(2+n)$, respectively.]
For example, for non-helical dragged MHD in the case of neutrino free
streaming we have, excluding factors of order unity,
\begin{eqnarray}
\label{Dis5} && B(T) \sim
B(T_i) \left( \frac{1}{G_F^2 \, m_{\rm Pl} \, T_i^3} \right)^{\!-\frac{1+p}{2(3+p)}}
\left( \frac{T}{T_i} \right)^{\!\frac{17+9p}{2(3+p)}},  \; T \gg T_i, \\
\label{Dis6} && \xi_B(T) \sim
\xi_B(T_i) \left( \frac{1}{G_F^2 \, m_{\rm Pl} \, T_i^3} \right)^{\!\frac{1}{3+p}}
\left( \frac{T}{T_i} \right)^{\!-\frac{5}{3+p}},  \; T \gg T_i,
\end{eqnarray}
where $G_F$ is the Fermi constant, $m_{\rm Pl}$ is the Planck mass, and we used the facts that
$t_i \sim m_{\rm Pl}/T_i^2$ (in radiation-dominated era) and $\alpha_i \sim G_F T_i^5$ (for the neutrino case).
Equations~(\ref{Dis5}) and (\ref{Dis6}) agree, respectively, with Eqs.~(48) and (47) of~\cite{Banerjee2}.
[Let us observe that in the notation of~\cite{Banerjee2}, $p=n-1$ and $B(T) \propto r(T)^{1/2} T^2$.]

{\it Inertial range}.-- According to the ``Kolmogorov hypothesis''(see, e.g.,~\cite{Biskamp}),
the cascade of energy in $k$-space is a quasi-local process independent on the particular
scale considered (although this is strictly possible only for scales larger than the
dissipation scale and smaller than the integral scale).
The $k$-space interval where the Kolmogorov hypothesis may apply is called ``inertial range''.
Here, the magnetic spectrum is expected to decay as ${\mathcal E}_B (k,t) \propto k^\beta$,
with $\beta$ being negative.
The value of $\beta$ could in principle depend on the particular phase (dragged or turbulent) and be different
for the non-helical and helical cases.
Indeed, the pure Kolmogorov spectrum, $\beta = -5/3$, has been observed in~\cite{Tevzadze}
for the turbulent non-helical case, while~\cite{Banerjee1,Banerjee2} found
$\beta \simeq -2$ both in helical and non-helical dragged case.

In our case, we do not find any inertial range where the magnetic spectrum
decays as a power law. This is probably to be ascribed to the mean-field approximation, which
neglects the small-scale fluctuating part of the velocity and magnetic fields (see section IVb).
It is the (quasi-linear) interaction of these small-scale fields with the corresponding large-scale ones
which ``opens'' this particular range in $k$-space.
However, if the inertial range is sufficiently narrow, we do not expect significant modifications to
the evolution laws we found for $B$ e $\xi_B$. Indeed, and interesting enough,
Ref.~\cite{Banerjee1,Banerjee2} found that the inertial range is almost absent in the
non-helical turbulent case, and that on intermediate scales the magnetic spectrum is
more consistent with an exponential law than with a power law. This seems to be in agreement
with the results we found in mean-field approximation, namely with Eq.~(\ref{psi}).

\section{XI. Conclusions and outlook}

{\it Conclusions}.-- The presence of large-scale magnetic fields in the universe is
still an open and unsolved mystery of modern cosmology. A
plethora of mechanisms able to generate cosmic magnetic fields in the
early universe, as for example during inflation or primeval phase transitions,
or during more recent eras by astrophysical mechanisms have been proposed
since the first attempt put forward by Harrison in 1970~\cite{Harrison}.

If on the one hand, astrophysical mechanisms seem to be ruled out by recent observations of
large-scale magnetic fields in cosmic voids~\cite{Blazars},
on the other hand inflationary mechanisms are able, at least in principle, to explain the
magnetization of the universe. However, these latter mechanisms
repose on the use of nonstandard physics,
an exception being the mechanism recently proposed in~\cite{Campanelli1}.
Nevertheless, if the scale of inflation is considerably below $10^{16}$GeV,
also this mechanism fails to explain galactic and galaxy cluster
magnetic fields.

If, instead, presently-observed magnetic fields originate from cosmological
phase transitions (such as QCD or electroweak phase transitions),
another unsolved question arise: how have these
magnetic fields evolved from the time of they generation
until today? The answer to this question can be given only in the
framework of magnetohydrodynamics.

In this paper, we have studied the evolution of phase-transition-generated magnetic fields
coupled the to primeval plasma by solving, analytically, the magnetohydrodynamic equations
in mean-field approximation.
The reduction of full MHD equations to simpler equations
due to mean-field approximation causes a lost of information about the transfer of magnetic energy
at intermediate scales (corresponding to the inertial range). Nevertheless, the main characteristics
of an evolving magnetic field are preserved.

In particular, we have analyzed the decay of primordial magnetic fields, both in radiation and matter eras,
in the two regimes which are relevant in a cosmological context, namely the turbulent and viscous free-streaming regimes.
During a viscous free-streaming phase, dissipation is ruled by a drag term in the Navier-Stokes
equation, the drag term depending on the free-steaming particle species, namely neutrinos, photons
or hydrogen atoms.

Our results can be summarized as follows.

If the initial magnetic field is not helical, then the magnetic
helicity remains null during the evolution of the system. There is
an initial phase in which the system is quiescent: magnetohydrodynamic effects do
not operate and the characteristic quantities of the system, such
as energy and correlation of the magnetic field, remain almost
constant. This phase persists for a period which is proportional
to the initial eddy turnover time. After that, the
evolution of the magnetic field proceeds through selective decay
of magnetic modes, that is there is no direct transfer of magnetic
energy from small scales to large scales, but simply modes with
larger wavenumbers are dissipated faster than those whose
wavenumbers are small. During this process, the magnetic correlation length
grows, while the magnetic energy
decays in time. The evolution laws for the magnetic field
depend on the initial magnetic power spectrum and on the particular regime.
They are summarized in Eqs.~(\ref{Result1})-(\ref{Result3}) and Table~II.

In the helical case, the magnetic helicity is an almost conserved
quantity. The system undergoes three different phases: a quiescent phase and
a selective-decay phase in which the system evolves irrespective of
the presence of magnetic helicity, and an inverse-cascade phase. As
in the case non-helical case, the quiescent phase ends
approximatively after a period of time equal to the initial eddy
turnover time. Then, the system enters into a selective-decay
phase characterized by a magnetic field evolution similar to the
non-helical one. This last phase ends due to the conservation of
magnetic helicity, favoring an inverse cascade of the
magnetic field during which the magnetic energy stored on small scales is partially
transferred to larger scales. This process of energy redistribution
weakly depends on the properties of the initial magnetic field, so that
the evolution laws of magnetic energy and correlation length do not depend
on the index of the initial magnetic spectrum power law, although they are different
in turbulent and free-streaming regimes [see Eqs.~(\ref{Result1})-(\ref{Result3}) and Table~II].
Moreover, the time when the system enters into the inverse-cascade regime
depends both on the form of the initial magnetic spectrum and on the
amount of initial magnetic helicity.

{\it Outlook}.-- Our analytical evolution laws in the different regimes and for helical and
non-helical magnetic fields, are in substantial agreement with the numerical results obtained
in~\cite{Banerjee1,Banerjee2}. Nevertheless, it is important to stress that
($i$) there is no yet convergence
in the literature on the results of numerical simulations of MHD turbulence in the
case of non-helical fields, and that ($ii$) the only available numerical results for the dragged phase
are those in~\cite{Banerjee1,Banerjee2}.
Therefore, it is desirable, in order to have a firm understanding of the evolution of primordial magnetic fields,
to solve this disagreement for turbulent non-helical fields
and, at the same time, to have an independent (and, possibly, full numerical) confirmation of the Banerjee and
Jedamzik results~\cite{Banerjee1,Banerjee2} for the evolution of magnetic fields
in dragged phase. Finally, ($iii$) we stress that
the study in~\cite{Banerjee1,Banerjee2} on the properties of phase-transition-generated magnetic fields
throughout the evolution of the universe assumed a linear dependence between the
initial magnetic correlation length and the initial magnetic field intensity.
It would be important to see if the conclusions of~\cite{Banerjee1,Banerjee2},
that phase-transition-generated magnetic fields may directly account for galactic and galaxy clusters
magnetic fields, are modified or even invalidated by relaxing the above ansatz on the
initial magnetic correlation length~\cite{Marrone}.


\begin{acknowledgments}
We would like to thank M. Giovannini for useful discussions.
\end{acknowledgments}



\section{Appendix A}

In this appendix, we write down the expressions for the coefficients $c_i$ and $\kappa_i$.
They are, respectively,
%
\begin{eqnarray}
\label{c0} && c_0 = \frac{(3+p) \zeta_B^2 \delta(0) \gamma(0)}{6 (1-a)} \, , \;
\label{c1} c_1 = \left\{ \frac{(3+p) \zeta_B^2 \delta(0) [\gamma(0)]^a}{6 (1-a)} \right\}^{\frac{2}{3+p}} , \;
\label{c2} c_2 = c_1^{-(1+p)/2} , \;
\label{c3} c_3 = c_1^{1/2} , \nonumber \\
\label{c4} && c_4 =  2 \ell_B^2 \, [\, \tau_{\rm eddy}(0)]^{-\frac{2(1-a)}{3+p}} c_1 , \;
\label{c5} c_5 = \left\{ \frac{\zeta_B^3}{2} \, \sqrt{\frac{p}{6}} \, \frac{\delta(0) [\gamma(0)]^a h_B}{\sqrt{1-a}} \right\}^{\!1/3} , \;
\label{c6} c_6 = \left\{\frac{\zeta_B^3}{9} \, p^2 \delta(0) [\gamma(0)]^a h_B (1-a) \right\}^{\!1/3} , \nonumber \\
\label{c7} && c_7 = \left\{\sqrt{\frac{p}{3 \, \delta(0) [\gamma(0)]^a}} \, h_B (1-a) \right\}^{2/3} , \;
\label{c8} c_8 = h_B \, c_7^{-1} , \nonumber \\
\label{c9} && c_9 = \left\{\!\! \left(\frac{\zeta_B}{\sqrt{2}}\right)^{\!\!\frac{2(1+p)}{3+p}} \!
\left\{\frac{(3+p)\delta(0) [\gamma(0)]^a}{3}\right\}^{\frac{\!2(3+2p)}{3(3+p)}} \!
(1-a)^{-\frac{2p}{3(3+p)}} \! \right\}^{\!\!-q} \!\! ,  \nonumber \\
\label{c10} && c_{10} = \left\{\! \left(\frac{\zeta_B}{\sqrt{2}}\right)^{\!\!\!-\frac{2}{3+p}} \!\!
\left(\frac{3+p}{3}\right)^{\!-\frac{6+p}{3(3+p)}} \!
\left\{\frac{\delta(0) [\gamma(0)]^a}{1-a}\right\}^{\frac{p}{3(3+p)}} \!\! \right\}^{\!\!-2q} \!\! , \;
\label{c11} c_{11} = (c_2 c_3)^{2q/3}, \nonumber \\
\label{c12} && c_{12} = \left\{ \!{\zeta_B^{-\frac{3(1+p)}{3+p}}} \frac{\sqrt{2} (2\ln2 + p)}{\sqrt{\ln2 + p}}
\!\left\{ \frac{(3+p) \delta(0)[\gamma(0)]^a}{6(1-a)} \right\}^{\!-\frac{p}{3+p}}  \!\right\}^{\!2q/3} \!\!, \nonumber \\
%
\label{c13} && c_{13} = \frac{2^{(11+3p)/4} \Gamma\!\left(\frac{1+p}{2}\right)}{\left[\Gamma\!\left(\frac{p}{2}\right)\right]^2 \Gamma\!\left(\frac{1+p}{4}\right)} \, , \;
\label{c14} c_{14} = \frac{2^{3(4+p)/4}}{\Gamma\!\left(\frac{p}{2}\right) \Gamma\!\left(\frac{p}{4}\right)} \, , \;
\label{c15} c_{15} = c_{14}/c_{13},
\end{eqnarray}
and
%
\begin{eqnarray}
\label{kappa0}  && \kappa_{0}  = \left( \frac{1+p}{12} \right)^{\!\!1/2} \zeta_B , \;
\label{kappa1}  \kappa_{1}  = \left[\frac{\zeta_B (3+p)}{(1-a)\sqrt{3(1+p)}} \right]^{-\frac{1+p}{2(3+p)}} , \;
\label{kappa2}  \kappa_{2}  = \kappa_{1}^{-2/(1+p)} , \;
\label{kappa3}  \kappa_{3}  = \kappa_{1}^{2(3+p)/(1-a)(1+p)} , \nonumber \\
\label{kappa4}  && \kappa_{4}  = \left(\!\frac{\zeta_B p \sqrt{1+p}}{6\sqrt{3}} \,\right)^{\!1/6} \! (1-a)^{1/3}, \;
\label{kappa5}  \kappa_{5}  = \kappa_{4}^{-2} , \;
\label{kappa6}  \kappa_{6}  = (\kappa_{4}/\kappa_{1})^{-2q}, \;
\label{kappa7}  \kappa_{7}  = (\kappa_{5}/\kappa_{2})^{-2q}, \nonumber \\
\label{kappa8}  && \kappa_{8}  = \left[-\frac{\zeta_B (3+p) 3^a}{a\sqrt{3(1+p)}} \right]^{-\frac{1+p}{2(3+p)}}, \;
\label{kappa9}  \kappa_{9}  = \kappa_{8}^{-2/(1+p)} , \;
\label{kappa10} \kappa_{10} = \kappa_{8}^{-\frac{4(3+p)}{3a(1+p)}} , \;
\label{kappa11} \kappa_{11} = \left(\!\frac{\zeta_B p \sqrt{1+p}}{6\sqrt{3}} \,\right)^{\!1/6} \! (-a)^{1/3} \, 3^{-a/6}, \nonumber \\
\label{kappa12} && \kappa_{12} = \kappa_{11}^{-2} , \;
\label{kappa13} \kappa_{13} = (\kappa_{11}/\kappa_{8})^{-2\bar{q}}, \;
\label{kappa14} \kappa_{14} = (\kappa_{12}/\kappa_{9})^{-2\bar{q}}, \;
\label{kappa15} \kappa_{15} = \left[\frac{(3+p)\zeta_B^2}{12\gamma_\infty} \right]^{-\frac{1+p}{2(3+p)}} , \;
\label{kappa16} \kappa_{16} = \kappa_{15}^{-2/(1+p)} , \nonumber \\
\label{kappa17} && \kappa_{17} = \kappa_{15}^{(3+p)/(1+p)} , \;
\label{kappa18} \kappa_{18} = (4p\gamma_\infty/3)^{1/6} , \;
\label{kappa19} \kappa_{19} = \kappa_{18}^{-2} , \;
\label{kappa20} \kappa_{20} = (\kappa_{18}/\kappa_{15})^{-2q}, \;
\label{kappa21} \kappa_{21} = (\kappa_{19}/\kappa_{16})^{-2q}, \nonumber \\
\label{kappa22} && \kappa_{22} = 2^{(1+p)/(3+p)} \kappa_{15}, \;
\label{kappa23} \kappa_{23} = 2^{-2/(3+p)} \kappa_{16}, \;
\label{kappa24} \kappa_{24} = \kappa_{22}^{(3+p)/(1+p)}, \;
\label{kappa25} \kappa_{25} = 2^{1/3} \kappa_{18}, \;
\label{kappa26} \kappa_{26} = 2^{-2/3} \kappa_{19}, \nonumber \\
\label{kappa27} && \kappa_{27} = (\kappa_{25}/\kappa_{22})^{-2q}, \;
\label{kappa28} \kappa_{28} = (\kappa_{26}/\kappa_{23})^{-2q}.
\end{eqnarray}


\section{Appendix B}

In this appendix, we find the asymptotic solution of
Eqs.~(\ref{differential1})-(\ref{differential2}), where $E_B$ and
$H_B$ as a function of $\zeta_{\rm diss}$ and $\zeta_{\alpha}$ are
given in Eqs.~(\ref{MF17}) and (\ref{MF18}), respectively.

In the limit $\zeta_\alpha/\sqrt{1 + \zeta_{\rm diss}^2} \ll 1$,
the magnetic energy and correlation length, Eqs.~(\ref{MF17}) and
(\ref{xi}), are given by Eq.~(\ref{Es0}) and (\ref{xis0}),
respectively. Inserting Eq.~(\ref{Es0}) in
Eq.~(\ref{differential1}), we get
\begin{equation}
\frac{d\zeta_{\rm diss}^2}{d\tau} =
                          \frac{\zeta_B^2}{\text{Re}_B(0)} +
                          \frac{\zeta_B^2}{3} \, \delta(0)
                          [1 + \tau/\gamma(0)]^{-a} \, \left( 1 + \zeta_{\rm diss}^2 \right)^{-\frac{1+p}{2}} \, ,
\end{equation}
whose solution, for very large initial magnetic Reynolds number,
${\text{Re}}_B(0) \gg 1$, is given by Eq.~(\ref{zetadiss}).
Consequently, in the case $\zeta_\alpha/\sqrt{1 + \zeta_{\rm
diss}^2} \ll 1$, the magnetic energy and correlation length are
given by Eqs.~(\ref{Es}) and (\ref{xis}), respectively, to wit, the same
as in the non-helical case (see section IV a).
In the limits $\zeta_{\rm diss} \gg 1$ and
$\zeta_\alpha/\zeta_{\rm diss} \gg 1$, the expressions for the
magnetic energy, helicity, and correlation length,
Eqs.~(\ref{MF17}), (\ref{MF18}), and (\ref{xi}),
take the form
\begin{eqnarray}
\label{E1} && \frac{E_B(\tau)}{E_B(0)}  \simeq  (1+h_B) \,
\frac{\sqrt{\pi}}{2^{p/2} \, \Gamma[(1+p)/2]} \,
\frac{\zeta_\alpha^p}{\zeta_{\rm diss}^{1+2p}} \, \exp \! \left(
\frac{\zeta_\alpha^2}{2\zeta_{\rm diss}^2} \right) \! ,
\\
\label{H1} && \frac{H_B(\tau)}{H_B(0)}  \simeq  \frac{1+h_B}{h_B} \,
\frac{\sqrt{\pi}}{2^{(p-1)/2} \, \Gamma(p/2)} \,
\frac{\zeta_\alpha^{p-1}}{\zeta_{\rm diss}^{-1+2p}} \, \exp \!
\left( \frac{\zeta_\alpha^2}{2\zeta_{\rm diss}^2} \right) \! ,
\\
\label{xi1} && \frac{\xi_B(\tau)}{\xi_B(0)}  \simeq
\frac{2}{\zeta_B} \, \frac{\zeta_{\rm diss}^2}{\zeta_\alpha} \, ,
\end{eqnarray}
respectively, where we used the asymptotic expansion of the Kummer confluent
hypergeometric function $_1 F_1 (a,b;z)$ for $z \gg 1$~\cite{Gradshteyn},
$_1 F_1 (a,b;z) \sim [\Gamma(b)/\Gamma(a)] \, z^{a-b} e^z$,
valid in the case $a>0$. From Eqs.~(\ref{E1})-(\ref{xi1}), one directly obtains Eq.~(\ref{Exi0}).
Because the magnetic helicity is quasi-conserved, $H_B(\tau)
\simeq H_B(0)$, from Eq.~(\ref{H1}) we get
\begin{equation}
\label{exp} \exp \! \left( \frac{\zeta_\alpha^2}{2\zeta_{\rm diss}^2} \right)  \simeq  \frac{h_B}{1+h_B} \, \frac{2^{(p-1)/2} \,
\Gamma(p/2)}{\sqrt{\pi}} \, \frac{\zeta_{\rm diss}^{-1+2p}}{\zeta_\alpha^{p-1}} \, .
\end{equation}
Inserting the above equation in Eq.~(\ref{E1}), we get
\begin{equation}
\label{E2} \frac{E_B(\tau)}{E_B(0)}  \simeq  h_B \, \frac{\zeta_B}{2} \, \frac{\zeta_\alpha}{\zeta_{\rm diss}^2} \, .
\end{equation}
For $\tau \gg 1$ and from Eq.~(\ref{exp}), we have
$\zeta_\alpha \simeq \sqrt{2} \zeta_{\rm diss} [ (2p-1) \ln \zeta_{\rm diss} + (1-p) \ln \zeta_\alpha]^{1/2}$.
%
%
From the above expression we get, neglecting terms of order $\ln (\ln \zeta_{\rm diss})$,
\begin{equation}
\label{zeta-approx} \zeta_\alpha \simeq (2p \ln \zeta_{\rm diss})^{1/2} \zeta_{\rm diss}.
\end{equation}
Inserting Eq.~(\ref{zeta-approx}) in Eqs.~(\ref{E2}) and
(\ref{xi1}), we obtain
\begin{eqnarray}
\label{E3} && \frac{E_B(\tau)}{E_B(0)} \simeq  h_B \sqrt{\frac{p}{2}} \: \zeta_B \, \frac{(\ln \zeta_{\rm diss})^{1/2}}{\zeta_{\rm diss}} \, ,
\\
\label{xi3} && \frac{\xi_B(\tau)}{\xi_B(0)} \simeq \sqrt{\frac{2}{p}} \:\frac{1}{\zeta_B} \, \frac{\zeta_{\rm diss}}{(\ln\zeta_{\rm diss})^{1/2}} \,,
\end{eqnarray}
respectively. In the limit of very large initial magnetic Reynolds number, and
taking into account Eq.~(\ref{E3}), equation~(\ref{differential1}) reads
\begin{equation}
\label{differential-approx} \frac{d\zeta_{\rm diss}^2}{d\tau} =
\frac{\zeta_B^3}{3} \, \sqrt{\frac{p}{2}} \, \delta(0) [\gamma(0)]^a h_B \, \tau^{-a} \, \frac{(\ln \zeta_{\rm diss})^{1/2}}{\zeta_{\rm diss}} \, .
\end{equation}
Note that the above equation is valid only for $\tau \gg 1$.
Introducing, for convenience, the reference time $\tau_*$ such that
$\tau_* \gg 1$, the solution of Eq.~(\ref{differential-approx}) is
given by
\begin{equation}
\label{sol-differential-approx1}
\mbox{erfi}\,[\sqrt{3\ln\zeta_{\rm diss}(\tau)}\:] -
\mbox{erfi}\,[\sqrt{3\ln\zeta_{\rm diss}(\tau_*)}\:]
= \sqrt{\frac{3}{\pi}} \, \frac{\zeta_B^3}{6} \, \sqrt{\frac{p}{2}} \, \frac{\delta(0) [\gamma(0)]^a h_B}{1-a} \,
(\tau^{1-a} - \tau_*^{1-a}) \, ,
\end{equation}
where $\mbox{erfi}(z)$ is the imaginary error function
\cite{Gradshteyn}. By using the asymptotic expansion of the
imaginary error function for $z \gg 1$~\cite{Gradshteyn},
$\mbox{erfi}(z) \sim [1/(\sqrt{\pi} z)] \, e^{z^2}$,
we can write Eq.~(\ref{sol-differential-approx1}), for $\tau \gg \tau_*$, as
\begin{equation}
\label{sol-differential-approx2} \zeta_{\rm diss}(\tau) \simeq
\left\{ \frac{\zeta_B^3}{2} \, \sqrt{\frac{p}{2}} \, \frac{\delta(0) [\gamma(0)]^a h_B}{1-a} \right\}^{\!1/3}
(\ln \zeta_{\rm diss})^{1/6} \, \tau^{(1-a)/3}.
\end{equation}
Neglecting terms of order $\ln (\ln \tau)$, we can write the above
expression as
\begin{equation}
\label{sol-differential-approx3} \zeta_{\rm diss}(\tau) \simeq
\left\{ \frac{\zeta_B^3}{2} \, \sqrt{\frac{p}{6}} \, \frac{\delta(0) [\gamma(0)]^a h_B}{\sqrt{1-a}} \right\}^{\!1/3}
(\ln \tau)^{1/6} \,  \tau^{(1-a)/3},
\end{equation}
which is indeed Eq.~(\ref{expansion-diss}). Now, inserting the above equation in Eqs.~(\ref{zeta-approx}), (\ref{E3}),
and (\ref{xi3}), and neglecting terms of order $\ln (\ln \tau)$, we get
\begin{eqnarray}
\label{zeta-approx2} && \zeta_{\alpha}(\tau) \simeq
\left\{\frac{\zeta_B^3}{9} \, p^2 \delta(0) [\gamma(0)]^a h_B (1-a) \right\}^{\!1/3} (\ln \tau)^{2/3} \, \tau^{(1-a)/3},
\\
\label{E-approx2} && \frac{E_B(\tau)}{E_B(0)} \simeq \left\{ \sqrt{\frac{p}{3 \, \delta(0) [\gamma(0)]^a}} \, h_B (1-a) \right\}^{2/3}
(\ln \tau)^{1/3} \, \tau^{-(1-a)/3},
\\
\label{xi-approx2} && \frac{\xi_B(\tau)}{\xi_B(0)} \simeq
\left\{\sqrt{\frac{p}{3 \, \delta(0) [\gamma(0)]^a h_B}} \, (1-a) \right\}^{-2/3}
(\ln \tau)^{-1/3} \, \tau^{(1-a)/3},
\end{eqnarray}
respectively, which are indeed Eqs.~(\ref{expansion-alpha}), (\ref{expansion-Energy}), and (\ref{expansion-correlation}).


\section{Appendix C}

In this appendix, we derive Eq.~(\ref{tauSpectrum}), namely the expression for the time $t_{\mathcal{E}}$
when the inverse cascade begins.
In section Vd, we defined it as the time when
the maximum of the magnetic spectrum meets the initial spectrum, namely
$\mathcal{E}_{B}(k_{\rm max},t_{\mathcal{E}}) = \mathcal{E}_{B}(k_{\rm max},0)$,
%
%
where $k_{\rm max}$ is determined by the condition
$\partial \mathcal{E}_{B}(k_{\rm max},t)/\partial k = 0$.
%
%
Introducing the quantity $x = k \ell_B$, the above two conditions read
\begin{eqnarray}
\label{B4} && \cosh (2x_{\rm max} \zeta_\alpha) + h_B \sinh (2x_{\rm max} \zeta_\alpha) =
\exp (2 x_{\rm max}^2 \zeta_{\rm diss}^2), \\
\label{B5} && 2x_{\rm max} \zeta_\alpha [h_B \cosh (2x_{\rm max} \zeta_\alpha) + \sinh (2x_{\rm max} \zeta_\alpha)]
+ [p - 4x_{\rm max}^2 (1+\zeta_{\rm diss}^2)] \exp (2 x_{\rm max}^2 \zeta_{\rm diss}^2) = 0,
\end{eqnarray}
respectively, where $x_{\rm max} = k_{\rm max} \ell_B$, and all quantities depending on the time
are evaluated in $\tau = \tau_{\mathcal{E}}$. In the limit $h_B \ll 1$ and
$2x_{\rm max} \zeta_\alpha \gg 1$,
Eqs.~(\ref{B4}) and (\ref{B5}) become
$2x_{\rm max} \zeta_\alpha = \ln 2 + 2 x_{\rm max}^2 \zeta_{\rm diss}^2$
and $2x_{\rm max}^2 (1 + \zeta_{\rm diss}^2) = \ln 2 + p$,
%
%
respectively. In the limit $\zeta_{\rm diss}(\tau_{\mathcal{E}}) \gg 1$,
%
%
these equations read
\begin{eqnarray}
\label{B10} && 2x_{\rm max} \zeta_\alpha = 2\ln 2 + p, \\
\label{B11} && 2x_{\rm max}^2 \zeta_{\rm diss}^2 = \ln 2 + p,
\end{eqnarray}
respectively. From Eq.~(\ref{B10}), we get that the condition $2x_{\rm max} \zeta_\alpha \gg 1$
is indeed satisfied for $p > 1$.
Moreover, since for $\tau = \tau_{\mathcal{E}}$ we have from Eqs.~(\ref{B10})-(\ref{B11}) that
\begin{equation}
\label{B12} \frac{\zeta_\alpha}{\zeta_{\rm diss}} = \frac{2\ln2 + p}{\sqrt{2(\ln2 + p)}} \ll 1,
\end{equation}
we can use Eq.~(\ref{zetadissapprox}), valid for $\zeta_\alpha/\zeta_{\rm diss} \ll 1$ and
$\tau \gg 1$ (see section Vd), evaluated at $\tau = \tau_{\mathcal{E}}$, to wit
$\zeta_{\rm diss} = c_1^{1/2} \tau_{\mathcal{E}}^{(1-a)/(3+p)}$.
%
%
This equation is then valid for $\tau_{\mathcal{E}} \gg 1$.
Using Eq.~(\ref{MF18}) in the limits $H_B(t) = H_B(0)$, $\zeta_\alpha/\zeta_{\rm diss} \ll 1$,
and $\zeta_{\rm diss} \gg 1$, we get
$\zeta_\alpha/\zeta_{\rm diss} = h_B \zeta_B \, c_1^{p/2} \tau_{\mathcal{E}}^{p\,(1-a)/(3+p)}/2$.
%
%
Comparing this equation with Eq.~(\ref{B12}), we find
\begin{equation}
\label{B16} \tau_{\mathcal{E}} =
\left\{ \!{\zeta_B^{-\frac{3(1+p)}{3+p}}} \frac{\sqrt{2} (2\ln2 + p)}{\sqrt{\ln2 + p}}
\!\!\left\{ \frac{(3+p) \delta(0)[\gamma(0)]^a}{6(1-a)} \right\}^{\!-\frac{p}{3+p}}  \!\!\right\}^{\!2q/3}
h_B^{-2q/3} \!,
\end{equation}
which is indeed Eq.~(\ref{tauSpectrum}). Since $h_B \ll 1$, the conditions $\tau_{\mathcal{E}} \gg 1$ and
$\zeta_{\rm diss}(\tau_{\mathcal{E}}) \gg 1$ are satisfied.

Finally and for the sake of completeness, we give the expression for $k_{\rm max}$. Taking into account
Eq.~(\ref{B11}), we have
$k_{\rm max} \ell_B = x_{\rm max} = (\ln2 + p)^{1/2} [2 c_1 (c_{12})^{\frac{3}{pq}}]^{-1/2} \: h_B^{1/p}$.
%
%
Note that $k_{\rm max} \ell_B$ is a small quantity in the limit of small $h_B$.






\end{document}